\documentclass{article}

\usepackage{blindtext}
\usepackage[toc,page]{appendix}
\usepackage{graphicx}
\usepackage{amsfonts} 
\usepackage{epsf,  amsmath, amssymb, graphicx}
\usepackage{dsfont}
\usepackage{xcolor}
\usepackage{hyperref}
\usepackage{sectsty}
\usepackage{amsmath}
\usepackage{amsthm}
\theoremstyle{plain}
\usepackage{fancyhdr}
\usepackage{subfigure}
\usepackage{multirow}
\usepackage{float}
 \usepackage{multirow}
\usepackage{bm}
\usepackage{mathtools}
\usepackage{natbib}
\usepackage{color}
\usepackage{bbm}
\usepackage[linesnumbered,lined,boxed,commentsnumbered]{algorithm2e}

\newcommand{\iid}{\overset{i.i.d.}{\sim}}
\newtheorem{definition}{Definition}[section]
\newtheorem{remark}{Remark} [section]

\newtheorem{prop}{Proposition}[section]
\newtheorem{lemma}{Lemma}[section] 

\newtheorem*{proof*}{Proof}

\newcommand{\1}{{\bm{1}}}

\newcommand{\myvol}{{\text{vol}}}
\newcommand{\myvec}{{\text{vec}}}
\newcommand{\myTr}{{\text{Tr}}}

\usepackage{natbib}
\usepackage{url} 

\newcommand{\blind}{0}

\usepackage{subfiles} 

\begin{document}
\graphicspath{images/}

\def\spacingset#1{\renewcommand{\baselinestretch}%
{#1}\small\normalsize} \spacingset{1}


\if0\blind
{
  \title{\bf Bayesian Kernel Two-Sample Testing}
  \author{Qinyi Zhang \\
    \vspace{0.3cm}
    Department of Statistics, University of Oxford \\
    Veit Wild \\
    \vspace{0.3cm}
    Department of Statistics, University of Oxford \\
    Sarah Filippi \\
    \vspace{0.3cm}
    Department of Mathematics, Imperial College London \\
    Seth Flaxman \\
    \vspace{0.3cm}
    Department of Computer Science, University of Oxford \\
    Dino Sejdinovic \\
    Department of Statistics, University of Oxford}
  \maketitle
} \fi

\if1\blind
{
  \bigskip
  \bigskip
  \bigskip
  \begin{center}
    {\LARGE\bf Title}
\end{center}
  \medskip
} \fi

\bigskip
\begin{abstract}
In modern data analysis, nonparametric measures of discrepancies between random variables are particularly important. The subject is well-studied in the frequentist literature, while the development in the Bayesian setting is limited where applications are often restricted to univariate cases. Here, we propose a Bayesian kernel two-sample testing procedure based on modelling the difference between kernel mean embeddings in the reproducing kernel Hilbert space utilising the framework established by \cite{FlaSejCunFil2016}. The use of kernel methods enables its application to random variables in generic domains beyond the multivariate Euclidean spaces. The proposed procedure results in a posterior inference scheme that allows an automatic selection of the kernel parameters relevant to the problem at hand. In a series of synthetic experiments and two real data experiments (i.e. testing network heterogeneity from high-dimensional data and six-membered monocyclic ring conformation comparison), we illustrate the advantages of our approach. 
\end{abstract}

\noindent%
\vfill

\newpage
\spacingset{1.5} 

\maketitle

\section{Introduction}
Nonparametric two-sample testing is an important branch of hypothesis testing with a
wide range of applications. For a paired two-sample testing problem, the data set under consideration is $\mathcal{D} = \{\{x_i\}^{n}_{i =1}, \{y_j\}^{n}_{j =1}\}$, where $\{x_i\}^{n}_{i =1} \iid P_X$ and $\{y_j\}^{n}_{j =1}\iid P_Y$. We wish to evaluate the evidence for the competing hypotheses 
\begin{equation}
    H_0: P_X = P_Y \ \text{v.s.} \ H_1: P_X \not = P_Y
\end{equation}
with the probability distributions $P_X$ and $P_Y$ unknown. In this work, we will pursue a Bayesian perspective to this problem. In this perspective, hypotheses can be formulated as models and hypothesis testing can therefore be viewed as a form of model selection, i.e. to identify which model is strongly supported by the data \citep{JeffreysBerger1992}. 

The classical Bayesian formulation of the two-sample testing problem is in terms of the Bayes factor \citep{JeffreyBF, JeffreyBF2, KassRafteryBF}. For any given data set $\mathcal{D}$ and two competing models/hypotheses $H_0$ and $H_1$, the Bayes factor is represented as the likelihood ratio of the samples given that they were generated from the same distribution (null hypothesis) to that they were generated from different distributions (alternative hypothesis): 
\begin{equation}
    BF = \frac{P(\mathcal{D}| H_0)}{P(\mathcal{D} | H_1)} \label{eq: BFdef}\;.
\end{equation}

The Bayes factor can be interpreted as the posterior odds on the null distribution when the prior probability on the null distribution is $\frac{1}{2}$ \citep{KassRafteryBF}. 
If the posterior probability of the model given the data is of interest, it can be easily written in terms of the Bayes factor:
\begin{align}
    P(H_0 | \mathcal{D}) &= \frac{BF}{1+BF} \label{eq6: BFposterior0} \\
    P(H_1 | \mathcal{D}) &=1- P(H_0|\mathcal{D})= \frac{1}{1+BF}\;. \label{eq6: BFposterior1}
\end{align}
When the prior probabilities on the models are not equal, the posterior probability of the null model can be written as:
  \begin{equation}
    P(H_0|\mathcal{D}) = \frac{BF}{\frac{P(H_1)}{P(H_0)} + BF}
\end{equation}
where $P(H_0)$ and $P(H_1)$ denote respectively the prior for models $H_0$ and $H_1$.


Instead of considering the observations directly, we propose to work with the difference between the two distributions' mean embeddings in the RKHS. In the kernel literature, this quantity is proportional to the witness function of the (frequentist) two-sample test statistic known as Maximum Mean Discrepancy \citep[Definition 2]{Gretton2sample}. 

Inspired by the work of \cite{FlaSejCunFil2016} where the kernel mean embedding is modelled with a Gaussian process prior and a normal likelihood, we use a similar model for the difference between the kernel mean embeddings. Intuitively, to model the kernel mean embedding for $X$ (or for $Y$) directly with a GP prior is not ideal as kernel mean embeddings for a non-negative kernel (like widely used Gaussian or Mat{\'e}rn kernels) are never negative, but the draws from any GP prior can be negative.
Hence, it seems more suitable to place a prior directly on the difference between the mean embeddings -- as we do in this contribution. A further advantage of modelling the difference directly is that we will no longer require the independence assumption between the random variables $X$ and $Y$. Such assumption is common in two-sample testing literature both in the frequentist and the Bayesian setting. In particular, a frequentist two-sample test based on MMD requires such an assumption.
The remainder of the paper is structured as follows. Section \ref{sec6: related work} overviews existing approaches to Bayesian nonparametric two-sample testing. Section \ref{sec: kme} recaps the formalism behind the main ingredient of our test, embeddings of distributions into RKHS. Section \ref{sec: method} introduces the testing methodology, defining the relevant quantities of interest and detailing their distributions under the two hypotheses, and finally giving a Metropolis Hasting within Gibbs type of approach for inferring the posterior distribution of hyperparameters $\theta$ and that of the model given the observed data. Section \ref{sec6: experiments} studies the performance of the proposed method on various synthetic data experiments. Section \ref{sec6: real_experiments} presents the results on two real data experiments where we test network heterogeneity from high dimensional data in the first experiment and compare six-membered monocyclic ring conformation under two different conditions in the second experiment.
\section{Related Work}
\label{sec6: related work}

Bayesian parametric hypothesis testing, i.e. when the probability distributions $P_X$ and $P_Y$ are of known form, is well developed and we refer the readers to \cite{Bernardo_Smith_2000} for a clear description of the setting.

Most Bayesian nonparametric approaches for hypothesis testing have been focusing on testing a parametric model versus a nonparametric one and a detailed summary has been provided by \cite{holmes2015}. 
\cite{chen2014bayesian} and \cite{holmes2015} concurrently proposed Bayesian nonparametric two-sample tests using P{\'o}lya tree priors for both the distributions of the pooled samples $ P_{X,Y}$
under the null and for the individual distributions $P_X$ and $P_Y$ under the alternative. The two approaches mainly differ in terms of the specific modelling choices in these priors -- while \cite{chen2014bayesian} used a truncated P{\'o}lya tree, \cite{holmes2015} showed that the computation of the Bayes Factor could be done analytically with a non-truncated P{\'o}lya tree -- and their centering distributions. Note that the method proposed by \cite{holmes2015} is restricted to one-dimensional data whereas the method by \cite{chen2014bayesian} is described in a multivariate setting. 


\cite{Borgwardt2009} also discuss a nonparametric test using Dirichlet process mixture models (DPMM) and exponential families. However, to the best of our knowledge, this test does not appear to lend itself to a practicable implementation, due to intractability of marginalizing the Dirichlet process.

Utilising the Bayes factor as a model comparison tool, \cite{Borgwardt2010} used Gaussian processes (GP) to model the probability of the observed data under each model in the problem of testing whether a gene is differentially expressed.
The values of the hyperparameters (i.e. the kernel hyperparameters and the variance of the noise distribution of the GP model) were set to those that maximise the log posterior distribution of the hyperparameters. While this approach is for detecting the genes that are differentially expressed, they proposed a mixture type of approach for detecting the intervals of the time series such that the effect is present. A binary switch variable was introduced at every observation time point to determine the model that describes the expression level at that time point. Posterior inference of such variable is achieved through variational approximation. 

Bayesian nonparametric approaches have also been proposed for independence testing. In particular, \cite{Sarah_dep} extended the work by \cite{holmes2015} to perform independence testing using P{\'o}lya tree priors while \cite{Filippi_DPMM_2016} proposed to model the probability distributions using Dirichlet process mixture models (DPMM) with Gaussian distributions for pairwise dependency detection in large multivariate datasets. Though in theory the Bayes factor with DPMM on the unknown densities can be computed via the marginal likelihood, this requires integrating over infinite dimensional parameter space which results in an intractable form \citep{Filippi_DPMM_2016}. Hence the problem was reformulated using a mixture modelling approach proposed by \cite{Kamary2014} where hypotheses are the components of a mixture model and the posterior distribution of the mixing proportion is the outcome of the test.  While in this paper we focus on the classical Bayes factor formalism, it is an interesting direction of future work to study its extensions to the muxture modelling framework.



\section{Kernel Mean Embeddings}
\label{sec: kme}
Before introducing the proposed model, we will first review the notion of a reproducing kernel Hilbert space (RKHS) and the corresponding reproducing kernel. This will enable us to introduce the key concept for our method -- the kernel mean embedding. For more detailed treatment of the subject, we refer the readers to \cite{BerTho04, Steinwart2008book,Bharath}.

\begin{definition} \label{def: Reproducing}
\citep[Definition 4.18]{Steinwart2008book}
Let $\mathcal{Z}$ be any topological space on which Borel measures can be defined. 
Let $\mathcal{H}$ be a Hilbert space of real-valued functions defined on $\mathcal{Z}$. A function $k:\mathcal{Z} \times \mathcal{Z} \rightarrow \mathbb{R}$ is called a {\bf reproducing kernel} of $\mathcal{H}$ if: 
\begin{enumerate}
\item $\forall z \in \mathcal{Z}, k(\cdot,z) \in \mathcal{H}$,
\item $\forall z \in \mathcal{Z}, \forall f \in \mathcal{H}, \langle f,k(\cdot,z) \rangle_\mathcal{H} = f(z).$ (The Reproducing Property)
\end{enumerate}
If $\mathcal{H}$ has a reproducing kernel, it is called a {\bf Reproducing Kernel Hilbert Space} (RKHS). 
\end{definition}
 The element $k(\cdot,z)\in\mathcal H$ is known as the canonical feature of $z$. Reproducing kernels can be defined on graphs, text, images, strings, probability distributions as well as Euclidean domains \citep{ShaweCristianini2004}. For Euclidean domain $\mathbb{R}^d$, the Gaussian RBF kernel $k(x, y) = \exp \left (-\frac{1}{2\sigma^2 } \Vert x - y \Vert^2 \right )$ with lengthscale $\sigma > 0$ is an example of a reproducing kernel. 

Probability distributions can be represented as elements of a RKHS and they are known as the kernel mean embeddings \citep{BerTho04, probembedding}. This setting has been particularly useful in the (frequentist) nonparametric two-sample testing framework \citep{Borgwardt2006,Gretton2sample} since discrepancies between two distributions can be written succinctly as the square Hilbert-Schmidt norm between their respective kernel mean embeddings. More formally, kernel mean embeddings can be defined as follows. 

\begin{definition} \label{def2: kernel_embedding}
Let $k$ be a kernel on $\mathcal{Z}$, and $\nu \in \mathcal{M}^1_+ (\mathcal{Z})$ with $\mathcal{M}^1_+ (\mathcal{Z})$ denoting the set of Borel probability measures on $\mathcal{Z}$. The {\bf kernel embedding} of probability measure $\nu$ into the RKHS $\mathcal{H}_k$ is $\mu_k (\nu) \in  \mathcal{H}_k$ such that 
\begin{equation}
\int f(z) d\nu(z) = \  \langle f,\mu_k(\nu) \rangle _{\mathcal{H}_k}, \  \forall f \in \mathcal{H}_k.
\end{equation}
\end{definition}
In other words, the kernel mean embedding can be written as $\mu_k (\nu) = \int k(\cdot, z) d\nu(z)$, i.e. any probability measure is mapped to the corresponding expectation of the canonical feature map $k(\cdot, z)$ through the kernel mean embedding. When the kernel $k$ is measureable on $\mathcal{Z}$ and $\mathbb{E}(\sqrt{k(z,z)}) < \infty$, the existence of the kernel mean embedding is guaranteed \citep[Lemma 3]{Gretton2sample}. Further, \cite{Fukumizu08kernelmeasures} showed that when the corresponding kernels are characteristic, the mean embedding maps are injective and hence preserve all information of the probability measure. An example of a characteristic kernel is the Gaussian kernel on the entire domain of $\mathbb{R}^d.$

\section{Proposed Method}
\label{sec: method}
Consider a paired data set $\mathcal{D} = \{(x_i, y_i)\}^n_{i=1}$ with $x_i, \ y_i \in \mathcal{X}$ for some generic domains $\mathcal{X}$. Further, let $x_i \iid P_X$ and $y_i \iid P_Y$ for some unknown distributions $P_X$ and $P_Y$. Let $k_\theta(\cdot, \cdot)$ be a positive definite kernel parameterised by $\theta$, with the corresponding reproducing kernel Hilbert space $\mathcal{H}$.

We wish to evaluate the evidence for the competing hypotheses 
\begin{equation}
    H_0: P_X = P_Y \ \text{v.s.} \ H_1: P_X \not = P_Y\;.
\end{equation} 
In this work, we develop a Bayesian two-sample test based on the difference between the kernel mean embeddings. We consider the empirical estimate of such difference evaluated at a set of locations and propose a Bayesian inference scheme so that the relative evidence in favour of $H_0$ and $H_1$ is quantified. The proposed test is conditional on the choice of the family of kernels parameterised by $\theta$. We focus on working with the Gaussian RBF kernel in this contribution, but other kernels are readily applicable to the framework developed here. To emphasize the dependence of the kernel function on the lengthscale parameter $\theta$, we write $k_\theta (\cdot, \cdot)$. Denote the respective kernel mean embeddings for $X$ and $Y$ as
\begin{align}
    \mu_X &= \mathbb{E}_X(k_\theta(\cdot,X)) = \int_\mathcal{X} k_\theta(\cdot, x) P_X (dx), \\
    \mu_Y &= \mathbb{E}_Y(k_\theta(\cdot,Y)) = \int_\mathcal{Y} k_\theta(\cdot, y) P_Y (dy),
\end{align}
with the empirical estimators and the corresponding estimates denoted by
\begin{align}
    \widehat \mu_X &= \frac{1}{n} \sum^n_{i =1} k_\theta(\cdot, X_i) \qquad \text{and} \qquad \widehat \mu_x = \frac{1}{n} \sum^n_{i =1} k_\theta(\cdot, x_i), \\
    \widehat \mu_Y &= \frac{1}{n} \sum^n_{i =1} k_\theta(\cdot, Y_i) \qquad \text{and} \qquad \widehat \mu_y = \frac{1}{n} \sum^n_{i =1} k_\theta(\cdot, y_i). 
\end{align}
We denote the witness function up to proportionality as $\delta = \mu_X - \mu_Y$, which is simply the difference between the kernel mean embeddings. Under the null hypothesis, the two distributions $P_X$ and $P_Y$ are the same and with the use of characteristic kernels, all information of the probability distribution is preserved through the kernel mean embeddings $\mu_X$ and $\mu_Y$. Hence, the null hypothesis is equivalent to $\delta \equiv 0 $ and the alternative is equivalent to $\delta \not = 0$.

 Given a set of evaluation points ${\bf z} = \{z_i\}_{i=1}^s \in \mathcal{X}$, we define the evaluation of $\delta$ at {\bf z} as 
\begin{align}
    \delta(z_i) &= \mu_X(z_i) - \mu_Y(z_i)\\
    &= \mathbb{E}_X(k_\theta (z_i, X)) -  \mathbb{E}_Y(k_\theta (z_i, Y)), \quad \forall \ i = 1, \cdots, s\\
    \delta({\bf z}) &= (\delta(z_1), ... \delta(z_s)))^{\top} \in \mathbb{R}^s.
\end{align}
\noindent Such evaluations $\delta({\bf z})$ will act as the quantity of interest 
of our proposed model, while the empirical estimate of $\delta({\bf z})$ on a given set of data $\mathcal{D} = \{(x_i, y_i)\}^n_{i =1}$ will be regarded as the observations. This will be made precise in the following sections. Ideally, half of the evaluation points ${\bf z} = \{z_i\}^s_{i=1}$ are sampled from $P_X$, while the other half are sampled from $P_Y$. When direct sampling is not possible (e.g. when we have access to the distributions only through samples), the evaluation points are subsampled from the given data set. We define the $s$-dimensional witness vector $\Delta$ as the empirical estimate of $\delta({\bf z})$:
\begin{align}
\Delta &= \{\widehat \mu_x (z_j) - \widehat \mu_y (z_j)\}^s_{j = 1}  \\
&= \left \{ \frac{1}{n} \sum_{i =1}^{n} k_\theta (x_i, z_j) - \frac{1}{n} \sum_{i=1}^{n} k_\theta(y_i, z_j) \right \}^s_{j = 1}  \\
&= \left \{ \frac{1}{n}\sum_{i = 1}^{n} (k_\theta (x_i, z_j) - k_\theta (y_i, z_j))  \right \}^s_{j=1}. 
\end{align}
and the corresponding random variable as $\Delta_{XY}:= \{\widehat \mu_X (z_j) - \widehat \mu_Y (z_j)\}^s_{j=1}.$ 

Following the classical Bayesian two-sample testing framework, we will quantify the evidence in favour of the two samples coming from the same distribution vs different distributions through Bayes factor:
\begin{align}
    BF_\theta = \frac{P(\Delta |H_0,\theta)}{P(\Delta |H_1,\theta)}\label{eq6: BF_Delta}
\end{align}
where $P(\Delta |H_0,\theta)$ and  $P(\Delta |H_1,\theta)$ are the marginal likelihood of $\Delta$ under each hypothesis for a given kernel hyperparameter $\theta$.


In sections \ref{sec6: AlterModel} and \ref{sec6: NullModel}, we describe how to compute $P(\Delta |H_0,\theta)$ and  $P(\Delta |H_1,\theta)$ for fixed kernel hyperparameter $\theta$. Similarly to the Bayesian kernel embeddings approach of \cite{FlaSejCunFil2016}, we propose to model $\delta$ with a Gaussian process (GP) prior under the alternative model. 
Assuming a Gaussian noise model, we derive the marginal likelihood of $\Delta$ for fixed kernel hyperparameter $\theta$.  Under the null hypothesis, the model simplifies significantly due to $\delta\equiv 0$, so we only need to pose a Gaussian noise model for $\Delta| \theta$. 

When the kernel hyperparameter is unknown, the framework of \cite{FlaSejCunFil2016} enables the derivation of the posterior distribution of the hyperparameter given the observations. This, however, requires heavy computation burden due to the need to compute the marginal likelihood of the dataset $\{(x_i, y_i)\}^n_{i=1} | \theta $. Hence, we propose an alternative formulation of the likelihood utilising the Kronecker product structure of our problem, as presented in the Supplementary Material.


\subsection{Alternative Model}
\label{sec6: AlterModel}
Under the alternative hypothesis, $\delta = \mu_X - \mu_Y \not = 0$. We propose to model the unknown quantity $\delta$ using a Gaussian Process (GP) prior. Draws from a naively defined prior $\mathcal{GP}(0, k_\theta (\cdot, \cdot))$ would almost surely fall outside of the RKHS $\mathcal{H}$ that corresponds to $k_\theta (\cdot, \cdot)$ \citep{Wahba1990}. Hence, \cite{FlaSejCunFil2016} proposed to define the GP prior as 
\begin{align}
    \delta| \theta \sim \mathcal{GP}(0, r_\theta(., .))
\end{align}
 with the covariance operator $r_\theta(.,.)$ defined as
 \begin{align}
     r_\theta (z,z') := \int k_\theta (z,u) k_\theta (u,z') \nu (du)
 \end{align}
 where $\nu$ is any finite measure on $\mathcal{X}$. Using results from \cite{LuckicBader2001} and Theorem 4.27 of \cite{Steinwart2008book}, \cite{FlaSejCunFil2016} showed that such choice of $r_\theta$ ensures that $\delta \in \mathcal{H}$ with probability 1 by the nuclear dominance for $k_\theta$ over $r_\theta$ for any stationary kernel $k_\theta$ and more generally $\int k_\theta (x,x) \nu (dx) < \infty$. Intuitively, the new covariance operator $r_\theta$ is a smoother version of $k_\theta$ since it is the convolution of $k_\theta$ with itself with respect to a finite measure $\nu$ \citep{FlaSejCunFil2016}. For our particular choice of $k_\theta$ being a Gaussian RBF kernel on $\mathcal{X} = \mathbb{R}^D$, \cite{FlaSejCunFil2016} showed (in A.3) that the covariance function $r_\theta$ of square exponential kernels 
\begin{align}
    k_\theta (x,y) = \exp \left (- \frac{1}{2} (x-y)^\top \tilde \Sigma^{-1}_\theta (x-y) \right )
\end{align}
with $x, y \in \mathbb{R}^D$  and diagonal covariance $\tilde \Sigma_\theta = \theta I = (\theta^{(1)}, \cdots, \theta^{(D)})^\top I$ can be written as 
\begin{align}
    r_\theta (x,y) = \pi^{D/2} \left (\prod^D_{d=1} \theta^{(d)} \right )^{1/2} \exp \left (- \frac{1}{2} (x-y)^\top (2\theta I_D)^{-1} (x-y) \right ).
\end{align}
When $\theta^{(1)} = \cdots = \theta^{(D)} = \theta$, the above can be further simplified as 
\begin{align}
    r_\theta (x,y) = \pi^{D/2} \theta^{D/2} \exp \left ( - \frac{1}{4 \theta} (x-y)^\top (x-y) \right ).
\end{align}
We will use this form of the covariance function in our experiments.

Given the set of evaluation points $\{z_j\}^s_{j=1}$, the prior translates into a $s$-dimensional multivariate Gaussian distribution
\begin{align}
\delta({\bf z})| \theta &\sim N(0, R_\theta) \label{eq: deltaNull}
\end{align}
with $[R_\theta]_{ij} = r_\theta(z_i, z_j)$. We link the empirical estimate $\Delta$ with the true differences $\delta$ evaluated at this set of evaluation points through a Gaussian likelihood of the model. This is an approximation of the true likelihood which hinges on the common ``Gaussianity in the feature space" assumption in the kernel method literature and is also utilised in \cite{FlaSejCunFil2016}. We write it as 
\begin{align}
\Delta | \delta, \theta \sim \mathcal{N} ([\delta(z_1), ..., \delta(z_s)]^{\top}, \frac{1}{n}\Sigma_\theta).
\end{align}

We details two ways to estimate $\Sigma_\theta$ empirically in the Supplementary Material. The constant $\frac{1}{n}$ is for notational convenience to be seen later. Integrating out the prior distribution of $\delta$, we obtain the marginal likelihood
\begin{align}
\Delta |  \theta \sim \mathcal{N}(0_s, R_\theta + \frac{1}{n}\Sigma_\theta).
\end{align}

When the kernel hyperparameter $\theta$ is unknown, we use the framework of \cite{FlaSejCunFil2016} to derive the posterior distribution of $\theta$ given the observation. This requires  access to the marginal \textit{pseudolikelihood} $p(\theta | \{(x_i, y_i)\}^n_{i=1}$. The term ``pseudolikelihood" is used since it relies on the evaluation of the empirical embedding at a finite set of inducing points and hence it is an approximation to the likelihood of the infinite dimensional empirical embedding \citep{FlaSejCunFil2016}. The derivation of the marginal pseudolikelihood is detailed in the supplementary materials \ref{sec6: EfficientCompute}.


 Although the derivations presented in this work follow essentially the same steps as \cite{FlaSejCunFil2016}, it is important to note that different from \cite{FlaSejCunFil2016}, we model the difference between the empirical mean embeddings of the two distributions of interest rather than the embedding of a single distribution. This has several implications. As discussed in \cite{FlaSejCunFil2016}, the marginal pseudolikelihood involves the computation of the inverse and the log determinant of an $ns$-dimensional matrix. A naive direct implementation would require a prohibitive computation of $\mathcal{O}(n^3s^3).$ Since we consider the difference between the empirical mean embeddings, the efficient computation utilising eigendecompositions of the relevant matrices  \citep[A.4]{FlaSejCunFil2016} cannot be applied directly. Fortunately, the special form of the corresponding $ns \times ns$ covariance matrix allows faster computation following Kronecker product algebra, the applications of matrix determinant lemma and Woodbury identity. This is detailed in the supplementary materials \ref{sec6: EfficientCompute}. Utilising the proposed efficient computation, the log marginal pseudolikelihood can be written as 
\begin{align}
    \log p(\{(x_i, y_i)\}^n_{i =1}| \theta)
    & \propto -\frac{1}{2} \log \det (\Sigma_\theta + n R_\theta) -\frac{1}{2} (n -1) \log \det (\Sigma_\theta) \nonumber \\
    &  \qquad -\frac{1}{2} Tr\left(\left(\Sigma_\theta +nR_\theta \right)^{-1}G_\theta G_\theta^{\top}+\left(\frac{1}{n}\Sigma_\theta R_\theta^{-1}\Sigma_\theta +\Sigma_\theta \right)^{-1}G_\theta HG_\theta^{\top}\right) \nonumber \\
    &  \qquad + \frac{1}{2} \sum^n_{i=1} \log \det (J_\theta (x_i, y_i)^\top J_\theta (x_i, y_i)). \label{eq6: loglik_altereff}
\end{align}

\subsection{Null Model}
\label{sec6: NullModel}
In the null model, calculations simplify, as we assume that $\delta=0$ holds. We propose to model the data directly with a Gaussian noise model, i.e. 
\begin{align}
\Delta | \theta \sim \mathcal{N}(0, \frac{1}{n} \Sigma_{\theta}),
\end{align}
where as before, we rewrite the covariance matrix as $\frac{1}{n} \Sigma _\theta$. 
In the supplementary materials \ref{sec6: EfficientCompute} we detail the derivation of the marginal pseudolikelihood in the null model and show that it can be written as
\begin{align}
    \log p(\{x_i, y_i\}^n_{i =1}|\theta) 
    \propto  -0.5n \log(\det(\Sigma_\theta)) -0.5\myTr(\Sigma_\theta^{-1}GG^\top) +  \sum^n_{i =1} \log \myvol(J_\theta(x_i, y_i)) \label{eq6: loglik_nulleff},
\end{align}
which avoids the prohibitive costs $\mathcal{O}(n^3s^3)$ of a naive evaluation.

\subsection{Posterior Inference}
\label{sec6: PosteriorInference}

When the kernel hyperparameter parameter $\theta$ is fixed, the computation of the posterior distribution $P(H_1|\mathcal{D})$ is straightforward. However, a wrong choice of the kernel hyperparameter can hurt the performance of the proposed Bayesian test (examples of which are presented in the Supplementary Material). Therefore we treat the parameter $\theta$ in a Bayesian manner and assign a Gamma(2,2) prior (under both model).
We propose to use a Metropolis Hasting within Gibbs type of approach for the joint posterior inference of $M\in\{H_0,H_1\}$ and $\theta$. In other words: We sample from $p(M, \theta | \mathcal{D})$ by sampling from $p(\theta| M, \mathcal{D}) $ and $p(M| \theta, \mathcal{D})$ iteratively. We can sample from  $p(\theta| M, \mathcal{D}) \propto p(\mathcal{D}| \theta, M) p(\theta) $ using No U-Turn Hamiltonian Monte Carlo (HMC) \citep{hoffman2014no}, since we know the marginal pseudolikelihood under $H_0$ and $H_1$ up to a constant (cp. section \ref{sec6: AlterModel} and \ref{sec6: NullModel}). To sample from $p(M| \theta, \mathcal{D})$, recall the relationship between Bayes factor and the posterior distribution of the null and alternative model respectively
\begin{align}
    p(H_1|\theta, \mathcal{D}) &= \frac{1}{1+BF_\theta}, \nonumber \\
    p(H_0| \theta, \mathcal{D}) & = \frac{BF_\theta}{1+BF_\theta} \nonumber
\end{align}
under the assumption that $P(H_0)=P(H_1)$.
We present the pseudocode of our posterior inference procedure in Algorithm \ref{alg6: MHGibbs}. The time complexity of Algorithm \ref{alg6: MHGibbs} is given as $\mathcal{O}\big(\tilde{m} \tilde{n} ( s^3+s^2n+D^2s + n^2s) \big) $.

We observe from our experiments, that increasing the number of HMC steps inside Gibbs $\tilde n$ improves the posterior convergence of the chain. The posterior marginal probability $P(\theta |\mathcal{D})$ is approximated by the posterior MCMC samples $\{\theta_1, \cdots, \theta_{\tilde m}\}$. Similarly, the posterior marginal probability $P(H_1 | \mathcal{D})$ can be estimated by the proportion of $M=H_1$ in all the samples $\{M_0, \cdots, M_{\tilde m}\}.$

\begin{algorithm}
\SetKwData{Left}{left}\SetKwData{This}{this}\SetKwData{Up}{up}
\SetKwInOut{Input}{Input}\SetKwInOut{Output}{Output}
\KwData{A paired sample $\mathcal{D} = \{x_i, y_i\}^n_{i=1}$; The number of inducing points $m$; The number of simulations $\tilde m$; The number of HMC steps $\tilde n$.}
\Output{A sample $\{\theta_i, M_i\}^{\tilde m}_{i=1}$ from the posterior distribution of $p(\theta, M | \mathcal{D})$.}
 Initialise $\theta_0$ = Median heuristic on the set $\{x_1, ..., x_n, y_1, ..., y_n\}$\;
Compute $BF_{\theta_0}$ and let $M_0=H_0$ with probability $(1+BF_{\theta_0})^{-1}$ and $M_0=H_1$ otherwise\;
 \For{ $i \leftarrow 1$ \KwTo $\tilde m$}{
    Simulate a chain $\{\theta_1, ... \theta_{\tilde n}\}$ from $p(\theta | M_{i-1}, \mathcal{D})$ using NUTS in Stan \citep{carpenter2017stan} \;
    %
    Set $\theta_i = \theta_{\tilde n}$\;
    Compute $BF_{\theta_i}$\;
    Let $M_i=H_0$ with probability $(1+BF_{\theta_i})^{-1}$ and $M_i=H_1$ otherwise.
    }
 \BlankLine
 \caption{Posterior inference of the kernel hyperparameter $\theta$ and the hypothesis $M\in\{H_0,H_1\}$.}
 \label{alg6: MHGibbs}
\end{algorithm}

\section{Synthetic Data Experiments} \label{sec6: experiments}
In this section, we investigate the performance of the proposed posterior inference scheme for $M$ and $\theta$ on synthetic data experiments. For each of the synthetic experiments, we generate 100 independent data sets of size $n$. We examine the distribution of the probability of the alternative hypothesis (i.e. $p(H_1| \mathcal{D})$) while varying the number of observed data points $n$. The number of evaluation points is fixed at $s = 40$, with half sampled from the distribution of $X$ and the other half sampled from the distribution of $Y$. 

For the posterior sampling, we run the algorithm for $\tilde m = 2000$. The initial 500 samples are discarded as burn-in and the thinning factor is 2. For every Gibbs sampling step, we take 9 steps in HMC which contains 3 warmup steps for step size adaption. Note, we have experimented with 1 HMC step for every step of Gibbs sampling, the convergence of the parameters $M$ and $\theta$ is much slower in that case. On the other hand, increasing the number of HMC steps beyond 9 does not seem to improve the performance by a significant amount. We used 9 steps for a balance between computational complexity and performance.

\subsection{Simple 1 Dimensional Distributions}
\subsubsection{Gaussian Distributions} 
\begin{figure}[h]
\begin{minipage}{0.45\textwidth}
    \centering
    \begin{subfigure}{}
        \includegraphics[width=\textwidth]{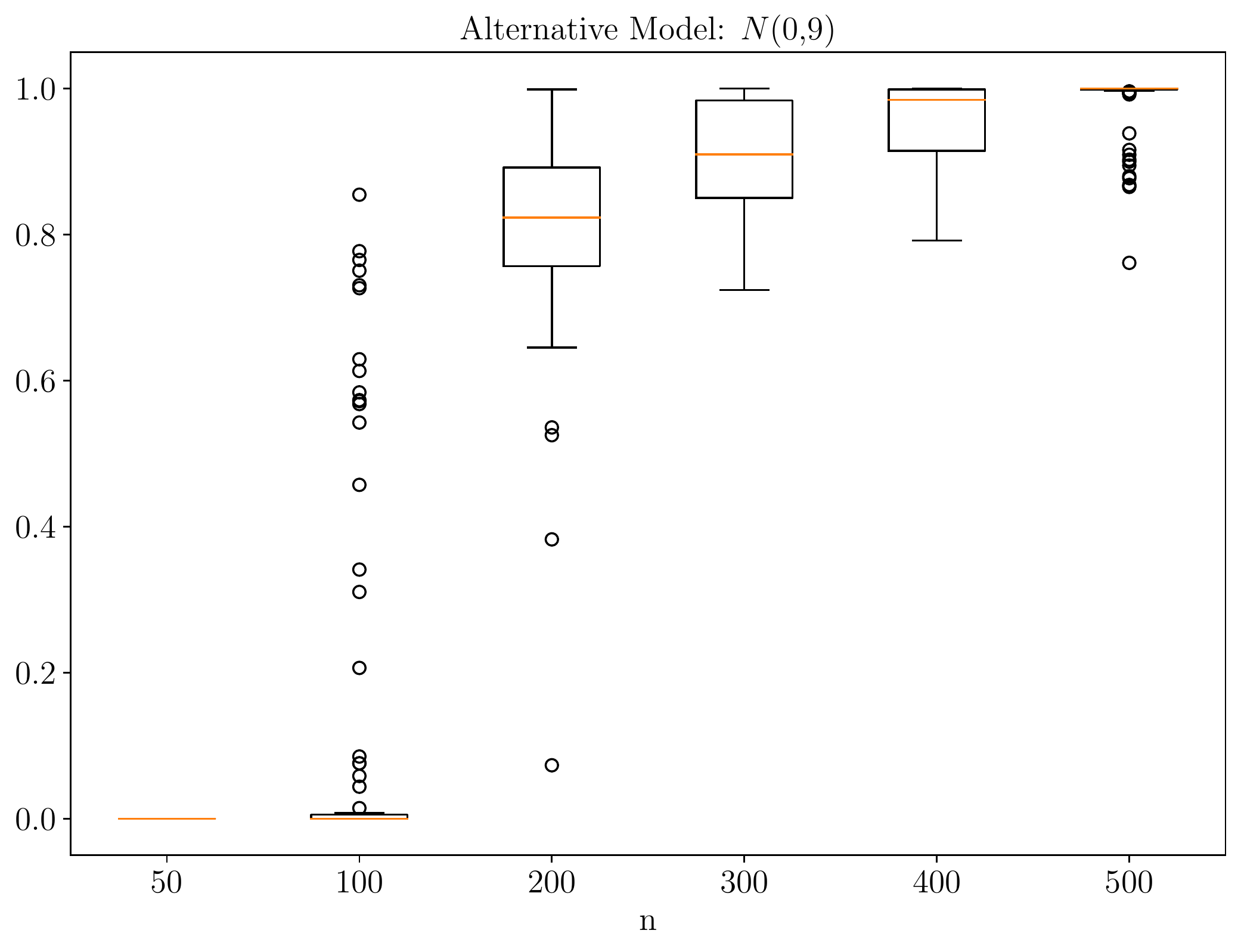}
    \end{subfigure}
\end{minipage}\hfill
\begin{minipage}{0.45\textwidth}
    \centering
    \begin{subfigure}{}
        \includegraphics[width=\textwidth]{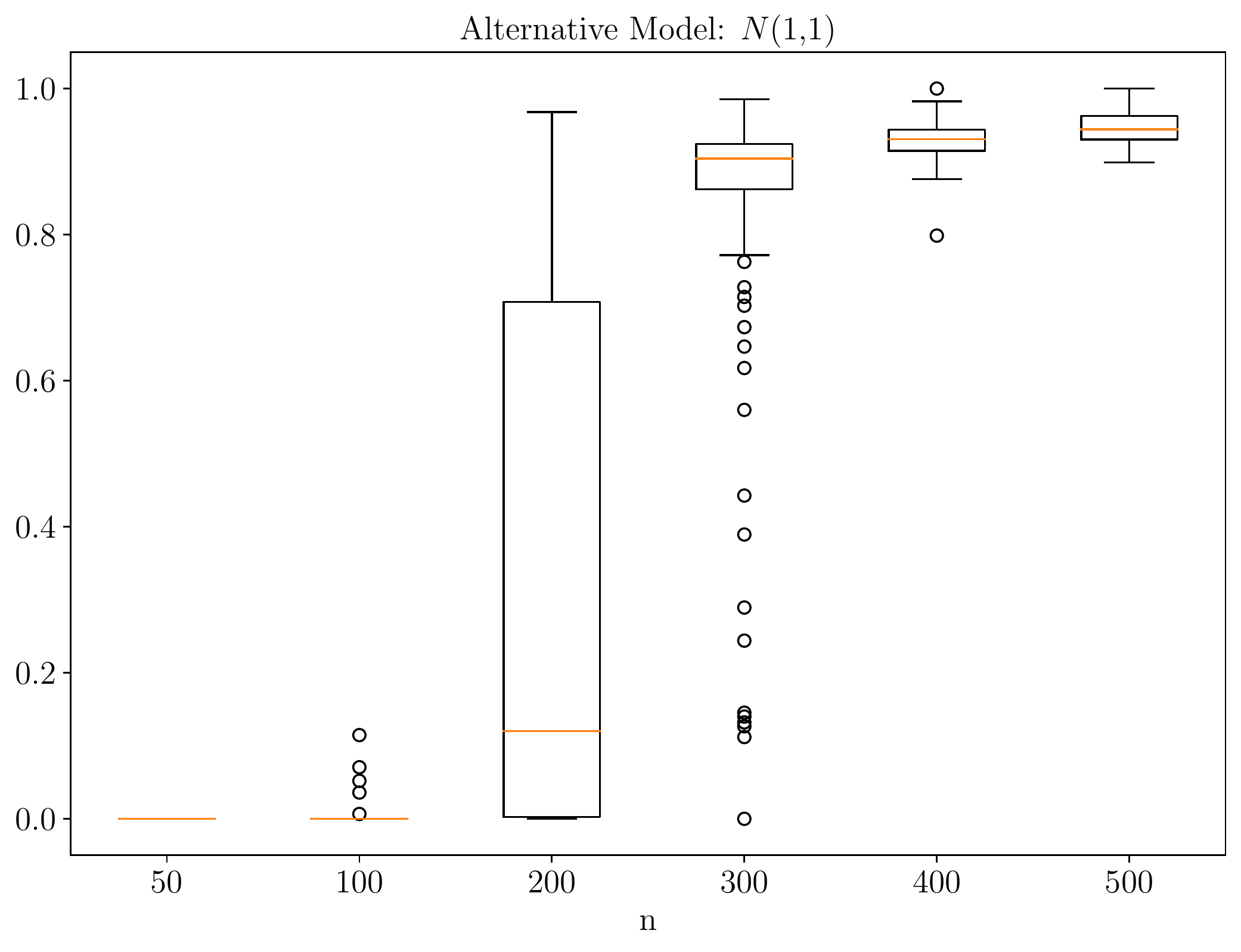}
    \end{subfigure}
\end{minipage}
\caption{1-dimensional Gaussian experiment: distribution (over 100 independent runs) of the probability of the alternative hypothesis $p(H_1|\mathcal{D})$ for a different number of observations $n$. Here, $P_X=\mathcal{N}(0,1)$ and $P_Y=\mathcal{N}(0, 9)$ (Left) or $P_Y=\mathcal{N}(1,1)$ (Right). }
\label{fig6: 1DGaussian}
\end{figure}

This section investigates if the proposed method is able to detect the change in mean or variance of simple 1-dimensional Gaussian distributions. We present the results of two cases in Figure \ref{fig6: 1DGaussian} when $X\iid P_X$ and $Y\iid P_Y$ where $P_X =\mathcal{N}(0,1)$ and $P_Y=\mathcal{N}(0,9)$ or $P_Y=\mathcal{N}(1,1).$ The null case and some other alternative cases are presented in the Supplementary Material for the interested reader.
We observe that for small sample sizes (50 or 100 samples) the posterior probability of $H_1$ is close to zero. As the sample size increases,  the posterior probability of $H_1$ concentrates towards a value close to 1.
This phenomena is observed for all the alternative models considered. 

Essentially, when the number of samples is small, there is not enough evidence to determine if the null hypothesis should be rejected. In such a case, the Bayes factor favours the simpler null hypothesis. This is to be expected since Bayesian modelling encompasses a natural Occam factor in the prior predictive \citep[Chapter 28]{Mackay2003}. We will see this reflected in all the synthetic experiments in this section. Note that for a Gaussian distribution, the difference in mean is easier to detect comparing to a difference in variance. This is reflected in the results presented here and in the Supplementary Material: we observe that the probability of the alternative hypothesis becomes very close to 1 at a much smaller sample size for the experiments with a difference in mean. 

\begin{figure}[h]
\begin{minipage}{0.45\textwidth}
    \centering
    \begin{subfigure}{}
        \includegraphics[width=\textwidth]{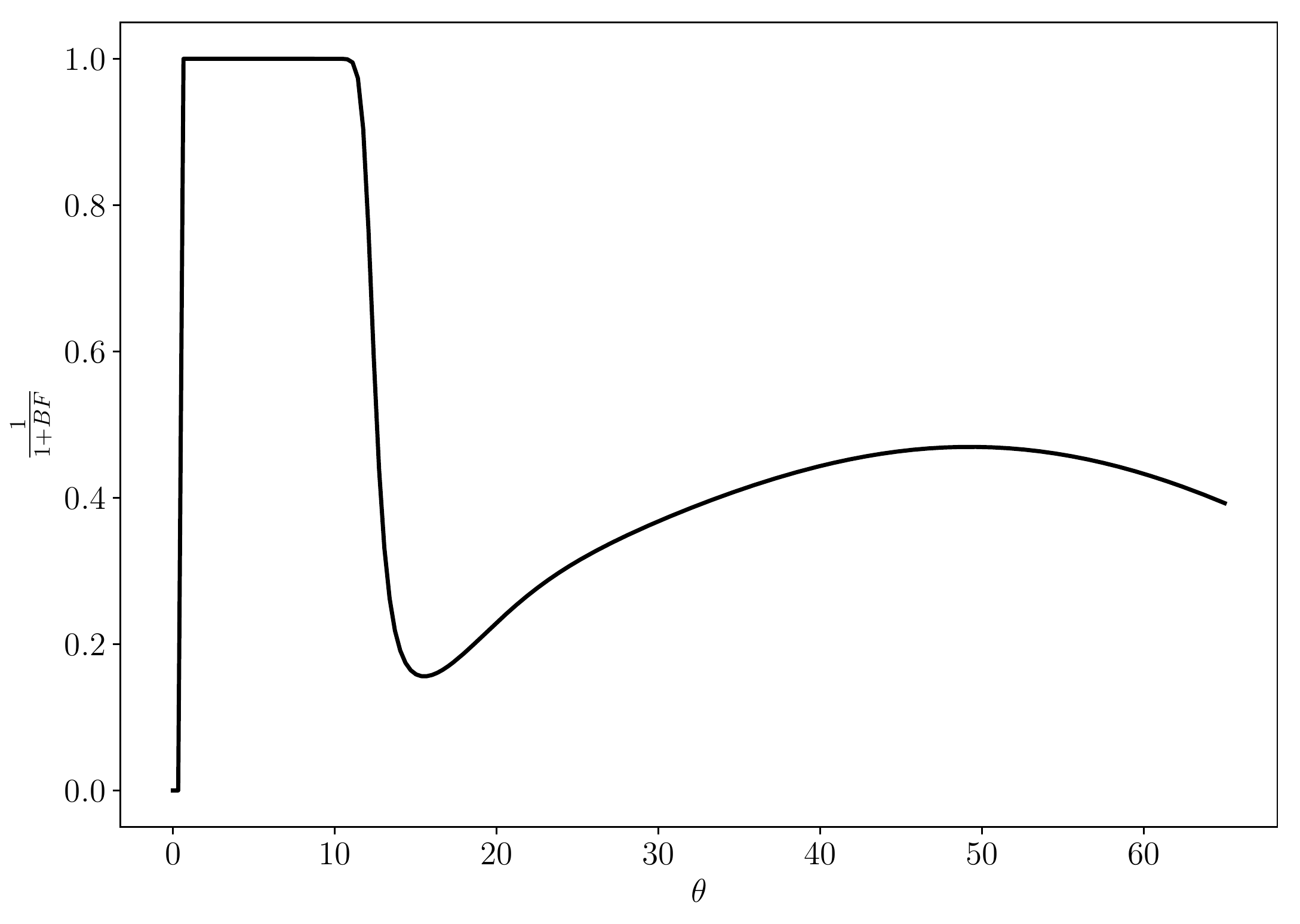}
    \end{subfigure}
\end{minipage}\hfill
\begin{minipage}{0.43\textwidth}
    \centering
    \begin{subfigure}{}
        \includegraphics[width=\textwidth]{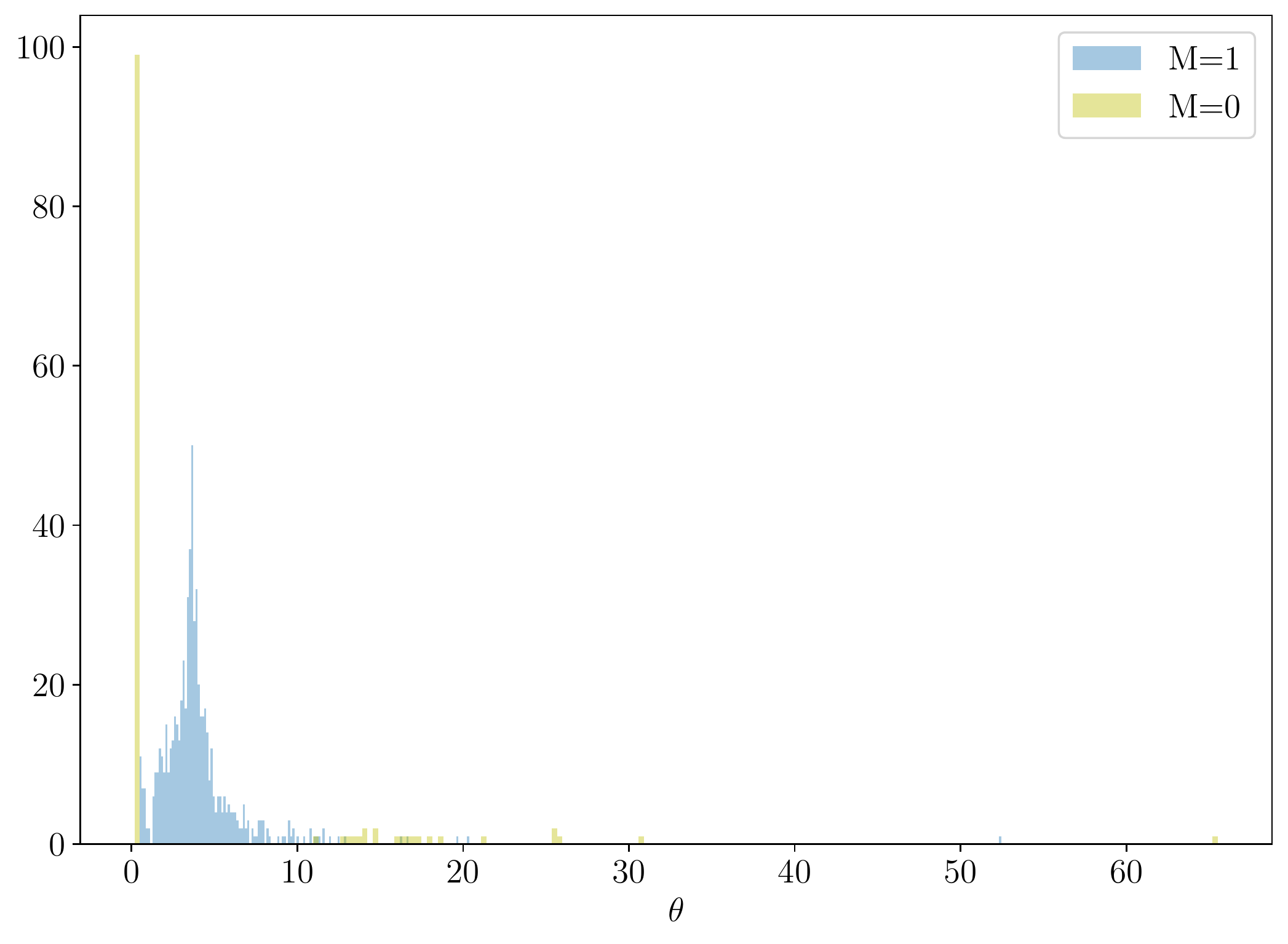}
       
    \end{subfigure}
\end{minipage}
 \caption{1-dimensional Gaussian experiment with for $P_X=\mathcal{N}(0,1)$ and $P_Y= \mathcal{N}(0,9)$ and 200 samples. Left: The plot illustrates $\frac{1}{1 + BF_\theta}$ as a function of $\theta$. Right: The histogram of $\theta | M, \mathcal{D}$ for $H_1$ and $H_0$.}
        \label{fig6: 1DGaussian_Marginal}
\end{figure}

We emphasize that, unlike the frequentist kernel two-sample test where a single value of the lengthscale parameter needs to be predetermined, the proposed Bayesian framework integrates over all possible $\theta$ values and alleviates the need for kernel lengthscale selection. However, some $\theta$ values are more informative in distinguishing the difference between the two distributions while others are less informative. As a specific example, we consider the case when $P_X=\mathcal{N}(0,1)$ and $P_Y= \mathcal{N}(0, 9)$ with 200 samples each. For this specific simulation, we obtain $P(H_1 | \mathcal{D})\approx 0.839$. Figure \ref{fig6: 1DGaussian_Marginal} (Left) illustrates the change of the probability of $H_1 | \theta, \mathcal{D}$ as a function of $\theta$. Clearly, the region of $\theta$ from approximately 0.05 to 11 is most informative for distinguishing these two distributions. This is also reflected in the marginal distribution of $\theta|H_1$ and $\theta|H_0$ from Figure \ref{fig6: 1DGaussian_Marginal} (Right). Rather than selecting a single lengthscale parameter, the proposed method is able to highlight the range of informative lengthscales. As we will see, this is more useful in cases when multiple lengthscale parameters are of interest for a single testing problem. 

\subsubsection{Laplace Distributions}
\begin{figure}[h]
\begin{minipage}{0.45\textwidth}
    \centering
    \begin{subfigure}{}
        \includegraphics[width=\textwidth]{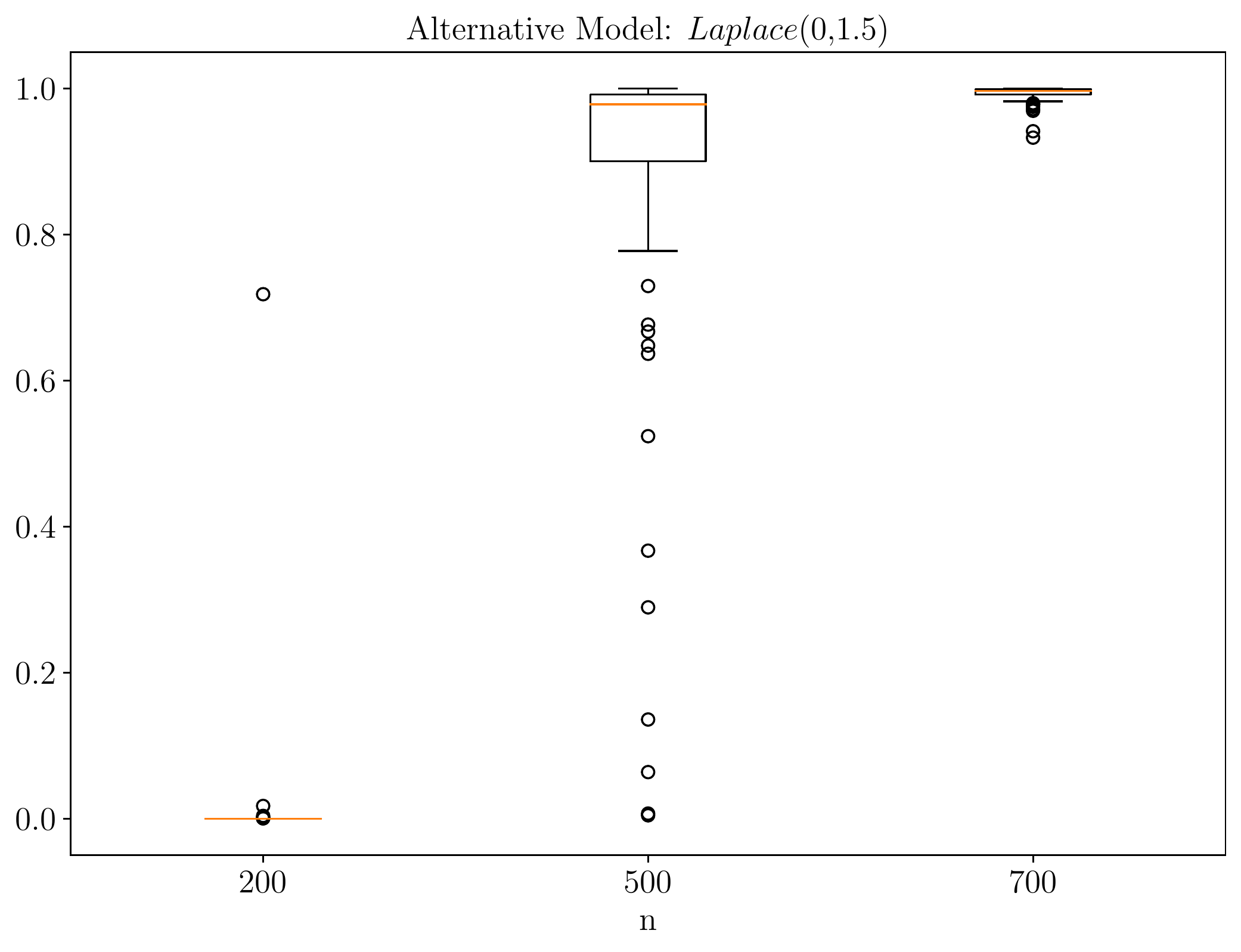}
    \end{subfigure}
\end{minipage}\hfill
\begin{minipage}{0.43\textwidth}
    \centering
    \begin{subfigure}{}
        \includegraphics[width=\textwidth]{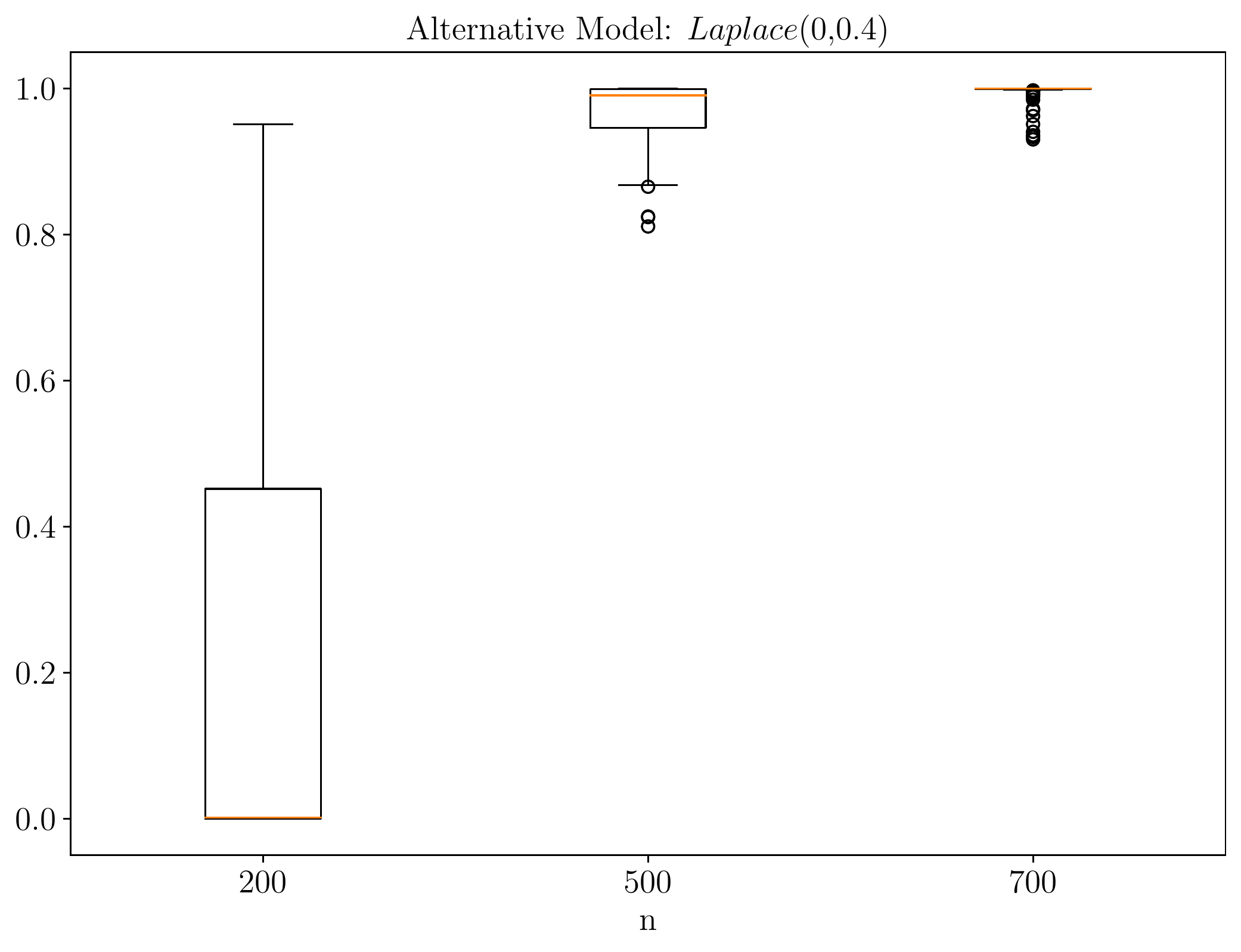}
    \end{subfigure}
\end{minipage}
\caption{1-dimensional Laplace experiment: 
distribution (over 100 independent runs) of the probability of the alternative hypothesis $p(H_1|\mathcal{D})$ for a different number of observations $n$. Here,  $P_X=\mathcal{N}(0,1)$ and $P_Y= Laplace (0, 1.5)$ (Left) or $P_Y= Laplace (0, 0.4)$ (Right).}
\label{fig6: Laplace_indep}
\end{figure}

We consider a scenario where the data are generated using the following distributions: $P_X=\mathcal{N}(0,1)$ and $P_Y=Laplace(0,1.5)$ or $P_Y=Laplace(0, 0.4)$ The results are presented in Figure \ref{fig6: Laplace_indep} which aligns with our expectation. As the number of samples increases the test is becoming increasingly certain of the difference between the null and alternative model and hence $P(H_1|\mathcal{D})$ concentrates  at 1. Since the proposed method is not restricted to two-sample testing between independent random variables, we also consider the same experiment with correlated standard Gaussian and Laplace distributions generated through copula transformation with correlation set to 0.5. The correlated structure has helped the discovery of the difference between the distributions. Results are presented in Supplementary Material illustrating that the method works equally well in correlated random variable cases. 

\subsection{Two by Two Blobs of 2-Dimensional Gaussian Distributions} \label{sec6: 2by2Blobs}

The performance of the frequentist kernel two-sample test using MMD depends heavily on the choice of kernel. When a Gaussian kernel is used, this boils down to choosing an appropriate lengthscale parameter. Often, median heuristic is used. However, \cite{optikernel12} showed that MMD with median heuristic failed to reject the null hypothesis when comparing samples from a grid of isotropic Gaussian v.s. a grid of non-isotropic Gaussian. The framework proposed by \cite{FlaSejCunFil2016} showed that, by choosing the lengthscale that optimise the Bayesian kernel learning marginal log likelihood (i.e. an empirical Bayes type of approach), MMD is able to correctly reject the null hypothesis at the desired significance level. Intuitively, the algorithm needs to look locally at each blob to detect the difference rather than at the lengthscale that covers all of the blobs which is given by the median distance between points. 

We repeat this experiment using the proposed Bayesian two-sample test with $P_X$ being a mixture of 2-dimensional isotropic Gaussian distributions and $P_Y$ a mixture of 2-dimensional Gaussian distributions centered at slightly shifted locations with rotated covariance matrix. Note, this is not the same dataset used in \cite{FlaSejCunFil2016} and \cite{optikernel12}, we shift the dataset to have multiple relevant lengthscales. We center the blobs of the 2-dimensional Gaussian distributions of $P_X$ at $\{(10,10)^\top, (10, 30)^\top,$ $(30, 10)^\top, (30,30)^\top\}$ and shift such locations by $(-1,-1)$ for $P_Y$. An equal number of observations is sampled from each of the blobs. The covariance matrix of $P_Y$ follows the same form as in Section \ref{sec6: 2DGaussian} with $\epsilon = \{2, 6, 10, 20\}.$ We present illustrations of the samples from these distributions in the Supplementary Material.  

\begin{figure}[htp]
\begin{minipage}{0.45\textwidth}
    \centering
    \begin{subfigure}{}
        \includegraphics[width=\textwidth]{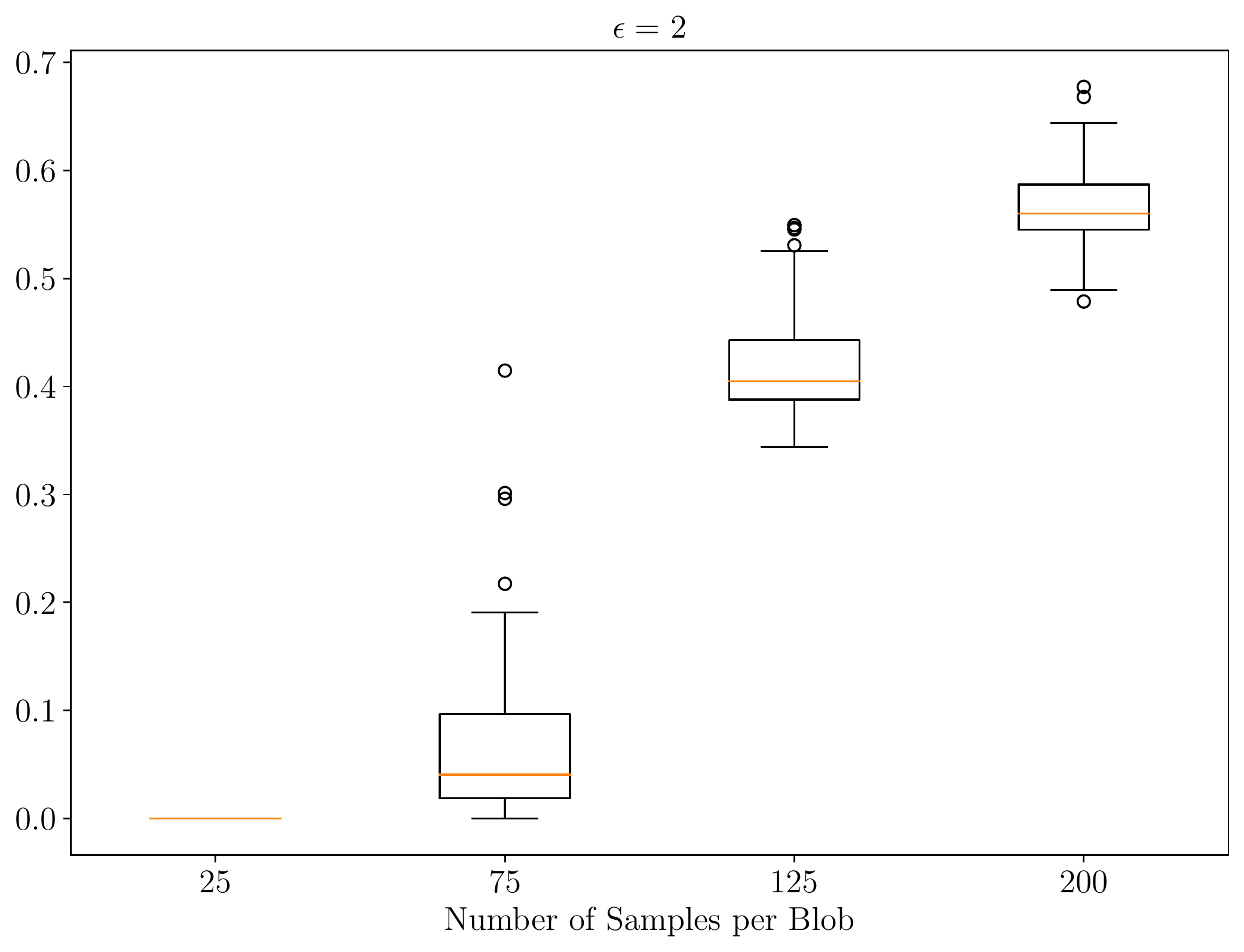}
    \end{subfigure}
    \begin{subfigure}{}
        \includegraphics[width=\textwidth]{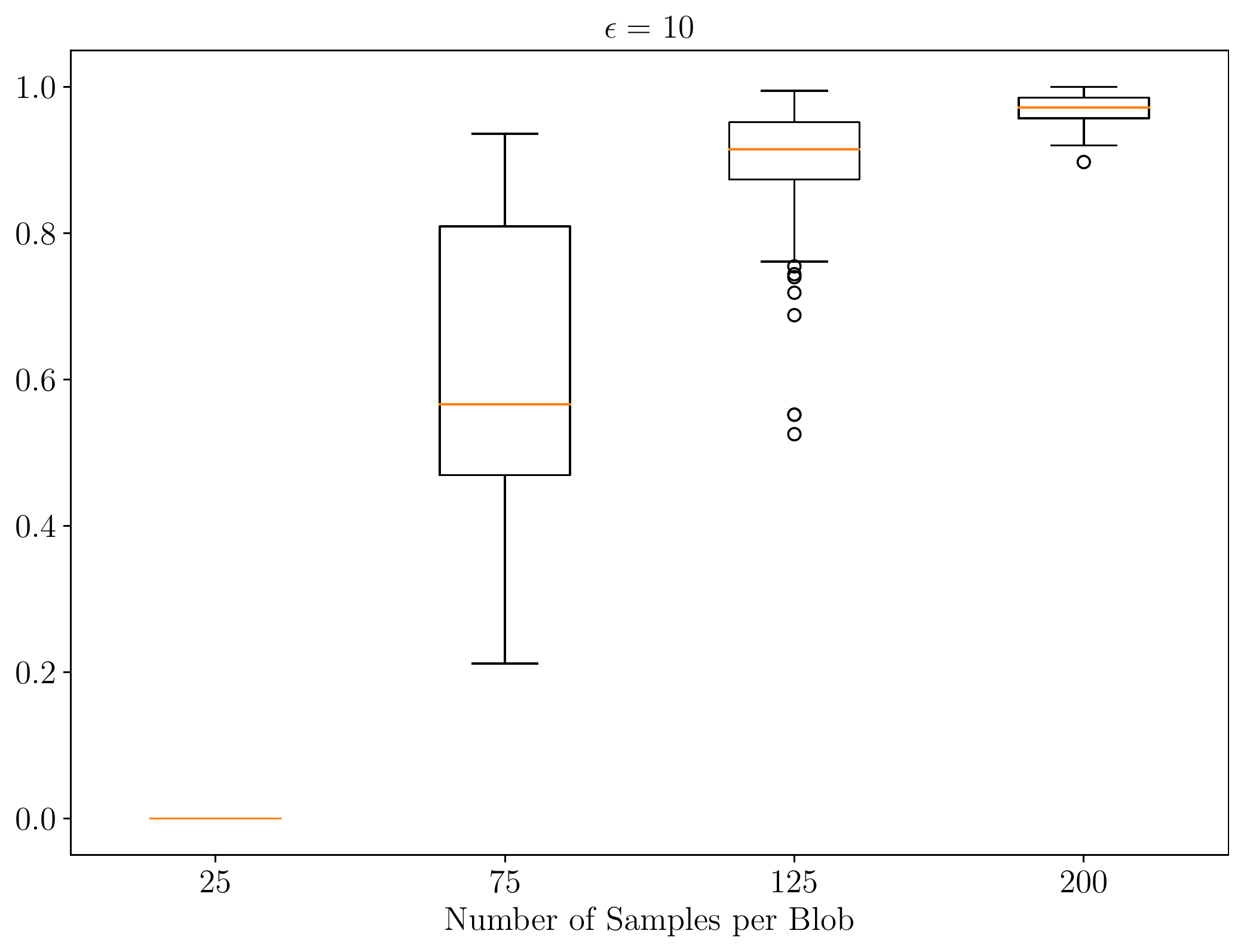}
    \end{subfigure}
\end{minipage}\hfill
\begin{minipage}{0.45\textwidth}
    \centering
    \begin{subfigure}{}
        \includegraphics[width=\textwidth]{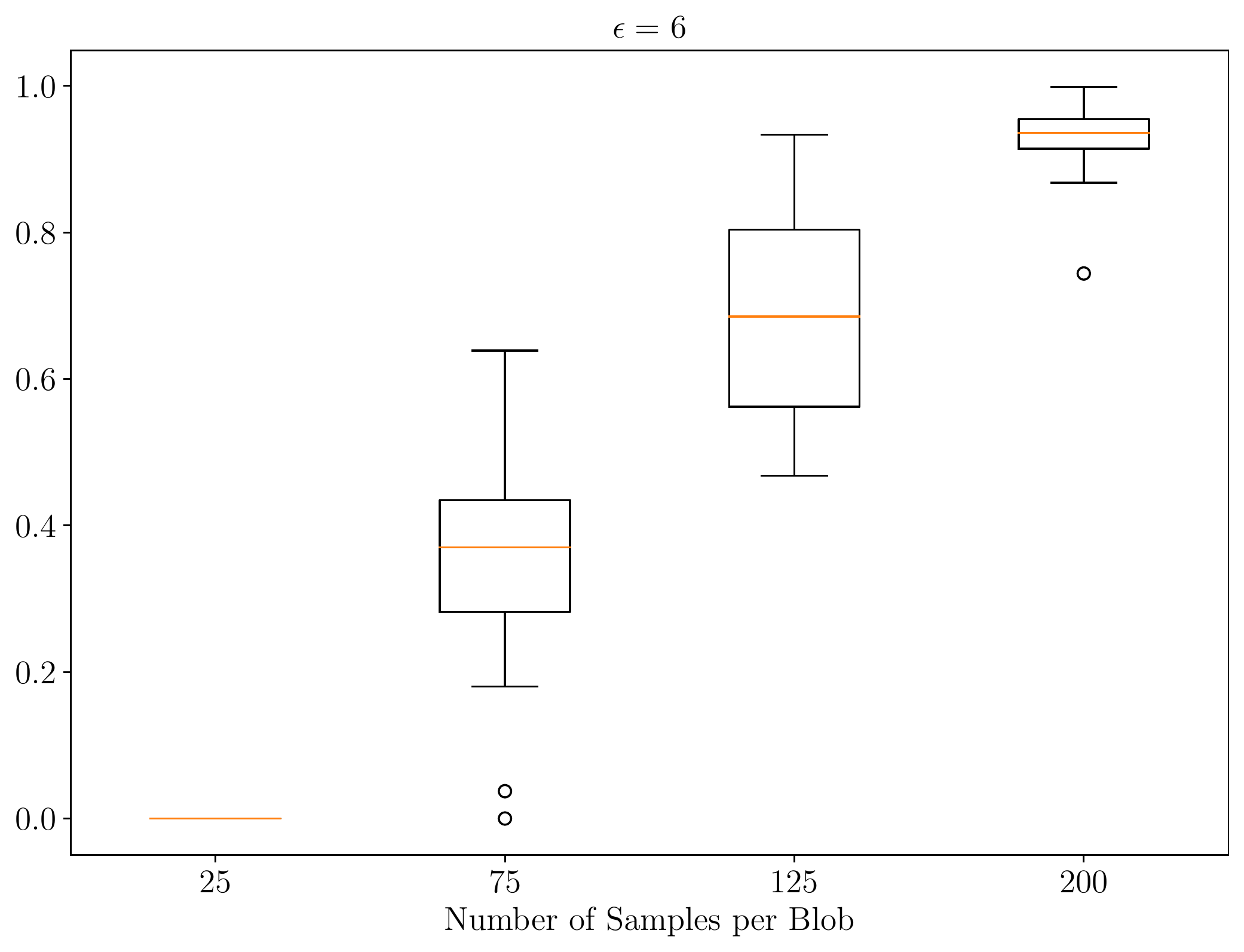}
    \end{subfigure}
    \begin{subfigure}{}
        \includegraphics[width=\textwidth]{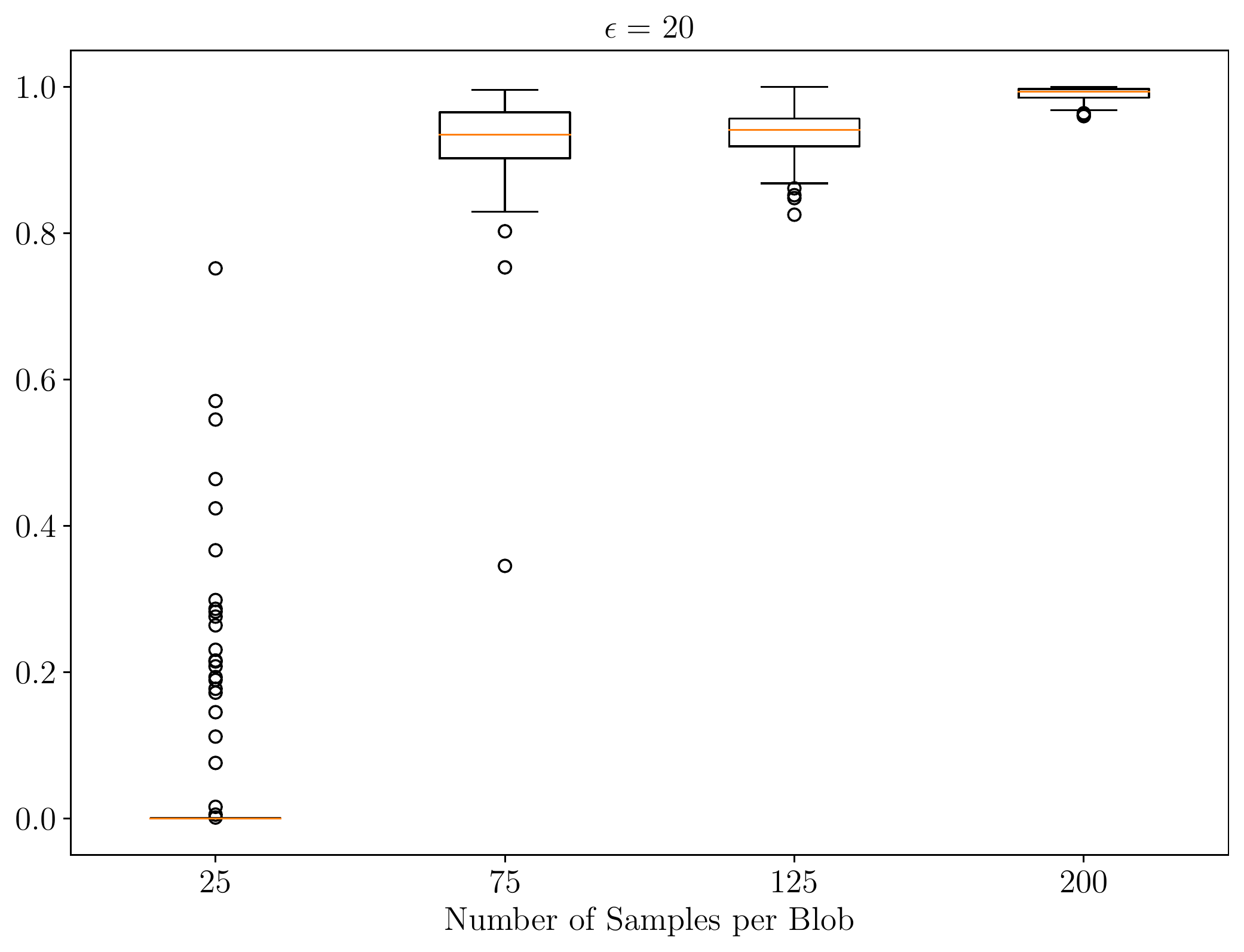}
    \end{subfigure}
\end{minipage}
\caption{2 by 2 blobs of bivariate Gaussian experiment: distribution (over 100 independent runs) of the probability of the alternative hypothesis $p(H_1|\mathcal{D})$ for a different number of observations $n$. The distribution of $X$ is a mixture of four bivariate Gaussian distributions with equal probability centered at $\{(10,10)^\top, (10, 30)^\top, (30, 10)^\top, (30,30)^\top\}$ and with $\epsilon = 1$ as in the set up from Section \ref{sec6: 2DGaussian}. The distribution of $Y$ is also a mixture of four bivariate Gaussian distributions with equal probability centered around the same locations but also shifted by $(-1,-1).$ In this experiment, we consider the cases when $\epsilon = \{2,6,10,20\}.$ } \label{fig6: 2by2Blobs}
\end{figure}

Figure \ref{fig6: 2by2Blobs} shows that our approach is able to detect the difference between the distributions since the probability of the alternative hypothesis becomes more concentrated around 1 as the number of samples increases. Note, when $\epsilon =2$, the distribution of the probability of the alternative hypothesis is around 0.6. We expect this to increase to 1 as we increase the number of samples to around 300 samples per blob given the pattern observed. 

\begin{figure}[htp]
\begin{minipage}{0.45\textwidth}
    \centering
    \begin{subfigure}{}
        \includegraphics[width=\textwidth]{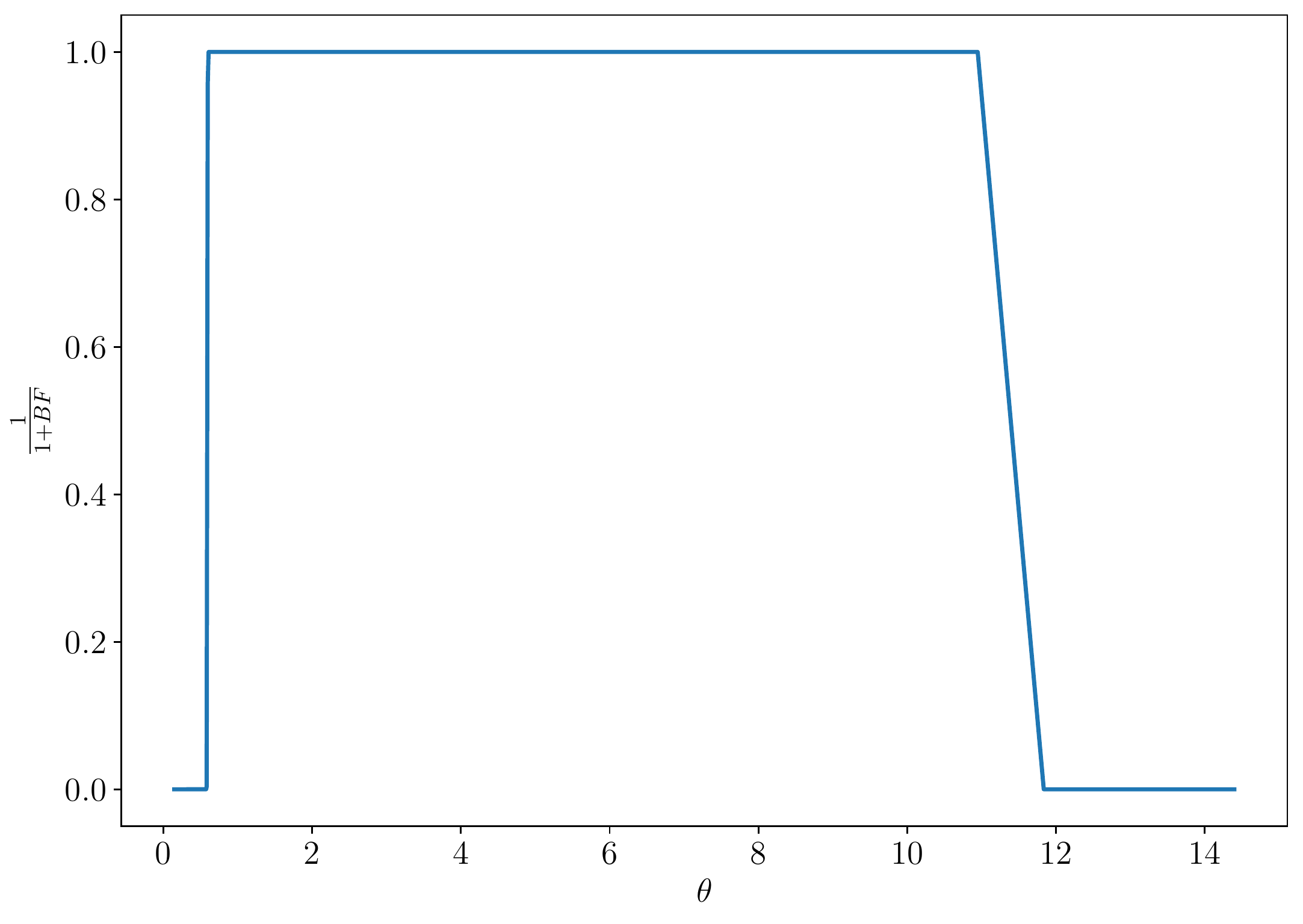}
    \end{subfigure}
\end{minipage}\hfill
\begin{minipage}{0.45\textwidth}
    \centering
    \begin{subfigure}{}
        \includegraphics[width=\textwidth]{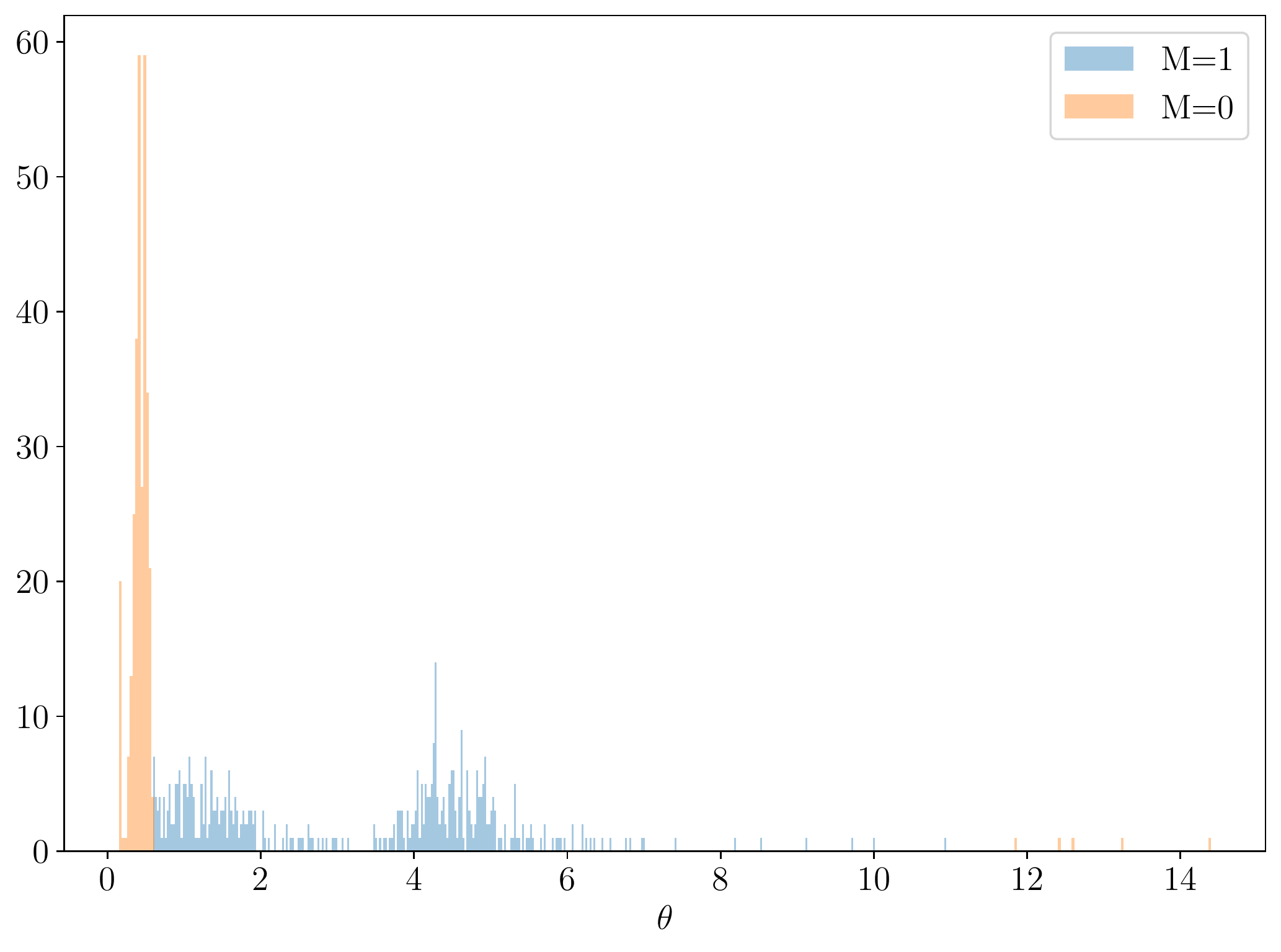}
       
    \end{subfigure}
\end{minipage}
 \caption{2 by 2 blobs of bivariate Gaussian experiment under the alternative model $\epsilon = 2$ and 200 samples per blob. Left: the plot illustrates $\frac{1}{1 + BF_\theta}$ against the value of $\theta$. Right: histogram of samples from the marginal distribution of $\theta|M, \mathcal{D}$ for $H_1$ and $H_0$.}\label{fig6: 2by2Gaussian_marginals}
\end{figure}

As an example, in Figure ~\ref{fig6: 2by2Gaussian_marginals} (Left), we see that a wide range of lengthscales is informative for this two-sample testing problem when $\epsilon =2$. If we further observe the marginal distributions $P(\theta |M =0, \mathcal{D})$ and $P(\theta | H_1, \mathcal{D})$ in Figure \ref{fig6: 2by2Gaussian_marginals} (Right), the method takes advantage of the large lengthscales to detect shift in location and the small lengthscales to detect the difference in covariance. But when the lengthscale is too small (approximately less than 0.5), the method regards the samples as identically distributed. 

\subsection{Higher Dimensional Gaussian Distributions} \label{sec6: HigherD}
\begin{figure}[htp]
\begin{minipage}{0.45\textwidth}
    \centering
    \begin{subfigure}{}
        \includegraphics[width=\textwidth]{figc6/2by2Blobs/epsilon_6.pdf}
    \end{subfigure}
    \begin{subfigure}{}
        \includegraphics[width=\textwidth]{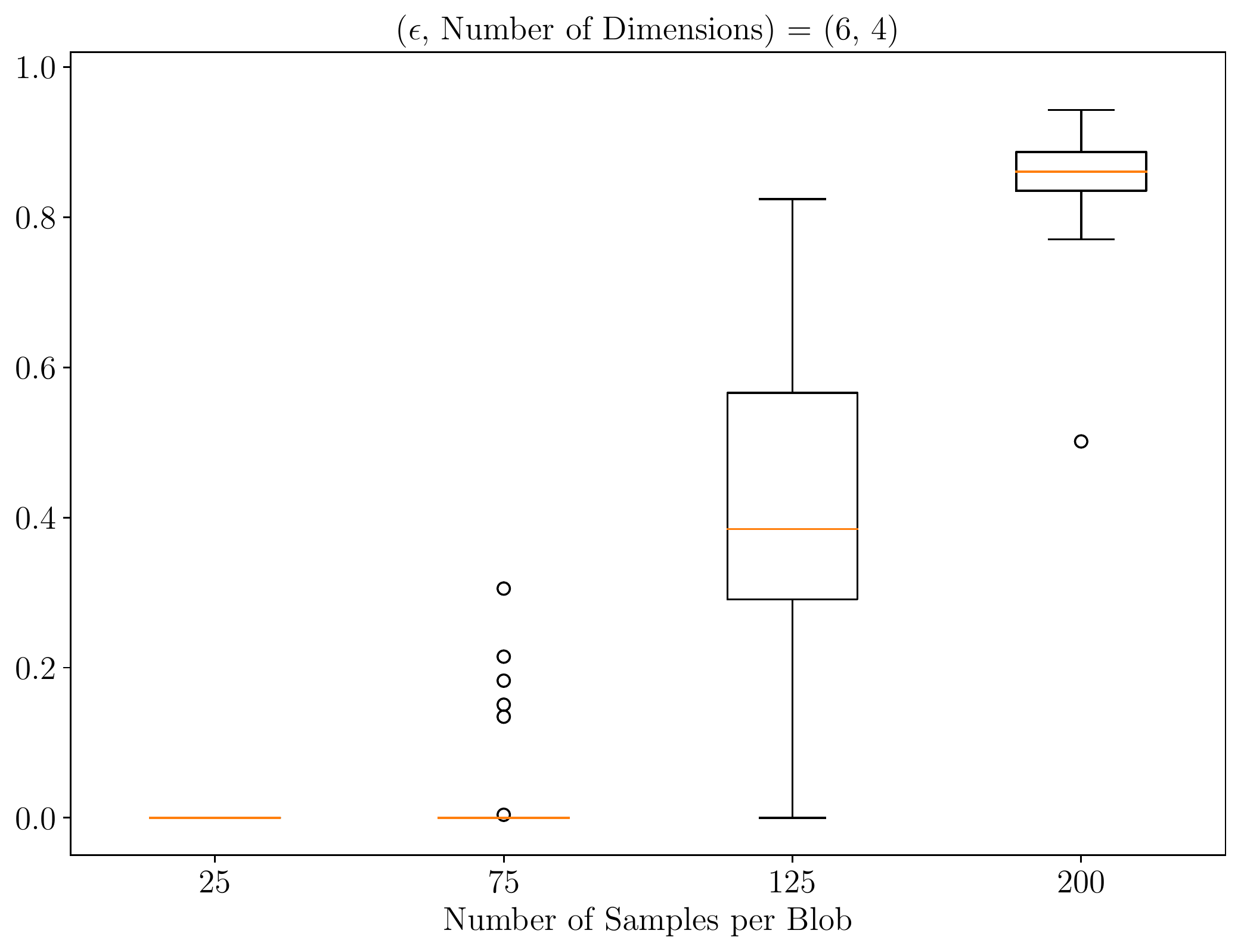}
    \end{subfigure}
    \begin{subfigure}{}
        \includegraphics[width=\textwidth]{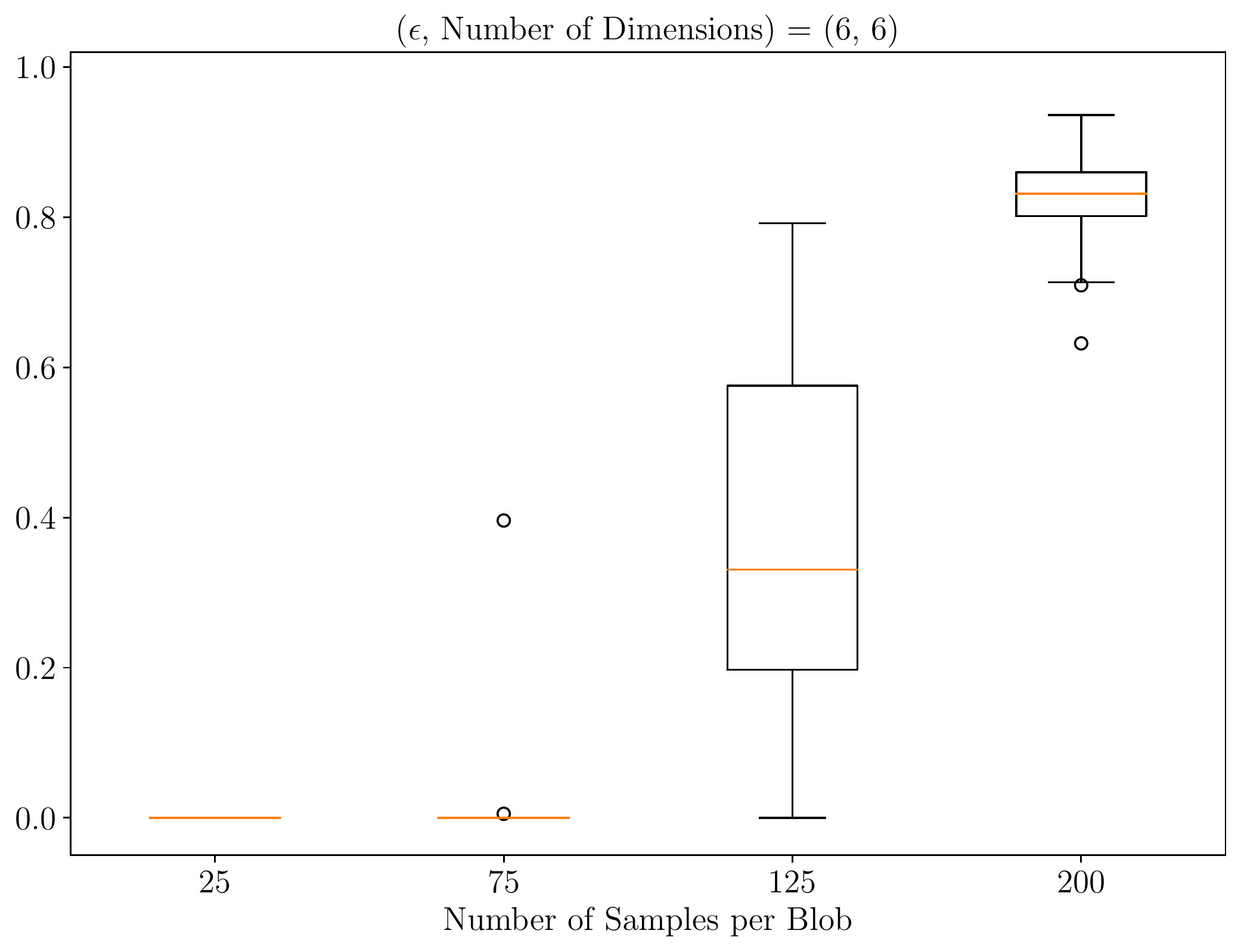}
    \end{subfigure}
\end{minipage}\hfill
\begin{minipage}{0.45\textwidth}
    \centering
    \begin{subfigure}{}
        \includegraphics[width=\textwidth]{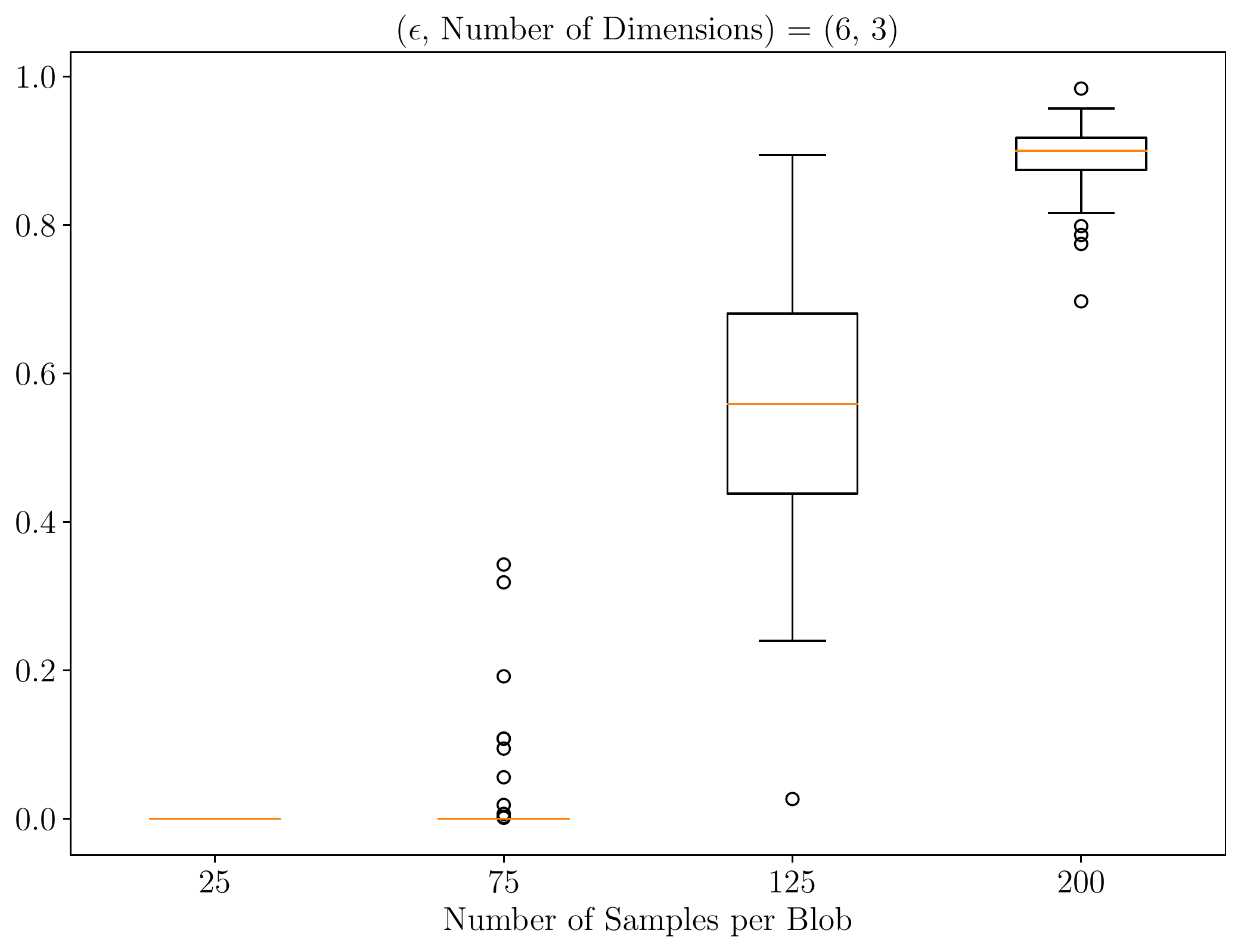}
    \end{subfigure}
    \begin{subfigure}{}
        \includegraphics[width=\textwidth]{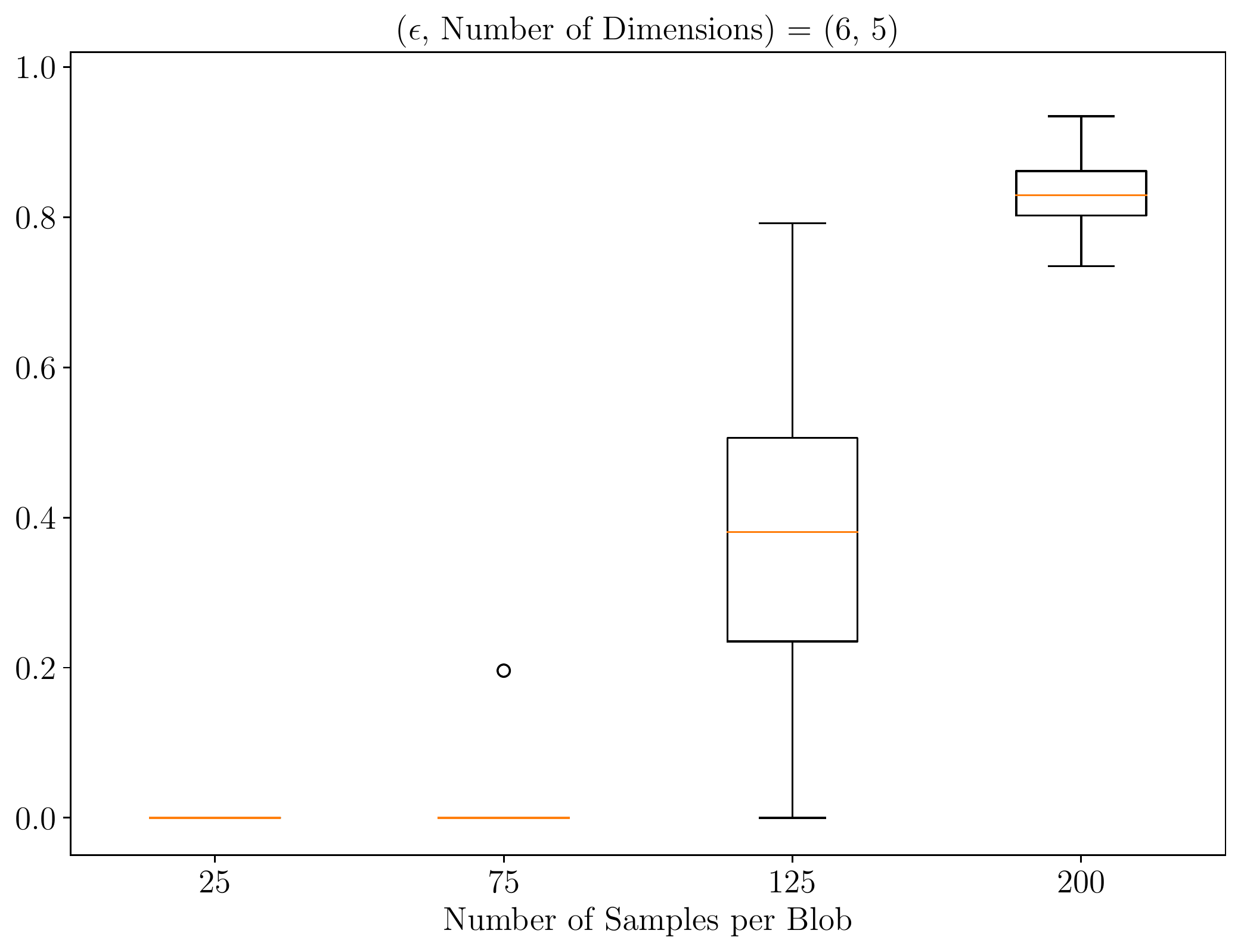}
    \end{subfigure}
    \begin{subfigure}{}
        \includegraphics[width=\textwidth]{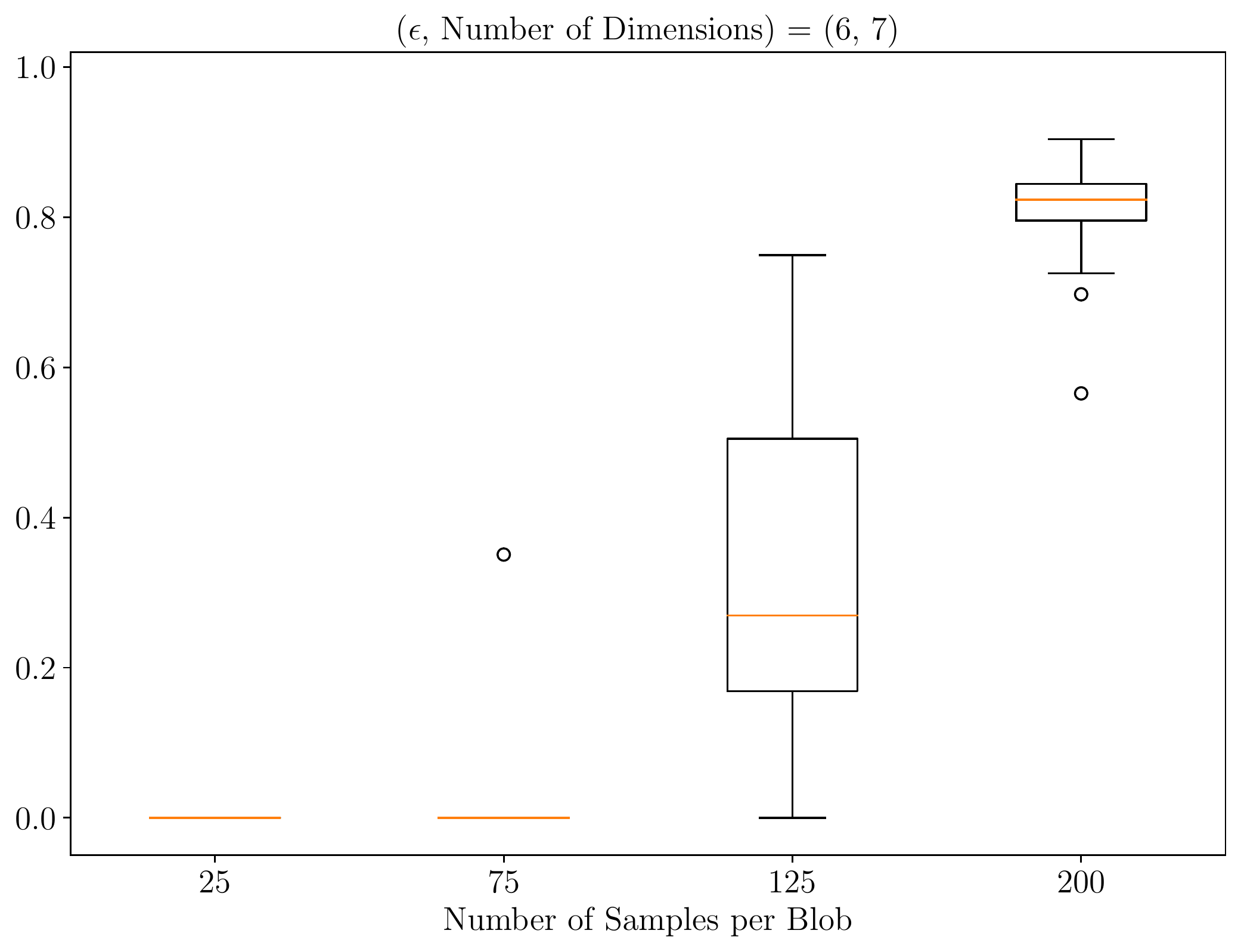}
    \end{subfigure}
\end{minipage}
\caption{Higher dimensional experiment: distribution (over 100 independent runs) of the probability of the alternative hypothesis $p(H_1|\mathcal{D})$ for a different number of observations $n$. For the first two dimensions, the data is generated as in Section \ref{sec6: 2by2Blobs} with $\epsilon=6$. Standard Gaussian noises are appended as the remaining dimensions. Top left figure is copied from Figure \ref{fig6: 2by2Blobs} for the ease of performance comparison. } \label{fig6: higherD_e6}
\end{figure}

We have seen that the proposed method is able to utilise informative value of the lengthscale parameter and make correct decisions about the probability of the alternative hypothesis given large enough samples. In this section, we investigate the effect of dimensionality of the given sample on the proposed two-sample testing method. We use the Gaussian blobs experiment from the previous section and append simple $\mathcal{N}(0,1)$ to both $X$ and $Y$ (i.e. the difference in distribution exists only in the first two dimensions). In particular, we consider the cases when the total number of dimensions are $\{3,4,5,6,7\}$. The results are presented in Figure \ref{fig6: higherD_e6}. 


We observe that the test requires more samples to detect the difference between the two distributions as the number of dimension increases. 
The noise in the additional dimensions has indeed made the problem harder for the given number of samples. But the proposed method still manages to discover the difference as the number of samples increases. For up to 8 dimensions, the method returns a posterior probability of $H_1$ higher than 0.8 as soon as there are more than 200 samples per blob. This illustrates the robustness of our proposed method to the dimensionality of the problem.

\subsection{Comparison to other Bayesian nonparametric methods}

\label{sec6: comparison}

In this section we compare our test to other Bayesian nonparametric two-sample tests in the literature. In particular we compare our method to the two approaches of \cite{chen2014bayesian} and \cite{holmes2015} based on P{\'o}lya tree priors. We consider the following setting: $X\overset{iid}{\sim} P_X=N(0,1)$ versus $Y \overset{iid}{\sim} P_Y=N(\mu,1)$ with $\mu \in \{0,1.5,3\}$ and a sample size $n=200$.  
\begin{figure}[htp]
\centering
\includegraphics[width=0.8\textwidth]{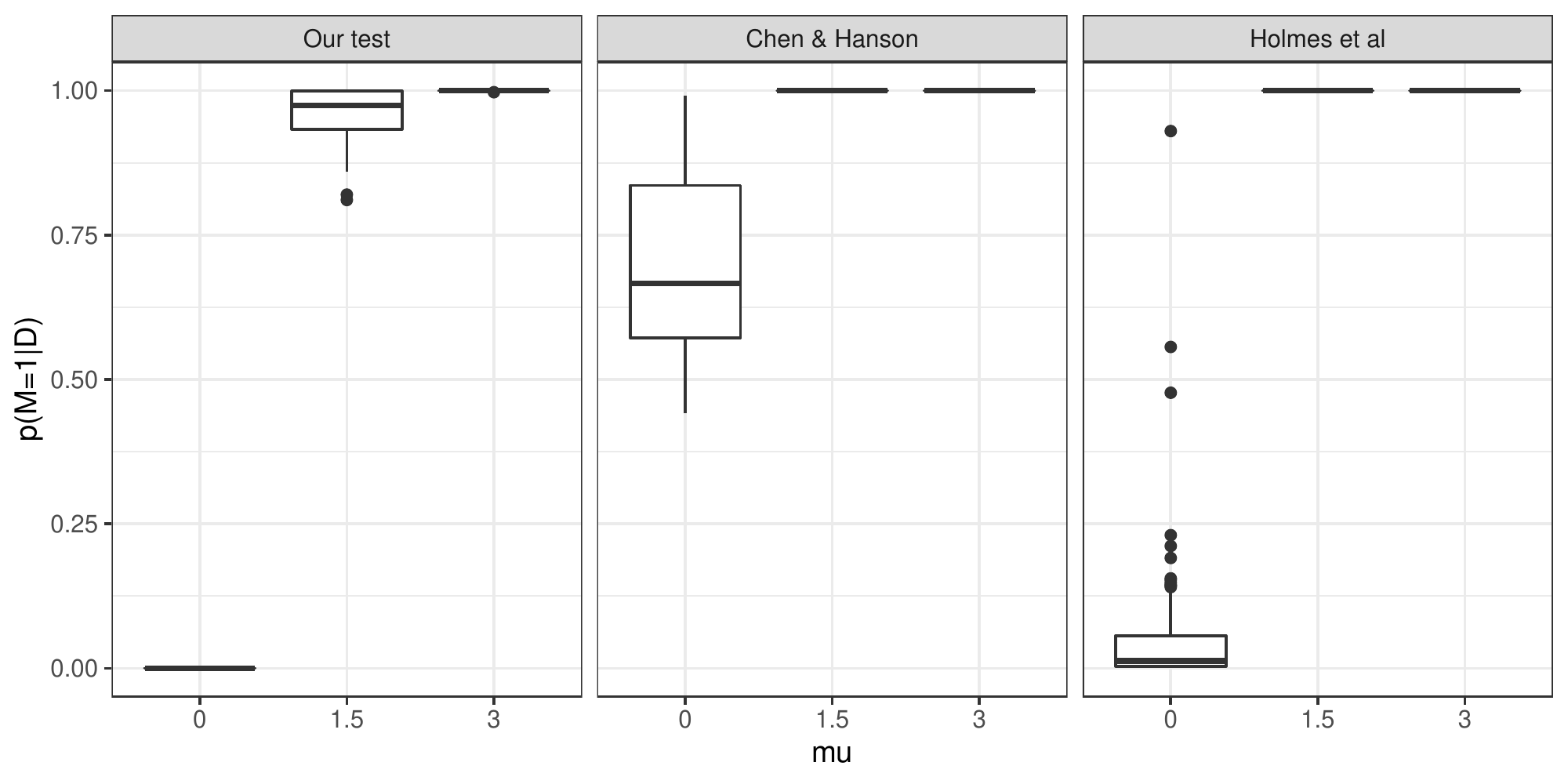}
\caption{Comparison of our approach (Left) to the methods proposed by \cite{chen2014bayesian} (Middle) and \cite{holmes2015} (Right): distribution (over 100 independent runs) of the  posterior probability $p(H_1|\mathcal{D})$ in a one-dimensional setting where $P_X=N(0,1)$ and $P_Y=N(\mu,1)$ for $\mu \in \{0,1.5,3\}$ with a sample size $n=200$. }
\label{fig: comparisson_mean}
\end{figure}  

\begin{figure}[htp]
\centering
\includegraphics[width=0.8\textwidth]{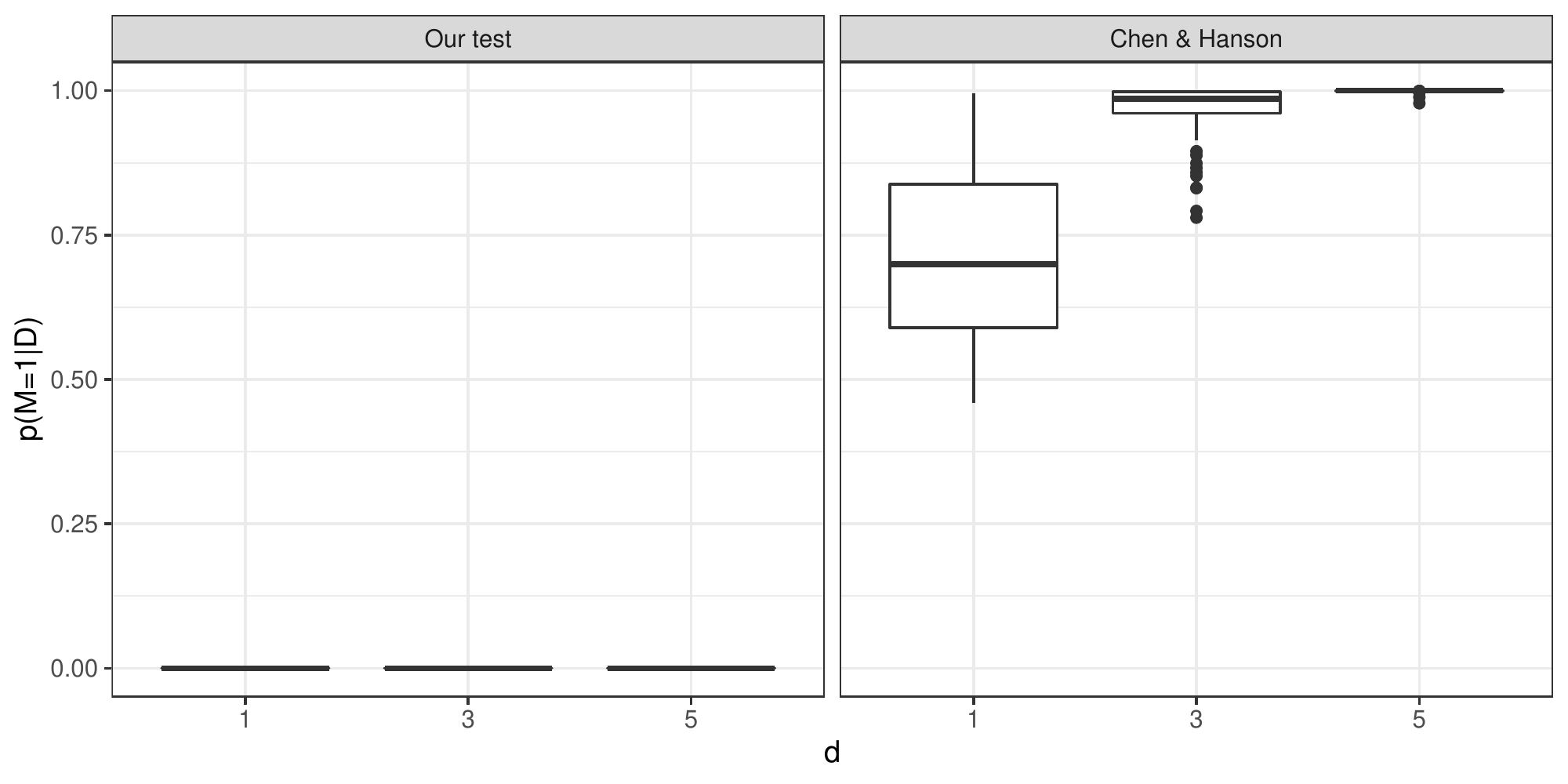}
\caption{Comparison of our approach (Left) to the methods proposed by \cite{chen2014bayesian} (Right): distribution (over 100 independent runs) of the  posterior probability $p(H_1|\mathcal{D})$ in a multi-dimensional setting where $P_X=N(0,I_D)$ and $P_Y=N(0,I_D)$ for $D \in \{0,1.5,3\}$ with a sample size $n=100$.}
\label{fig: comparisson_Ddim}
\end{figure}

As depicted in Figure \ref{fig: comparisson_mean}, one can see that our  method correctly returns a posterior probability of $H_1$ close to $0$ under $H_0$ ($\mu=0$) and a posterior probability close to 1 under $H_1$, when $\mu=1.5$ and $\mu=3$. Similar behaviour is displayed for the method of \cite{holmes2015}.
However, the test of \cite{chen2014bayesian} appears to have posterior probabilities for $H_1$ which are systematically biased upwards. The problem becomes aggravated  with increasing dimension. In Figure \ref{fig: comparisson_Ddim} we simulate $X \overset{iid}{\sim}{P_X}=N(0,I_D)$ versus $Y \overset{iid}{\sim}P_Y=N(0,I_D)$ with $D\in \{1,3,5\}$ and a sample size of $n=100$ to see if \cite{chen2014bayesian} can correctly identify $H_0$ in higher dimensions. 

As shown in Figure \ref{fig: comparisson_Ddim} with increasing dimension, the posterior probability of $H_1$ becomes closer to one, which means that the test is unable to identify $H_0$. However, our test correctly identifies $H_0$ in all dimensions.
Presumably to overcome the issue of biased posterior probabilities, \cite{chen2014bayesian} do not use their Bayes factor in the classical Bayesian sense. Instead, they use it as a test statistic for a (frequentist) permutation test. The utility of power comparisons between Bayesian and frequentist paradigms is questionable so we do not pursue that direction here. 
We recall that the test of \cite{holmes2015} can only handle one-dimensional data.

\section{Real Data Experiments}
\label{sec6: real_experiments}

\subsection{Network Heterogeneity From High-Dimensional Data}
In system biology and medicine, the dynamics of the data under analysis can often be described as a network of observed and unobserved variables, for example a protein signalling network in a cell \citep{stadler_Rcode}. One interesting problem in this area is to investigate if the signalling pathways (networks) reconstructed from two subtypes are statistically different. 

\begin{figure}[ht]
\begin{minipage}{0.45\textwidth}
    \centering
    \begin{subfigure}{}
        \includegraphics[width=\textwidth]{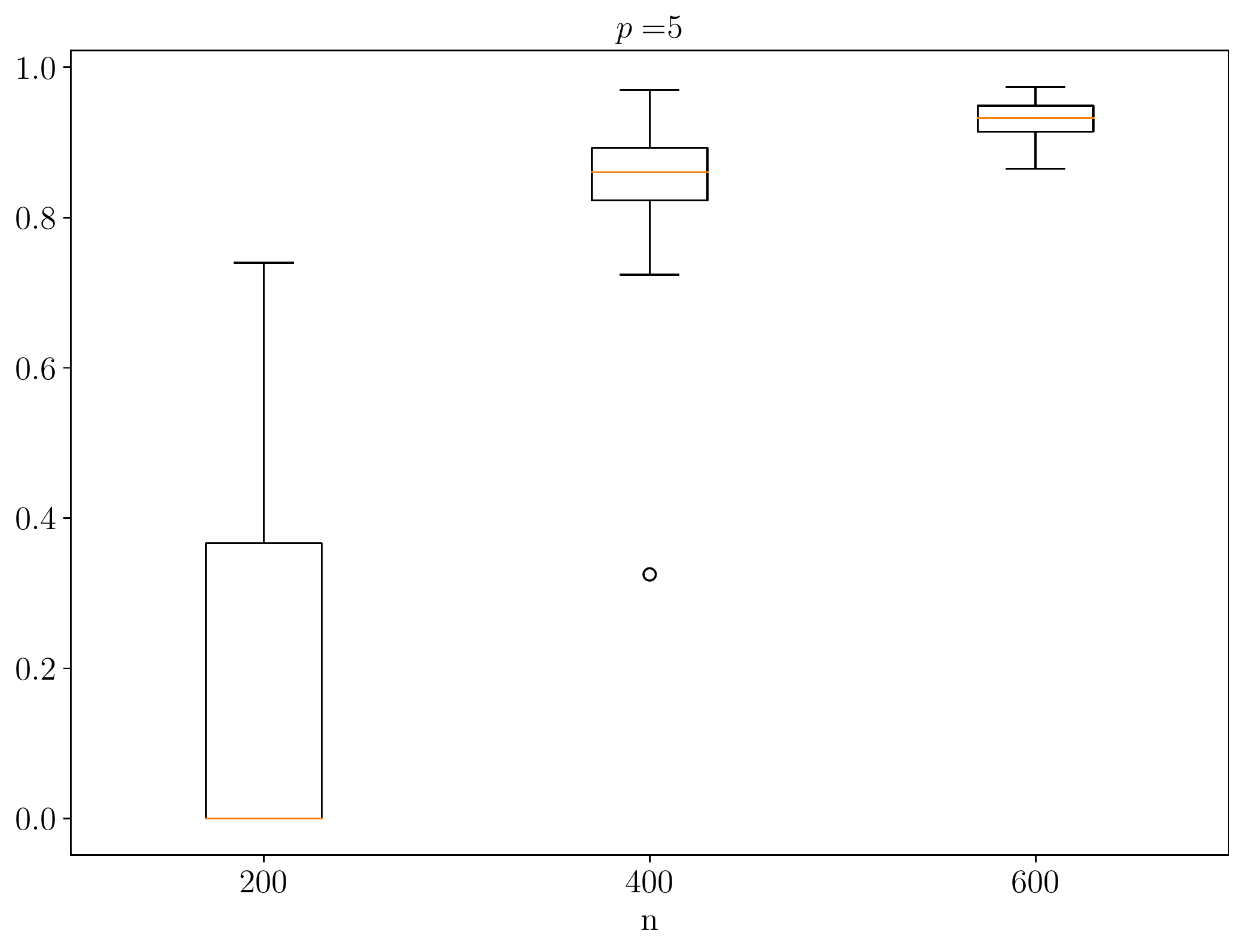}
        \caption{\textcolor{black}{Random networks heterogeneity testing: distribution (over 100 independent runs) of the probability of $H_1 | \mathcal{D}$ for a different number of samples.} }\label{fig6: randomp5_ProbM1}
    \end{subfigure}
\end{minipage}\hfill
\begin{minipage}{0.45\textwidth}
    \centering
    \begin{subfigure}{}
        \includegraphics[width=\textwidth]{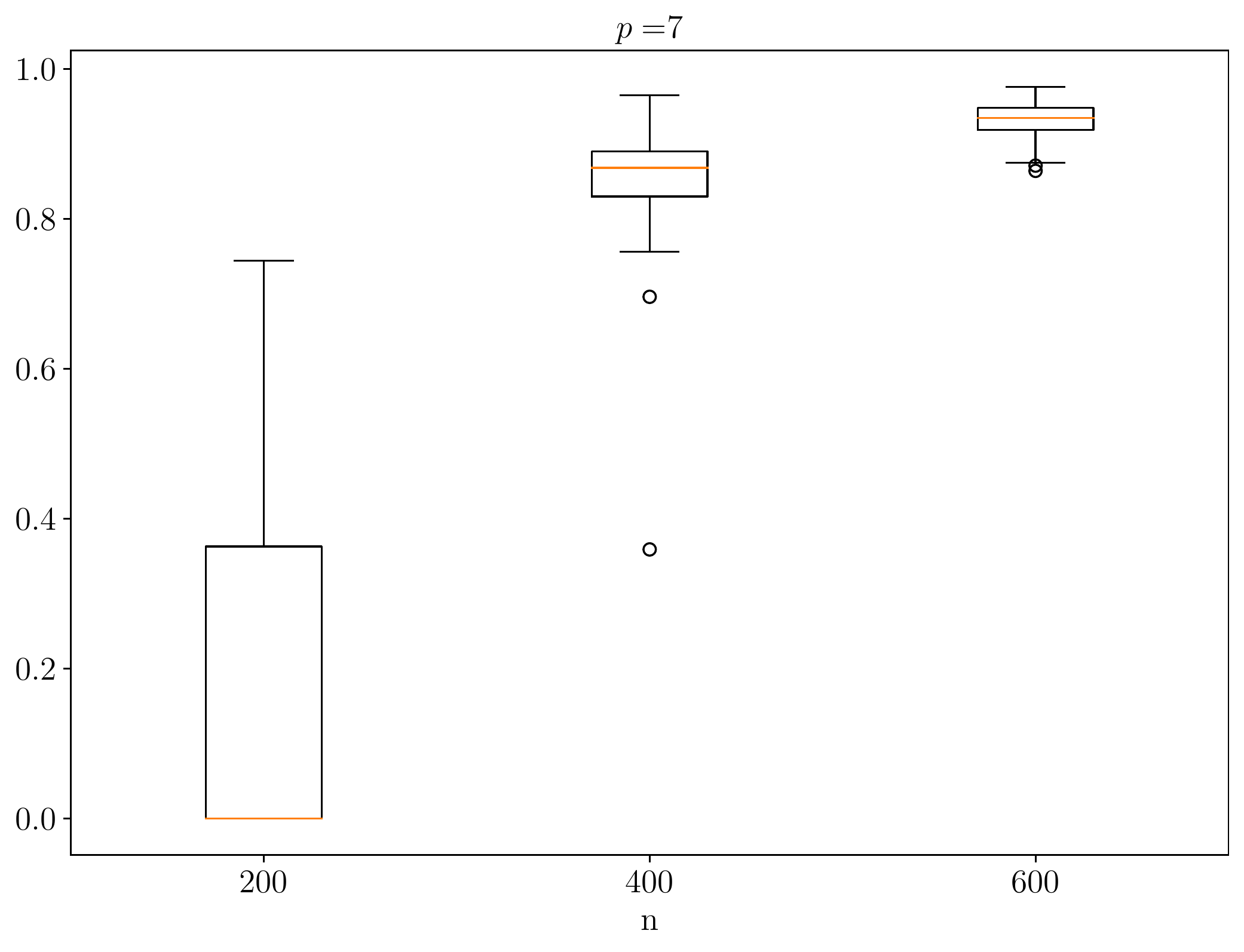}
        \caption{\textcolor{black}{Hub network heterogeneity testing: distribution (over 100 independent runs) of the probability of $H_1 | \mathcal{D}$ for a different number of samples.}}\label{fig6: hubp7_ProbM1}
    \end{subfigure}
\end{minipage}
\end{figure}

In this section, we follow the statistical setup given in \cite{Stadle_GGM, stadler_Rcode} and describes the networks by Gaussian graphical models (GGMs) which use an undirected graph (or network) to describe probabilistic relationships between $p$ molecular variables. Assume that each sample $X_i$ (and similarly for $Y_i$) is sampled from a multivariate Gaussian distribution with zero mean and some concentration matrix $\Omega$ (i.e. the inverse of a covariance matrix). The concentration matrix defines the graph $G$ via $$(j, j') \in E(G) \Leftrightarrow \Omega_{jj'} \not = 0$$
for $j \not = j' \in \{1, \cdots, p\}$ and $E(G)$ denotes the edge set of graph $G$. Network homogeneity problem presented in the previous paragraph can be formulated as a two-sample testing problem in statistics where we are interested in testing the null hypothesis 
\begin{align}
    H_0: G_1 = G_2
\end{align}
In the first experiment, we use the code from \cite{stadler_Rcode} to generate a pair of networks with 5 nodes that present 4 common edges and then obtain the corresponding correlation matrices to use as the covariance matrices for the multivariate Gaussian distribution. The results are presented in Figure \ref{fig6: randomp5_ProbM1}. In the second experiment, we again use the code from \cite{stadler_Rcode} to generate hub networks with 7 nodes that are divided into 3 hubs with 1 hub that is different and use the obtained correlation matrices in the multivariate Gaussian distribution as with the first experiment. The results are presented in Figure \ref{fig6: hubp7_ProbM1}. Both tests were able to recover the ground truth as the number of observed sample increases.

\subsection{Real Data: Six-Membered Monocyclic Ring Conformation Comparison}

In this section, we consider a real world application of our proposed method to detect if the conformation observed in crystal structures differ from its lowest energy conformation in gas phase. Qualitative descriptions are often given to show that the two are distributed differently due to the crystal packing effect. While no quantitative analysis has been provided in the chemistry literature, we aim to perform such analysis through the use of the proposed Bayesian two-sample test. 

We utilise the Cremer-Pople puckering parameters \citep{Chemistry_realdata} to describe the six-membered monocyclic ring conformation and compare their shapes under the two different conditions described above. This coordinate system first defines a unique mean plane for a general monocyclic puckered ring. Amplitudes and Phases coordinates are then used to describe the geometry of the puckering relative to the mean plane. For a six-membered monocyclic ring, there are three puckering degrees of freedom, which are described by a single amplitude-phase pair $(q_2, \phi_2)$ and a single puckering coordinate $q_3$. As we consider general six-membered rings, we can omit the phase parameters $\phi_2$ for simplicity and compare the degree of puckering (maximal out-of-plane deviation) under different conditions. 
The crystal structures of 1936 six-membered monocyclic rings are extracted from the Crystallography Open Database (COD) and the associated puckering parameters are calculated. Independently, we calculate the lowest energy conformations of a diverse set of 26405 molecules using a semi-empirical method GFN2 and record the puckering parameters. 
We consider 100 random samples of size $n=\{200, 400, 600, 800\}$ from each of the datasets and conduct our Bayesian two-sample test 100 times while inferring the kernel bandwidth parameter $\theta$.

\begin{figure}[htp]
\begin{minipage}{0.45\textwidth}
    \centering
    \begin{subfigure}{}
        \includegraphics[width=\textwidth]{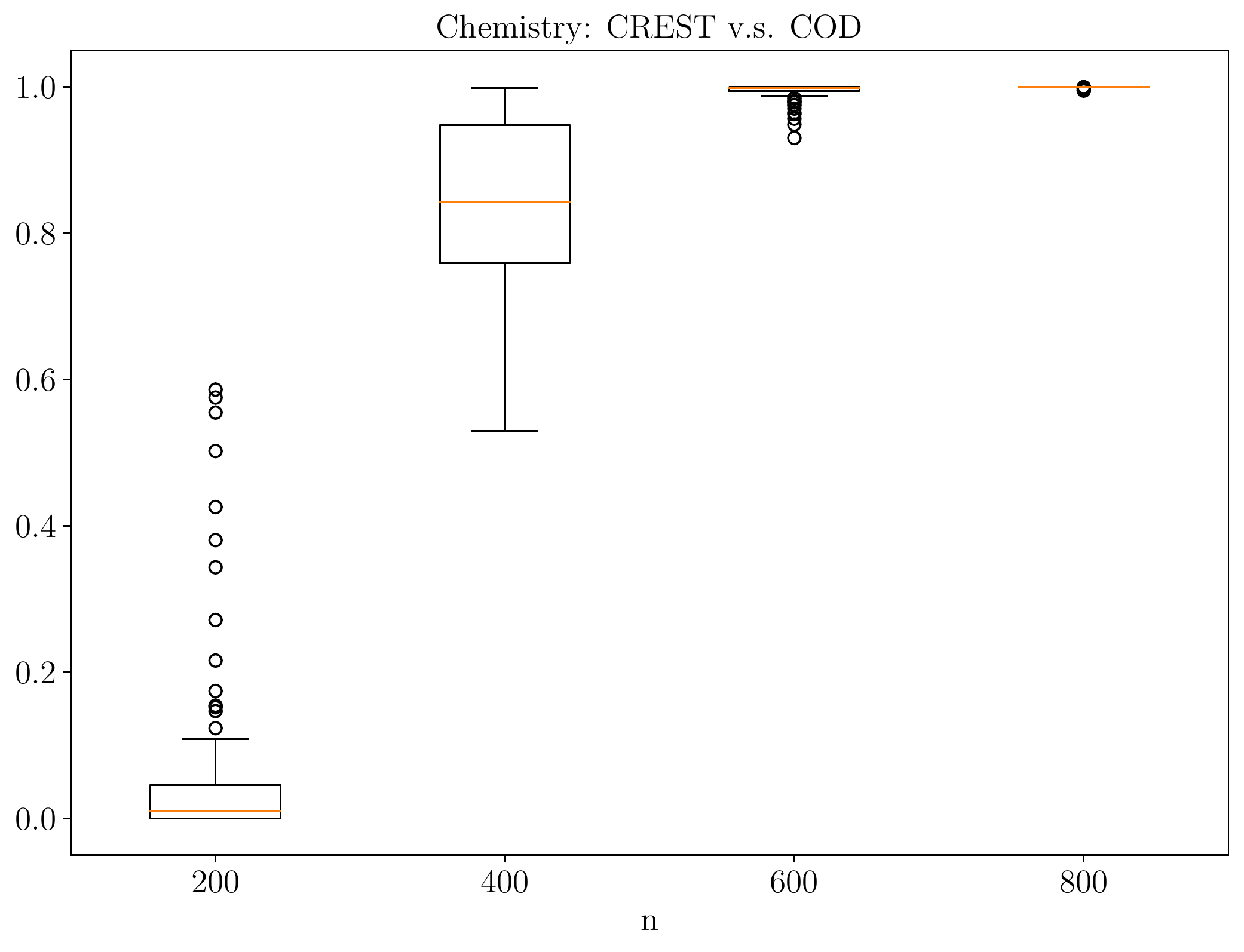}
        \caption{\textcolor{black}{Six-membered monocyclic ring conformation comparison: distribution (over 100 independent runs) of the probability of $H_1 | \mathcal{D}$ for a different number of samples.} }\label{fig6: Chemistry_ProbM1}
    \end{subfigure}
\end{minipage}\hfill
\begin{minipage}{0.45\textwidth}
    \centering
    \begin{subfigure}{}
        \includegraphics[width=\textwidth]{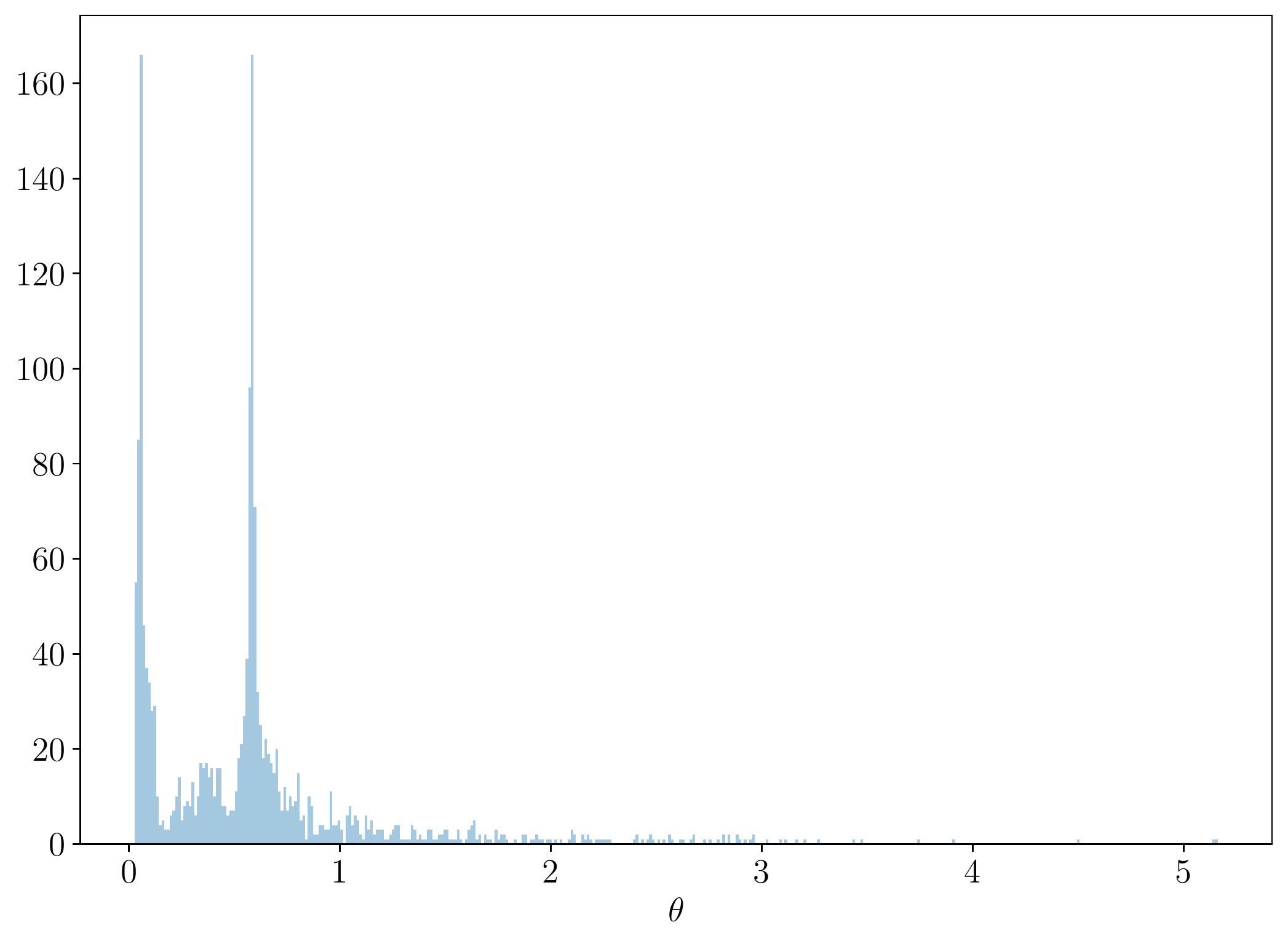}
        \caption{\textcolor{black}{Histogram of samples from the marginal distribution $\theta | M = 1, \mathcal{D}$ for the  experiment on six-membered monocyclic ring conformation comparison with 800 samples.}}\label{fig6: Chemistry_theta}
    \end{subfigure}
\end{minipage}
\end{figure}

Figure \ref{fig6: Chemistry_ProbM1} illustrates the results of our test. The proposed method is becoming more certain that the lowest energy conformation in gas phase of six-membered monocyclic rings is distributed differently from its crystal structures as the number of samples increases. At 800 samples, the proposed method gives the probability of $H_1|\mathcal{D}$ equals to 1 which aligns with expert opinions that the two are indeed distributed differently due to the crystal packing effect. In Figure \ref{fig6: Chemistry_theta}, we provide the posterior histogram of $\theta | H_1, \mathcal{D}$ when 800 samples are observed. The frequency distribution is multi-model indicating that multiple lengthscale is of interest for this problem at hand. 

\section{Conclusion} \label{sec6: futurework}
In this work, we have proposed a Bayesian two-sample testing framework utilising the Bayes factor. Rather than directly considering the observations, we have proposed to consider the differences between the empirical kernel mean embeddings (KME) evaluated at a set of inducing points. Following the learning procedure of the empirical KME \citep{FlaSejCunFil2016}, we have derived the Bayes factors when the kernel hyperparameter is given as well as when it is treated in a fully Bayesian way and marginalised over. Further, we have obtained efficient computation methods for the marginal pseudolikelihood utilising the Kronecker structure of the covariance matrices. The posterior inference of the model label and the kernel hyperparameter is done by HMC within Gibbs. We have showed in a range of synthetic and real experiments that our proposed Bayesian test is able to simultaneously utilising multiple lengthscales and correctly uncover the ground truth given sufficient data.

Following this work, there are several possible directions for future research. We have seen in Section \ref{sec6: experiments} that larger sample sizes are required for more challenging problems. A random Fourier feature approximation of the above framework can be easily developed to enable the use of large sample size without having prohibitive runtime. In this case, explicit finite dimensional feature maps are available, the difference between the mean embeddings $\delta = \mu_X - \mu_Y$ can be written more explicitly as $\delta = \mathbb{E}(\phi(X)) - \mathbb{E} (\phi(Y))$. Assume a GP prior with an appropriate covariance matrix for $\delta$, $\phi(x_i) - \phi(y_i)$ can be modelled by a Gaussian distribution with mean $\mathbb{E}(\phi(X)) - \mathbb{E}(\phi(Y))$ and covariance estimated as presented in the Supplementary Material. While the rest of the inference procedure can follow similarly as presented here, this large scale approximation requires careful specification of the covariance matrix for the GP model of $\delta$ to ensure that draws from such GP lie in the correct RKHS.

Recently, \cite{Kamary2014} proposed a mixture modelling framework for Bayesian model selection. The authors argues that the mixture modelling framework provides a more thorough assessment of the strength of the support of one model against the other and allows the use of noninformative priors for model parameters when the two competing hypotheses share the same set of parameters which is prohibitive in the classical Bayesian two-sample testing approach like Bayes factor \citep{DeGroot1973}. The proposed Bayesian two-sample testing framework using Bayes factor can be equivalently formulated as a mixture model: $$\Delta | \theta, \pi \sim \pi N(0, \frac{1}{n}\Sigma_\theta) + (1 - \pi) N(0, R_\theta +\frac{1}{n}\Sigma_\theta).$$ The posterior distribution of the mixture proportion $0 \leq \pi \leq 1$ indicates the model preferred. 
A joint inference of $\pi$ and the kernel bandwidth parameter $\theta$ can be easily done through MCMC. It would be interesting to see if there is a difference in performance between the mixture approach and the Bayes factor approach proposed here. Lastly, the Bayesian testing framework developed here and the directions for future work can all be applied to independence testing. 

Code for the proposed method and experiments is available at: \url{ https://github.com/qinyizhang/BayesianKernelTesting} \\

\noindent {\bf \large Supplementary Material} \\
The supplementary material provides details on derivation, proofs and additional synthetic data experiments. \\

\noindent {\bf \large Acknowledgements} \\
Q. Z. was supported by the Engineering and Physical Sciences Research Council (EPSRC) (EP/M50659X/1). The authors thank Lucian Chan for helping with the chemistry data and Chris Holmes whose advice in the early stages of the project was instrumental in shaping its direction.

\newpage

\bibliographystyle{plainnat}
\bibliography{ref.bib}
\clearpage
\newpage


\begin{appendix}

\section{Efficient Computation of Marginal Pseudolikelihood} \label{sec6: EfficientCompute}

In this section, we illustrate the efficient computation of the multivariate Gaussian distribution utilising the Kronecker product structure of our problem. Note, \cite{FlaSejCunFil2016} used the identity matrix as $\Sigma_\theta$ which results in their particular derivation of the fast computation of the marginal pseudolikelihood in their section A.4. However, we propose a different approach considering the full covariance matrix $\Sigma_\theta$. 

\begin{prop}
 Under the assumption of the alternative model, described in section \ref{sec6: AlterModel}, the marginal pseudolikelihood is given as 
\begin{align}
    \log \  p(\{(x_i, y_i)\}_{i=1}^n| \theta) 
    \propto & -\frac{1}{2} \log \det (\mathbbm{1} \mathbbm{1}^\top \otimes R_\theta + I_n \otimes \Sigma_\theta) \nonumber \\
     &\quad  -\frac{1}{2} \myvec (G_\theta)^\top (\mathbbm{1} \mathbbm{1}^\top \otimes R_\theta + I_n \otimes \Sigma_\theta)^{-1} \myvec (G_\theta) \nonumber \\
     & \quad \quad +\frac{1}{2} \sum^n_{i=1} \log \det (J_\theta (x_i,y_i)^\top J_\theta (x_i,y_i)).  \label{eq6: loglik_alter}
\end{align}
\end{prop}
\begin{proof*}
For every fixed pair of $(x_i, y_i)$, define the finite dimensional feature map $\phi_\theta: \mathbb{R}^D \rightarrow \mathbb{R}^s$ as 
\begin{align}
\phi_\theta(x) - \phi_\theta(y) = [k_\theta(x, z_1) - k_\theta(y, z_1), ..., k_\theta(x, z_s) - k_\theta(y, z_s)]^{\top} \in \mathbb{R}^s.
\end{align}
Note, $k_\theta(X_i, z_j)- k_\theta(Y_i, z_j)$ are independent for all $i = 1, ..., n$ given $\delta.$ 
This implies that $n\Delta$ can be written as a sum of independent random variables: $\Delta = \frac{1}{n} \sum^n_{i=1} \left (\phi_\theta (X_i) - \phi_\theta (Y_i) \right )$. By Cramer's decomposition theorem, we obtain the distribution of each of the contributing components
\begin{align}
\phi_\theta(X_i) - \phi_\theta(Y_i) |\delta, \theta \sim \mathcal{N} ([\delta_\theta(z_1), ..., \delta_\theta(z_s)]^{\top}, \Sigma_\theta). 
\end{align}

Applying the change of variable $(x, y) \mapsto g_\theta (x, y):= (\phi_\theta(x) - \phi_\theta(y))$ and using the generalisation of the change-of-variable formula to non-square Jacobian matrices \citep{BenIsrael1999} to obtain a distribution for $(x,y)$ conditionally on $\delta$
\begin{align}
p(x, y | \delta, \theta) = p(g_\theta (x, y) |\delta, \theta) \myvol [J_\theta(x, y)].
\end{align}
The Jacobian $J_\theta$ is a $s \times 2D$ matrix with 
\begin{align}
    J_\theta(x,y) =
    \begin{cases}
      \frac{\partial g_\theta(x, y)_{z_i}}{\partial x^j }, & \text{for}\ j = 1, ..., D \\
      \frac{\partial g_\theta(x, y)_{z_i}}{\partial y^j }, & \text{for}\ j = D+1, ..., 2D
    \end{cases}
\end{align}
and $\myvol[J_\theta(x, y)] := \sqrt{\det (J_\theta(x, y)^{\top} J_\theta(x,y))}.$ This can be further simplified into 
\begin{equation}
    J_\theta(x,y) =
    \begin{cases}
      \frac{\partial k_\theta(x, z_i)}{\partial x^j }, & \text{for}\ j = 1, ..., D \\
      \frac{\partial k_\theta(y, z_i)}{\partial y^j }, & \text{for}\ j = D+1, ..., 2D.
    \end{cases}
\end{equation}
The detailed computation of the Jacobian matrix will be discussed in below. By the conditional independence of $g(x_i, y_i)$ for each $i$ given $\delta$ and noting that $\theta$ represents the hyperparameter of the kernel $k_\theta (\cdot, \cdot)$, then 
\begin{align}
p(\{x_i, y_i\}^n_{i = 1}| \delta, \theta) 
= & \ \prod^n_{i =1} p(g_\theta (x_i, y_i) | \delta, \theta) \myvol[J_{\theta}(x_i, y_i)] \nonumber \\
= & \ \prod^n_{i =1} \mathcal{N} (g_\theta(x_i, y_i); \delta_\theta ({\bf z}), \Sigma_{\theta}) \myvol [J_{\theta}(x_i, y_i)] \nonumber \\
= & \ \mathcal{N} (\myvec(G_\theta); m_\theta ({\bf z}), I_n \otimes \Sigma_\theta)   \prod^n_{i = 1} \myvol[J_{\theta} (x_i, y_i)]
\end{align}
where $G_\theta = [g_\theta (x_1, y_1), ... g_\theta (x_n, y_n)]$ is a $s \times n$ matrix and $vec(G_\theta) \in \mathbb{R}^{ns}$ is the vectorisation of the matrix $G_\theta.$ The mean vector $ m_\theta ({\bf z}) :=  [\delta_\theta ({\bf z})^{\top} ... \delta_\theta ({\bf z})^{\top}]^\top$ where each $\delta_\theta({\bf z}) = [\delta(z_1), ..., \delta(z_s)]^\top.$ To obtain the marginal pseudolikelihood of 
$p(\{(x_i, y_i)\}^n_{i = 1}|  \theta)$, we compute the integral 
\begin{align}
p(\{(x_i, y_i)\}_{i=1}^n| \theta) &= \int p(\{(x_i, y_i)\}^n_{i = 1}| \delta,  \theta) p(\delta| \theta) d\delta \nonumber \\
&= \int \mathcal{N} (\myvec(G_\theta); m_\theta (z), I_n \otimes \Sigma_\theta) \mathcal{ N} ( \delta; 0, R_\theta) \prod^n_{i = 1} \myvol[J_{\theta} (x_i, y_i)] d\delta \nonumber  \\
& = \mathcal{ N}(\myvec(G_{\theta}); 0, \mathbbm{1} \mathbbm{1}^{\top} \otimes R_\theta + I_n \otimes \Sigma_\theta) \prod^n_{i = 1} \myvol[J_{\theta} (x_i, y_i)].  
\end{align} \label{eq6: marginal_pseudolikelihood_alter}

\end{proof*}

We propose the efficient computation method through the following proposition.

\begin{prop} \label{sec6: proporsition_Wtheta}
 Denote $W_\theta:=\1\1^T\otimes R_\theta+I_n\otimes\Sigma_\theta$. Then,
 \begin{enumerate}
  \item[(a)] $\det(W_\theta)=\det(\Sigma_\theta +nR_\theta)\det(\Sigma_\theta)^{n-1}$.
  \item[(b)] $\myvec(G_\theta)^\top W^{-1} \myvec(G_\theta)= \myTr\left(\left(\Sigma_\theta +nR_\theta \right)^{-1}G_\theta G_\theta^{\top}+\left(\frac{1}{n}\Sigma_\theta R_\theta^{-1}\Sigma_\theta +\Sigma_\theta \right)^{-1}G_\theta HG_\theta^{\top}\right)$
 \end{enumerate}
 where $H = I - \frac{1}{n} \1 \1^\top$.
\end{prop}

\begin{proof*}
 (a) Using matrix determinant lemma, and note that for any $m\times m$ matrix,
 $$\1\1^\top\otimes A = \left(\1\otimes I_m\right)A\left(\1 \otimes I_m\right)^\top.$$
 Hence, the first term of $W$ can be written as
 \begin{eqnarray*}
 \1\1^\top\otimes R_\theta = \left(\1\otimes I_m\right)R_\theta \left(\1 \otimes I_m\right)^\top.
 \end{eqnarray*}
For the second term of $W$, 
 \begin{eqnarray*}
 \det(I_n \otimes \Sigma_\theta ) = \det (I_n)^m \det (\Sigma_\theta)^n = \det(\Sigma_\theta)^n
 \end{eqnarray*}
by the property of the determinant of Kronecker product.   
We can then write
\begin{align}
&\det(\1\1^\top \otimes R_\theta + I_n \otimes \Sigma_\theta)  \nonumber \\
&= \det((\1\otimes I_m)R_\theta (\1 \otimes I_m) + I_n \otimes \Sigma_\theta) \nonumber \\
&= \det(I + (\1 \otimes I_m)^{\top}(I_n \otimes \Sigma_\theta)^{-1} (\1 \otimes I_m)R_\theta)\det(\Sigma_\theta)^n
\end{align}
 The last equation follows from the property that $det(A + UV^{\top}) = det(I + V^{\top} A^{-1} U)det(A)$ given conformable matrices.  Recall the properties of Kronecker products: $(\1 \otimes I_m)^{\top} = \1^{\top} \otimes I_m^{\top}$; $(I_n \otimes \Sigma_\theta)^{-1} = I_n^{-1} \otimes \Sigma_\theta^{-1} $ and $(A\otimes B)(C\otimes D) = (AC)\otimes (BD)$ for conformable matrices. By repetitive applications of these properties, the above can be simplified into
 \begin{align}
 &  \det(I + (\1^{\top} \otimes  \Sigma_\theta^{-1})(\1 \otimes I_m)R_\theta) \det (\Sigma_\theta)^n  \nonumber \\ 
 &=  \det(I + (\1^{\top} \1) \otimes (\Sigma_\theta^{-1} I_m) R_\theta) \det (\Sigma_\theta)^n  \nonumber \\  
 &= \det(I + n\Sigma_\theta^{-1}R_\theta) \det(\Sigma_\theta)^n  \nonumber \\ 
 &= \det(\Sigma_\theta^{-1}(\Sigma_\theta + nR_\theta)) \det(\Sigma_\theta)^n   \nonumber\\ 
 &=  \det(\Sigma_\theta)^{-1} \det(\Sigma_\theta + nR_\theta) \det (\Sigma_\theta)^n   \nonumber \\  
 &= \det(\Sigma_\theta + nR_\theta)\det (\Sigma_\theta)^{n-1}  \nonumber
 \end{align}
 Hence the desired equation is obtained. \\

 (b) Using Woodbury identity and the properties of the Kronecker product,
 \begin{eqnarray}
  W_\theta^{-1} =  I_n\otimes \Sigma_\theta^{-1} - \left(\1\otimes \Sigma_\theta^{-1}\right)\left(\left(R_\theta\right)^{-1}+n\Sigma_\theta^{-1}\right)^{-1}\left(\1^\top\otimes \Sigma_\theta^{-1}\right).
 \end{eqnarray}
 Note that $\left(I_n\otimes \Sigma_\theta^{-1}\right)\myvec(G_\theta)=\myvec(\Sigma_\theta^{-1}G_\theta)$ and
 $$\left\{\myvec(G_\theta)^\top \left(\1\otimes \Sigma_\theta^{-1}\right)\right\}^\top=\left(\1^\top\otimes \Sigma_\theta^{-1}\right)\myvec(G_\theta)=\myvec(\Sigma_\theta^{-1}G_\theta\1)=\Sigma_\theta^{-1}G_\theta\1,$$
 and rearranging products under the trace, we have
 \begin{align}
 &\myvec(G_\theta)^\top W_\theta^{-1}\myvec(G_\theta) \nonumber \\
 &=  \myvec(G_\theta)^\top \myvec(\Sigma^{-1}_\theta G_\theta) - (\Sigma^{-1}_\theta G_\theta \1 )^\top \left(\left(R_\theta\right)^{-1}+n\Sigma_\theta^{-1}\right)^{-1} \Sigma^{-1}_\theta G_\theta \1  \nonumber \\
 &=  \myTr(G_\theta^\top \Sigma^{-1}_\theta G_\theta) - \myTr(\1^\top G_\theta^\top \Sigma^{-1}_\theta \left(\left(R_\theta\right)^{-1}+n\Sigma_\theta^{-1}\right)^{-1} \Sigma^{-1}_\theta G_\theta \1) \nonumber \\
 &= \myTr\left(\Sigma_\theta^{-1}G_\theta G_\theta^\top- \Sigma_\theta^{-1}\left(\left(R_\theta \right)^{-1}+n\Sigma_\theta^{-1}\right)^{-1}\Sigma_\theta^{-1}G_\theta\1{\1}^\top G_\theta^\top\right).
 \end{align}
 Denoting the matrix under the trace by $A$, we can further simplify 

 \begin{eqnarray*}
A & = & \Sigma_\theta^{-1}G_\theta G_\theta^{\top}-\Sigma_\theta^{-1}\left(\left(nR_\theta\right)^{-1}+\Sigma_\theta^{-1}\right)^{-1}\Sigma_\theta^{-1}G_\theta \left(\frac{1}{n}\1\1^{\top}\right)G_\theta^{\top}\\
 & = & \left(\left(nR_\theta\right)^{-1}+\Sigma_\theta^{-1}\right)\left(\left(nR_\theta\right)^{-1}+\Sigma_\theta^{-1}\right)^{-1}\Sigma_\theta^{-1}G_\theta G_\theta^{\top}\\
 &  & \quad-\Sigma_\theta^{-1}\left(\left(nR_\theta\right)^{-1}+\Sigma_\theta^{-1}\right)^{-1}\Sigma_\theta^{-1}G_\theta \left(\frac{1}{n}\1\1^{\top}\right)G_\theta^{\top}\\
 & = & \left(nR_\theta\right)^{-1}\left(\left(nR_\theta\right)^{-1}+\Sigma_\theta^{-1}\right)^{-1}\Sigma_\theta^{-1}G_\theta G_\theta^{\top}\\
 &  & \quad+\Sigma_\theta^{-1}\left(\left(nR_\theta\right)^{-1}+\Sigma_\theta^{-1}\right)^{-1}\Sigma_\theta^{-1}G_\theta\left(I_{n}-\frac{1}{n}\1\1^{\top}\right)G_\theta^{\top}\\
 & = & \left(\Sigma_\theta+nR_\theta \right)^{-1}G_\theta G_\theta^{\top}+\left(\frac{1}{n}\Sigma_\theta R_\theta^{-1}\Sigma_\theta+\Sigma_\theta\right)^{-1}G_\theta HG_\theta^{\top}.
\end{eqnarray*}

\end{proof*}

\begin{remark}
The above simplifies further when $\hat\Sigma_\theta$ is the empirical covariance of $G_\theta$'s, i.e. $\hat\Sigma_\theta=\frac{1}{n}G_\theta HG_\theta^\top$. Indeed,
\begin{align*}
&\myTr\left(\left(\hat\Sigma_\theta+nR_\theta\right)^{-1}G_\theta G_\theta^{\top}+\left(\frac{1}{n}\hat\Sigma_\theta R_\theta^{-1}\hat\Sigma_\theta+\hat\Sigma_\theta\right)^{-1}G_\theta HG_\theta^{\top}\right)\\
&=\myTr\left(\left(\hat\Sigma_\theta+nR_\theta\right)^{-1}G_\theta G_\theta^{\top}+\left(\hat\Sigma_\theta +nR_\theta\right)^{-1}nR_\theta\;\hat\Sigma_\theta^{-1}G_\theta HG_\theta^{\top}\right)\\
&=\myTr\left(\left(\hat\Sigma_\theta+nR_\theta\right)^{-1}G_\theta G_\theta^{\top}+\left(\hat\Sigma_\theta +nR_\theta\right)^{-1}n^2R_\theta\right)\\
&=\myTr\left(\left(G_\theta HG_\theta^\top+n^2R_\theta\right)^{-1}nG_\theta G_\theta^{\top}+\left(G_\theta HG_\theta^\top +n^2R_\theta \right)^{-1}n^3R_\theta \right)\\
&=\myTr\left(n\left(G_\theta HG_\theta^\top+n^2R_\theta \right)^{-1}(G_\theta G_\theta^{\top}+n^2R_\theta)\right).
\end{align*}
\end{remark}

Under the assumption of the null model in \ref{sec6: NullModel} one obtains by a similar argument as before that 
\begin{align}
g_{\theta} (x_i,y_i) | \theta \sim \mathcal{N}(0, \Sigma_\theta)
\end{align}
for $i = 1, \cdots, n$. This implies that the marginal pseudolikelihood can be written as
\begin{align}
    p(\{x_i, y_i\}^n_{i =1}|\theta)
    &= \prod^n_{i =1} p(g_\theta(x_i, y_i)|\theta) \myvol(J_\theta (x_i, y_i)) \\
    &= \mathcal{N}(\myvec(G_\theta); 0, I_n \otimes \Sigma_\theta) \prod^n_{i =1} \myvol(J_\theta(x_i, y_i)). 
\end{align}
The following lemma shows how the most costly term $\myvec(G_\theta)^\top (I_n \otimes \Sigma_\theta)^{-1} \myvec(G_\theta)$ can be computed more efficiently.
\begin{lemma}
$\myvec(G_\theta)^\top (I_n \otimes \Sigma_\theta)^{-1} \myvec(G_\theta) = \myTr(G^\top \Sigma^{-1} G)$.
\end{lemma}
\begin{proof*}
By properties of matrix inversion:
$$(I_n \otimes \Sigma_\theta)^{-1} = I_n^{-1} \otimes \Sigma^{-1}_\theta = I_n \otimes \Sigma_\theta^{-1}.$$
Note that $(I_n \otimes \Sigma_\theta^{-1})\myvec(G) = \myvec(\Sigma_\theta^{-1} G)$, we have 
\[
\myvec(G)^\top (I_n \otimes \Sigma_\theta^{-1}) \myvec(G) = \myvec(G)^\top \myvec(\Sigma_\theta^{-1} G) = \myTr(G^\top \Sigma_\theta^{-1} G).
\]
\end{proof*}


\section{Covariance Matrix $\Sigma_\theta$ Estimations}
\label{sec6: CovEstSig}
In this section, we consider two approaches of estimating the empirical covariance matrix $\Sigma_\theta$. This is common to both the null and the alternative models. It is important to note that the inducing points ${\bf z} = \{z_i\}^s_{i=1}$ are treated as fixed. The first approach computes the covariance of the random variable $\Delta_{XY}$ directly and assumes independence between the random variables $X$ and $Y$. The second approach, on the other hand, computes the empirical covariance matrix of $g_\theta (X, Y)$ and does not require the independence assumption.  \\

\noindent {\bf Method 1} \\
In this approach, we consider the covariance of the random variable $\Delta_{XY}$. The $ij^{th}$ component of the estimated covariance matrix $\hat \Sigma_{ij} = Cov([\Delta_{XY}]_i, [\Delta_{XY}]_j)$ can be written as
\begin{align}
    &\mathbb{E}_{XY}((\widehat \mu_X (z_i) - \widehat \mu_Y (z_i))(\widehat \mu_X (z_j) - \widehat \mu_Y (z_j)))  \\
    &= \mathbb{E}_{XY} (\widehat \mu_X(z_i) \widehat \mu_X(z_j) - \widehat \mu_X(z_i) \widehat \mu_Y(z_j) + \widehat \mu_Y(z_i) \widehat \mu_Y(z_j) - \widehat \mu_Y(z_i) \widehat \mu_X(z_j)). \label{eq: M1cov}
\end{align}
The first term of \eqref{eq: M1cov} can be expanded as
\begin{align}
    &\mathbb{E}_X (\widehat \mu_X (z_i) \widehat \mu_X (z_j)) \\
    &=  \mathbb{E}_X \left (\frac{1}{n^2} \sum^n_{a=1} \sum^n_{b=1} k(X_a, z_i) k(X_b, z_j)\right ) \\
    &=  \frac{1}{n^2} \left [ \sum^n_{a =1} \mathbb{E}_X (k(X_a,z_i) k(X_a, z_j)) + \sum^n_{b=1} \sum_{b \not = a} \mathbb{E}_X( k(X_a, z_i) k(X_b, z_j) ) \right ] \\
    &= \frac{1}{n^2} \left [n \mathbb{E}_X(k(X, z_i) k(X, z_j)) + n(n-1) \mathbb{E}_X (k(X,z_i)k(X, z_j)) \right ] 
\end{align}
Given observations $\{(x_i, y_i)\}^n_{i=1}$, the above is now readily estimated as 
\begin{align}
    &\approx  \frac{1}{n^2} \sum^n_{a =1} k(x_a, z_i)k(x_a, z_j) + \frac{n-1}{n} \left (\frac{1}{n} \sum^n_{a=1} k(x_a, z_i)\right ) \left ( \frac{1}{n} \sum^n_{b = 1} k(x_b, z_j) \right ) \\
    &=  \frac{1}{n^2} K_{z_i {\bf x}} K_{{\bf x} z_j} + \frac{n-1}{n} \left (\frac{1}{n} K_{z_i {\bf x}} \mathbbm{1} \right ) \left (\frac{1}{n} K_{z_j {\bf x}} \mathbbm{1} \right) \\
    &=  \frac{1}{n^2} K_{z_i {\bf x}} K_{{\bf x} z_j} + \frac{n-1}{n} \widehat \mu_x(z_i) \widehat \mu_x (z_j),
\end{align}
where $\mathbbm{1}$ is a vector of 1s in $\mathbb{R}^n$ and ${\bf x} = [x_1 \cdots x_n]^\top \in \mathbb{R}^{n \times D}$. Similarly, the third term in \eqref{eq: M1cov} can be approximated as 
\begin{align}
    \mathbb{E}_Y (\widehat \mu_Y (z_i) \widehat \mu_Y(z_j)) \approx \frac{1}{n^2} K_{z_i {\bf y}} K_{{\bf y} z_j} + \frac{n-1}{n} (\widehat \mu_y(z_i)) (\widehat \mu_y (z_j))
\end{align}
where ${\bf y} = [y_1 \cdots y_n]^\top \in \mathbb{R}^{n \times D}.$
Assuming that $X$ and $Y$ are independent, the second term in \eqref{eq: M1cov} can be empirically estimated as 
\begin{align}
    &\mathbb{E}_{XY} (\widehat \mu_X (z_i) \widehat \mu_Y (z_j)) \\
    &= \left (\frac{1}{n} \sum^n_{a=1} \mathbb{E}_X( k(X_a, z_i)) \right ) \left ( \frac{1}{n} \sum^n_{b=1} \mathbb{E}_Y( k(Y_b, z_j)) \right ) \\ 
    &\approx  (\widehat \mu_x(z_i)) (\widehat \mu_y (z_j))
\end{align}
and similarly for the fourth term of \eqref{eq: M1cov}. When we consider all ${\bf z} = \{z_1, ..., z_s\}$, \eqref{eq: M1cov} can be written as 
\begin{align}
    &\mathbb{E}_{XY}((\widehat \mu_X ({\bf z}) - \widehat \mu_Y ({\bf z}))(\widehat \mu_X ({\bf z}) - \widehat \mu_Y ({\bf z}))^\top) \label{eq: cov_first_step_LHS}\\
    &\approx  \frac{1}{n^2} K_{{\bf z}{\bf x}} K_{{\bf x}{\bf z}} + \frac{n-1}{n} \widehat \mu_x ({\bf z}) \widehat \mu_x ({\bf z})^\top - \widehat \mu_x({\bf z}) \widehat \mu_y ({\bf z})^\top  \nonumber \\ 
    & \quad  + \frac{1}{n^2} K_{{\bf z}{\bf x}} K_{{\bf x}{\bf z}} + \frac{n-1 }{n} \widehat \mu_y ({\bf z}) \widehat \mu_y ({\bf z})^\top  - \widehat \mu_y ({\bf z}) \widehat \mu_x ({\bf z})^\top . \label{eq: cov_first_step_RHS}
\end{align}
The second term of $\hat \Sigma = Cov( \Delta_{XY})$ can be written as
\begin{align}
    &[\mathbb{E}_{XY}(\widehat \mu_X({\bf z}) - \widehat \mu_Y({\bf z}))][\mathbb{E}(\widehat \mu_X({\bf z}) - \widehat \mu_Y({\bf z}))]^\top  \\
    &\approx  \widehat \mu_x ({\bf z}) \widehat \mu_x({\bf z})^\top + \widehat \mu_y({\bf z}) \widehat \mu_y ({\bf z})^\top - \widehat \mu_x ({\bf z}) \widehat \mu_y ({\bf z})^\top - \widehat \mu_y ({\bf z}) \widehat \mu_x ({\bf z})^\top
\end{align}
Overall the empirical estimate of $\hat \Sigma$ can be computed as 
\begin{align}
    & Cov (\Delta_{XY}) \\
    & =  \mathbb{E}_{XY}( (\widehat \mu_X ({\bf z}) - \widehat \mu_Y ({\bf z}))(\widehat \mu_X ({\bf z}) - \widehat \mu_Y ({\bf z}))^\top) \nonumber \\ 
     & \qquad \qquad - \mathbb{E} (\widehat \mu_X ({\bf z})-  \widehat \mu_Y ({\bf z}))\mathbb{E} (\widehat \mu_X ({\bf z})-  \widehat \mu_Y ({\bf z}))^\top  \\
    & \approx  \frac{1}{n^2} K_{{\bf zx}} K_{\bf xz} - \frac{1}{n} \widehat \mu_x({\bf z}) \widehat \mu_x ({\bf z})^\top + \frac{1}{n^2} K_{\bf zy} K_{\bf yz} - \frac{1}{n} \widehat \mu_y ({\bf z}) \widehat \mu_y ({\bf z})^\top \\
    & =  \frac{1}{n^2} K_{\bf zx} H K_{\bf xz} + \frac{1}{n^2} K_{\bf zy} H K_{\bf yz} \label{eq: Method1}
\end{align}
Since $\widehat \mu_x(z) \widehat \mu_x (z)^\top = \frac{1}{n^2} K_{\bf zx} \mathbbm{1} \mathbbm{1}^\top K_{\bf xz}$ and $H:= I - \frac{1}{n} \mathbbm{1} \mathbbm{1}^\top$, the final estimation equation \eqref{eq: Method1} follows. \\

\noindent {\bf Method 2} \\
In this approach, we consider the random variable $$g_\theta (X, Y) = (k(X, z_1) - k(Y,z_1), ... , k(X, z_s) - k(Y, z_s))^\top.$$ The $ij^{th}$ component of the empirical covariance matrix of $g_\theta (X, Y)$ can be written as 
\begin{align}
    & Cov(k(X, z_i) - k(Y, z_i), k(X,z_j) - k(Y, z_j)) \\
    &  = \mathbb{E}_{XY} \left( \left(k(X,z_i) - k(Y, z_i)\right)\left( k(X,z_j) - k(Y, z_j) \right)\right) \nonumber \\
    & \quad \quad \quad  - \mathbb{E}_{XY} \left( k(X,z_i) - k(Y, z_i) \right) \mathbb{E}_{XY} \left( k(X,z_j) - k(Y, z_j) \right) \label{eq: M2cov}
\end{align}

\noindent The first term of \eqref{eq: M2cov} can be written as
\begin{align}
    &\mathbb{E}_{XY} ((k(X,z_i) - k(Y, z_i))(k(X,z_j) - k(Y, z_j))) \\
    &  =  \mathbb{E}_{XY} (k(X,z_i)k(X, z_j) - k(X, z_i)k(Y, z_j) - k(Y, z_i) k(X, z_j) + k(Y, z_i)k(Y, z_j)) \\
    &  \approx  \frac{1}{n} K_{z_i {\bf x}} K_{{\bf x} z_j} - \frac{1}{n} K_{z_i {\bf x}} K_{{\bf y} z_j} -\frac{1}{n} K_{z_j {\bf x}} K_{{\bf y} z_i} + \frac{1}{n} K_{z_i {\bf y}} K_{{\bf y} z_j} 
\end{align}
While the second term of \eqref{eq: M2cov} can be written as 
\begin{align}
    &\mathbb{E}_{XY}(k(X, z_i) - k(Y, z_i)) \mathbb{E}_{XY}(k(X,z_j) - k(Y, z_j)) \\
    & =  (\mu_X(z_i) - \mu_Y (z_i))( \mu_X(z_j) - \mu_Y(z_j)) \\
    & \approx  \frac{1}{n^2} K_{z_i {\bf x}} \mathbbm{1} \mathbbm{1}^\top K_{{\bf x} z_j} - \frac{1}{n^2} K_{z_i {\bf x}} \mathbbm{1} \mathbbm{1}^\top K_{{\bf y} z_j} - \frac{1}{n^2} K_{z_i {\bf y}} \mathbbm{1} \mathbbm{1}^\top K_{{\bf x} z_i} + \frac{1}{n^2} K_{z_i {\bf y}} \mathbbm{1} \mathbbm{1}^\top K_{{\bf y} z_j}
\end{align}
Hence, combining these two terms and consider all ${\bf z} = \{z_1, ..., z_s\} $:
\begin{align}
    &Cov(g_\theta(X, Y)) \\
    & \approx  \frac{1}{n} K_{\bf zx} H K_{\bf xz} + \frac{1}{n} K_{\bf zy} H K_{\bf yz} - \frac{1}{n} K_{\bf zx} H K_{\bf yz} - \frac{1}{n} K_{\bf zy} H K_{\bf xz} \label{eq: cov_expand}\\
    & =  \frac{1}{n} G_\theta HG_\theta^\top. \label{eq: GHG}
\end{align}
To see how \eqref{eq: GHG} follows,  recall $G_\theta := [g_\theta (x_1, y_1), \cdots, g_\theta(x_n, y_n)] \in \mathbb{R}^{s \times n}$ which can be equivalently written in terms of kernel matrices as $G_\theta = K_{\bf zx} - K_{\bf zy}$. \\ 

Note that the distribution of $g_\theta (x_i, y_i)$ arises from the distribution of $\Delta$ due to Cramer's decomposition theorem, this implies $Cov(\Delta_{XY}) = \frac{1}{n}Cov(g_\theta(X, Y))$. If in addition, we assume the independence between $X$ and $Y$, the computation of $Cov(g_\theta (X,Y))$ will be simplified and the cross terms in \eqref{eq: cov_expand} will be zero. However, even in the case where independence assumption is satisfied, numerical instability leads to slight difference of the covariance computation between the two estimation methods. Nonetheless, we did not observe any difference of the resulting posterior inference in that case. 

\section{Computation of the Jacobian Matrix}
\label{sec6: Computation_JacobianMatrix}
In this section, we detail the computation of the Jacobian term. As the variable under consideration is the same for the null and the alternative models, the Jacobian term remains the same. The variables under consideration is
$$g_\theta: (x,y) \mapsto (k_\theta(x, z_1) - k_\theta(y,z_1), ..., k_\theta(x,z_s)- k_\theta(y,z_s)$$
where $g_\theta: \mathbb{R}^{2D} \rightarrow \mathbb{R}^s.$ For each $i=1, ...,n$, the Jacobian matrix $J_\theta (x_i,y_i) \in \mathbb{R}^{s \times 2D}$:
$$
J_\theta (x_i, y_i) = \bigg[\begin{array}{c|c}
J_x (x_i) & J_y(y_i)
\end{array}\bigg]
$$
where $J_x (x_i) \in \mathbb{R}^{s \times D}$ and $J_y (y_i) \in \mathbb{R}^{s \times D}$ are block matrices separating the terms related to X and those that are related to Y. The $\theta$ dependence is omitted here for notational simplicity. $J_x (x_i)$ can be written as 
\begin{align}
J_x(x_i) = \left[\begin{array}{ccc}
   \frac{\partial}{\partial x_{\cdot 1}}g_\theta(x_i,y_i)_1&\cdots &\frac{\partial}{\partial x_{\cdot D}}g_\theta(x_i,y_i)_1\\
   \vdots & \ddots &\vdots \\
   \frac{\partial}{\partial x_{\cdot 1}}g_\theta(x_i,y_i)_s &\cdots &\frac{\partial}{\partial x_{\cdot D}}g_\theta(x_i,y_i)_s\\
   \end{array}\right] \label{eq: Jx}
\end{align}
and $J_y (y_i)$ can be written as 
\begin{align}
J_y(y_i) = \left[\begin{array}{ccc}
   \frac{\partial}{\partial y_{\cdot 1}}g_\theta(x_i,y_i)_1&\cdots &\frac{\partial}{\partial y_{\cdot D}}g_\theta(x_i,y_i)_1\\
   \vdots & \ddots &\vdots \\
   \frac{\partial}{\partial y_{\cdot 1}}g_\theta(x_i,y_i)_s &\cdots &\frac{\partial}{\partial y_{\cdot D}}g_\theta(x_i,y_i)_s\\
   \end{array}\right] \label{eq: Jy}
\end{align}
where $g_\theta(x_i, y_i)_l$ denotes the $l^{th}$ component of $g_\theta (x_i, y_i)$, i.e. $g_\theta (x_i, y_i)_l = k_\theta(x_i, z_l) - k_\theta(y_i, z_l)$ and $\frac{\partial}{\partial x_{\cdot d}}$ is the partial derivative with respect to the $d^{th}$ dimension of the random variable $x$.\\

For observations of $D$ dimension, the Gaussian RBF kernel under consideration is of the form:
\begin{align}
    k_\theta (x_i, z_l) &= \exp \left (-\frac{1}{2\theta^2} \Vert x_i-z_l \Vert^2_2 \right ) \\
    &= \exp \left (-\frac{1}{2\theta^2} \left [(x_{i1} - z_{l1})^2 + \cdots (x_{iD} - z_{lD})^2 \right ] \right ) 
\end{align}
Let the Gram matrix between $\{x_i\}^n_{i =1}$ and $\{z_j\}^s_{i=1}$ be $K_{xz}$ such that $[K_{xz}]_{il} = k_\theta (x_i, z_l).$ Similarly, $[K_{yz}]_{il} = k_\theta (y_i, z_l)$.\\

For each dimension $d = 1, \cdots, D$, we denote the difference between $x_{id}$ and $z_{jd}$ for $i = 1, \cdots, n$ and $j = 1, \cdots, s$ as $D^x_d \in \mathbb{R}^{n \times s}$:
\begin{align}
    D^x_d & = \left[\begin{array}{ccc}
   x_{1d}- z_{1d}&\cdots &x_{1d} - z_{sd}\\
    & \vdots & \\
   x_{nd} - z_{1d}&\cdots &x_{nd}-z_{sd}\\
   \end{array}\right].
\end{align}
Similarly, we have $D^y_d$ for $d = 1, \cdots D$. \\

Given the form of the kernel function, each term in \eqref{eq: Jx} can be written as 
\begin{align}
\left [ J_x(x_i) \right ]_{lm} & = \frac{\partial g_\theta (x_i, y_i)_l}{\partial x_{\cdot m}} = \frac{\partial k_\theta(x_i, z_l)}{\partial x_{\cdot m}} \\
&= \exp \left (-\frac{1}{2\theta^2} \left [(x_{i1} - z_{l1})^2 + \cdots (x_{iD} - z_{lD})^2 \right ] \right ) \left (-\frac{2}{2\theta^2} (x_{im} - z_{lm}) \right) \\
&= k_\theta(x_i,z_l) \left (-\frac{1}{\theta^2} (x_{im} - z_{lm}) \right)
\end{align}
Similarly for the Jacobian term associated with Y: 
$$\left [ J_y(y_i) \right ]_{lm} = \frac{\partial g_\theta (x_i, y_i)_l}{\partial y_{\cdot m}} = -\frac{\partial k_\theta(y_i, z_l)}{\partial y_{\cdot m}}.$$
In computing the probability of the observations $\{x_i, y_i\}^n_{i=1}$ given the hyperparameter $\theta$, we are interested in computing the log determinant of the Jacobian term  
$\log \det (J_\theta (x_i,y_i)^\top J_\theta(x_i, y_i))$. \\

Note, for a fixed $i$, the square matrix can be written as 
\begin{align}
   J_\theta (x_i,y_i)^\top J_\theta(x_i, y_i)=\left[
\begin{array}{c|c}
 J_x (x_i)^\top J_x(x_i) & J_x(x_i)^\top J_y(y_i) \\ \hline
 J_y (y_i)^\top J_x(x_i) & J_y (y_i)^\top J_y (y_i)
\end{array}\right]  \label{eq: BlockJacobian}
\end{align}
where each $J_\cdot (\cdot_i)^\top J_\cdot (\cdot_i)$ is a block matrix of dimension $D \times D$. For notational convenient, we will omit the $\theta$ in the kernel function $k_\theta(\cdot, \cdot)$ when the context is clear. 
\begin{align}
    J_x(x_i)^\top J_x (x_i) =
    \left[\begin{array}{ccc}
   \sum_{j=1}^s \frac{\partial k(x_i, z_j)}{\partial x_{\cdot 1}} \frac{\partial k(x_i, z_j)}{\partial x_{\cdot 1}} &\cdots& \sum_{j=1}^s \frac{\partial k(x_i, z_j)}{\partial x_{\cdot 1}} \frac{\partial k(x_i, z_j)}{\partial x_{\cdot D}}\\
   &\vdots & \\
   \sum_{j=1}^s \frac{\partial k(x_i, z_j)}{\partial x_{\cdot D }} \frac{\partial k(x_i, z_j)}{\partial x_{\cdot 1}}&\cdots & \sum_{j=1}^s \frac{\partial k(x_i, z_j)}{\partial x_{\cdot D}} \frac{\partial k(x_i, z_j)}{\partial x_{\cdot D}}\\
   \end{array}\right]
\end{align}
The term $J_y (y_i)^\top J_y(y_i)$ is similar, while the cross-term can be computed as
\begin{align}
    J_x(x_i)^\top J_y(y_i) = 
    \left[\begin{array}{ccc}
   -\sum_{j=1}^s \frac{\partial k(x_i, z_j)}{\partial x_{\cdot 1}} \frac{\partial k(y_i, z_j)}{\partial y_{\cdot 1}} &\cdots& -\sum_{j=1}^s \frac{\partial k(x_i, z_j)}{\partial x_{\cdot 1}} \frac{\partial k(y_i, z_j)}{\partial y_{\cdot D}}\\
   &\vdots & \\
   -\sum_{j=1}^s \frac{\partial k(x_i, z_j)}{\partial x_{\cdot D }} \frac{\partial k(y_i, z_j)}{\partial y_{\cdot 1}}&\cdots & -\sum_{j=1}^s \frac{\partial k(x_i, z_j)}{\partial x_{\cdot D }} \frac{\partial k(y_i, z_j)}{\partial y_{\cdot D}}\\
   \end{array}\right]
\end{align}
with 
\begin{align}
     &\left [J_x(x_i)^\top J_x (x_i) \right ]_{uv} \nonumber \\
     &=  \sum_{j=1}^s \frac{\partial k(x_i, z_j)}{\partial x_{\cdot u}} \frac{\partial k(x_i, z_j)}{\partial x_{\cdot v}} \\
     &= \sum_{j=1}^s k_\theta(x_i,z_j) \left (-\frac{1}{\theta^2} (x_{iu} - z_{ju}) \right)k_\theta(x_i,z_j) \left (-\frac{1}{\theta^2} (x_{iv} - z_{jv}) \right) \\
     &= \sum_{j=1}^s \frac{1}{\theta^4} k_\theta(x_i,z_j)^2 (x_{iu} - z_{ju}) (x_{iv} - z_{jv}) \\
     & = \sum_{j=1}^s \frac{1}{\theta^4} [K_{xz}]_{ij}^2 [D^x_u]_{ij}[D^x_v]_{ij}. \label{eq: JxTJx}
\end{align}
Essentially, the above equation indicates that we compute the row sum of the elements-wise product of these three matrices of interest. Similarly, 
\begin{align}
    \left [J_x(x_i)^\top J_y (y_i) \right ]_{uv}
    &= -\sum_{j=1}^s \frac{\partial k(x_i, z_j)}{\partial x_{\cdot u}} \frac{\partial k(y_i, z_j)}{\partial y_{\cdot v}} \\
    & = -\sum_{j=1}^s \frac{1}{\theta^4} [K_{xz}]_{ij} [K_{yz}]_{ij} [D^x_u]_{ij} [D^y_v]_{ij} \label{eq: JxTJy}
\end{align}
Equations \eqref{eq: JxTJx} and \eqref{eq: JxTJy} enable the computation of the square matrix $J_\theta (x_i, y_i)^\top J_\theta (x_i, y_i)$ for each $i =1 , \cdots, n$.

\section{Bayes Factor with Fixed $\theta$} \label{sec: fix_theta_1DGaussian}
Given the framework set out before and assume equal prior probabilities for the null and alternative model, when the lengthscale parameter $\theta$ is fixed, the Bayes factor can be computed directly by utilizing Gaussian distributions of $\Delta$ as 
\begin{align}
BF &= \frac{P(\Delta|M =0, \theta)}{P(\Delta|M =1,\theta)} \label{eq6: BF}\\
    \Delta | M &= 0, \theta \sim N(0, \frac{1}{n}\Sigma_\theta) \\
    \Delta | M &= 1, \theta \sim N(0, R_\theta + \frac{1}{n}\Sigma_\theta).
\end{align}

We consider some simple one dimensional experiments where under the null, both samples come from $N(0,1)$ and under the alternative, the data sample $\{y_i\}^n_{i=1}$ comes from distributions listed in Table~\ref{tab: 1DGaussian}.
\begin{table}[h]
\centering
\begin{tabular}{c|l|l}
Alternative Model & Bayes Factor & $P(M=0|\Delta, \theta)$\\ \hline
$\mathcal{N}(0,1)$ & $1.79 \times 10^{25}$ & $1.00$\\
$\mathcal{N}(0,2^2)$ & $5.44 \times 10^{-5}$ & $5.44 \times 10^{-5}$\\
$\mathcal{N}(0,3^2)$ & $2.12 \times 10^{-51}$ & $2.12 \times 10^{-51}$\\
$\mathcal{N}(1,1)$ & $2.31 \times 10^{-13}$ & $2.31 \times 10^{-13}$\\
$\mathcal{N}(2,1)$ & $3.55 \times 10^{-185}$ & $3.55 \times 10^{-185}$\\
$\mathcal{N}(3,1)$ & $0$ & $0$  \\
$Laplace (0,\sqrt{1/2})
$ & $5.86 \times 10^{22}$ & $1.00$\\
$Laplace (0, 1.5)$ & $2.13 \times 10^{7}$ & $9.99 \times 10^{-1}$\\
$Laplace(0, 0.4)$ & $6.51 \times 10^{-2} $ & $6.11 \times 10^{-2}$
\end{tabular}
\caption{Table of Bayes factors and posterior probability of the null hypothesis. Recall a Bayes factor with a value greater than 1 is supporting the null hypothesis while a value smaller than 1 is supporting the alternative hypothesis. The larger the value of the Bayes factor the stronger the support for the null. Results  obtained using 500 samples with 40 evaluation points and averaged across 100 simulations.}
\label{tab: 1DGaussian}
\end{table}

In this experiment, we fix the $\theta$ parameter to the median heuristic. The results shown in Table~\ref{tab: 1DGaussian} were obtained using 500 samples with 40 evaluation points and averaged across 100 simulations. Equivalently, by the relationship between the posterior probability of $M|\Delta = 0$ and the BF, the model clearly recovers the ground truth for distinguishing Gaussian distributions that differ in mean or variance. On the other hand, when comparing the null distribution with Laplace distributions, the model is struggling to detect the difference at 500 samples. However, when we set $\theta$ by searching over a grid of 60 values in $[0.01,40]$ that minimises the Bayes factor, the proposed method is able to do better as presented in Table \ref{tab: 1DLaplace}. Although in the case of comparing $\mathcal{N}(0,1)$ against $Laplace(0, \sqrt{1/2})$ (i.e. a Laplace distribution with mean 0 and variance 1) at 500 samples, the model still prefers the null hypothesis, we do observe that the value of the Bayes factor is orders of magnitude smaller. On the other hand, when comparing $\mathcal{N}(0,1)$ with $Laplace(0, 1.5)$ and $Laplace(0,0.4)$, we see that the model is more certain that the two distributions are different. This highlights the fact that median heuristic is not the best method to select the lengthscale parameter for this hypothesis testing problem. In what follows, we will present a Bayesian approach that infers the posterior distribution of $\theta$ and hence alleviate the need to choose a fixed lengthscale parameter whose value is crucial for the performance of our test. 

\begin{table}[h]
\centering
\begin{tabular}{c|l|l|c}
Alternative Model & Bayes Factor & $P(M=0|\Delta, \theta)$ & $\theta$ \\ \hline
$Laplace (0, \sqrt{1/2})$ & $4.64 \times 10^{13}$ & $1.00$ & $31.19$ \\
$Laplace (0, 1.5)$ & $2.13 \times 10^{-2} $ & $2.13 \times 10^{-2}$ & $4.75$\\
$Laplace(0, 0.4)$ & $3.12 \times 10^{-7} $ & $3.12 \times 10^{-7}$ & 2.04
\end{tabular}
\caption{Table of Bayes factors with $\theta$ obtained by grid search 60 values over $[0.001, 40]$.}
\label{tab: 1DLaplace}
\end{table}

\section{1-Dimensional Mixture of Gaussian Distributions} \label{sec6: 1DMixture}

We have seen that the proposed Bayesian two-sample testing method works well in distinguishing 1 and 2 dimensional Gaussian distributions. In this section, we provide an additional experiment where we consider 1-dimensional mixture of Gaussian distributions as our null and alternative models. More specifically, we consider the following list of null and alternative models where Figure~\ref{fig6: visual_1Dmixture} provides illustration of samples from these models. 
\begin{itemize}
    \item Null Model: $X \sim 0.5\mathcal{N}(0,1) + 0.5\mathcal{N}(4,1)$
    \item Alternative Model: $Y \sim 0.5\mathcal{N}(0,4) + 0.5\mathcal{N}(4,4)$
    \item Alternative Model: $Y \sim 0.5\mathcal{N}(0,1) + 0.5\mathcal{N}(8,1)$
    \item Alternative Model: $Y \sim 0.5\mathcal{N}(0,4) + 0.5\mathcal{N}(8,4)$
    \item Alternative Model: $Y \sim 0.5\mathcal{N}(2,1) + 0.5\mathcal{N}(6,1)$
    \item Alternative Model: $Y \sim 0.5\mathcal{N}(2,4) + 0.5\mathcal{N}(8,4)$
    \item Alternative Model: $Y \sim 0.5\mathcal{N}(-4,1) + 0.5\mathcal{N}(8,1)$
    \item Alternative Model: $Y \sim 0.5\mathcal{N}(-4, 4) + 0.5\mathcal{N}(8,4)$.
\end{itemize}

\begin{figure}
\begin{minipage}{0.45\textwidth}
    \centering
    \begin{subfigure}{}
        \includegraphics[width=\textwidth]{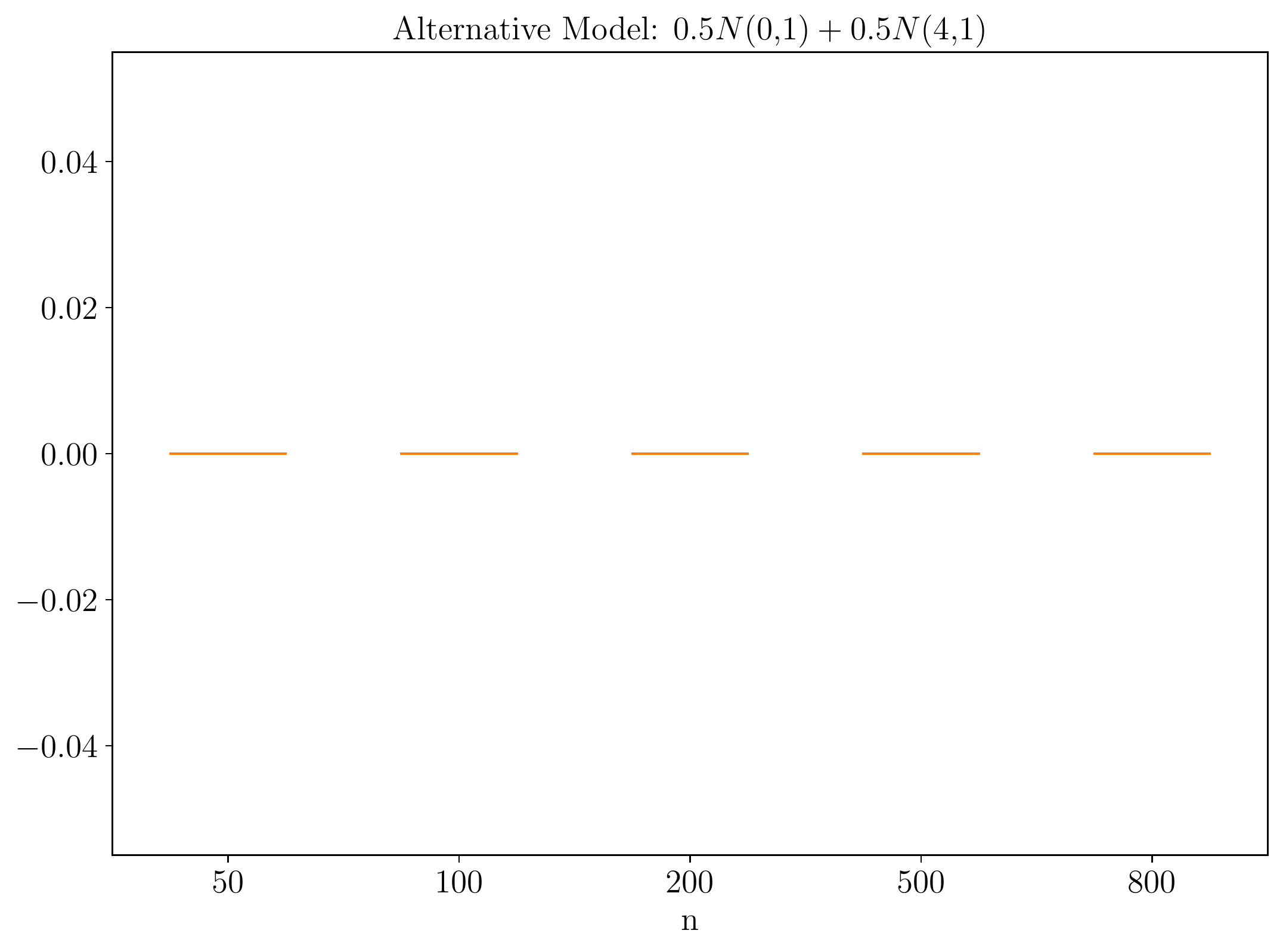}
    \end{subfigure}
    \begin{subfigure}{}
        \includegraphics[width=\textwidth]{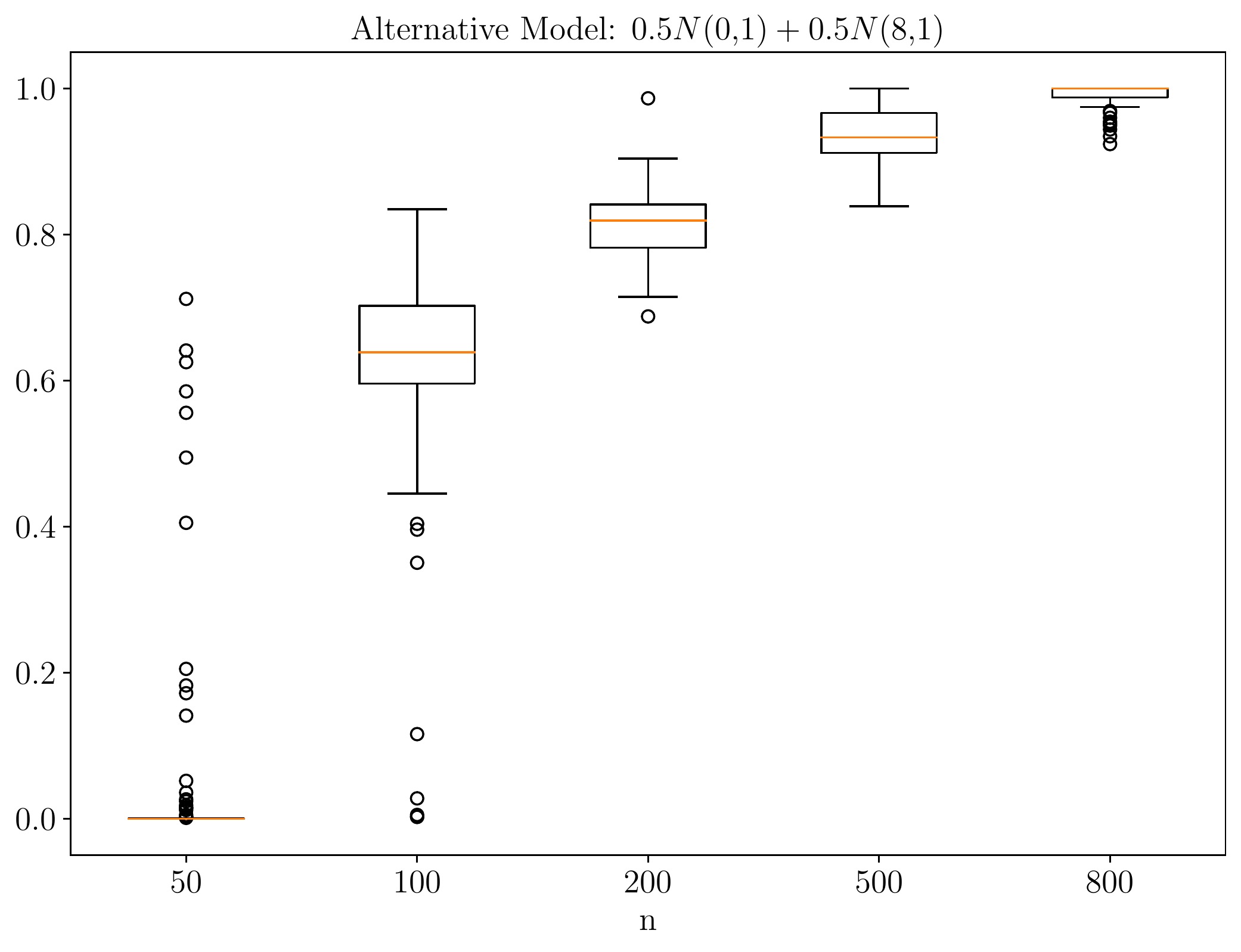}
    \end{subfigure}
\end{minipage}\hfill
\begin{minipage}{0.45\textwidth}
    \centering
    \begin{subfigure}{}
        \includegraphics[width=\textwidth]{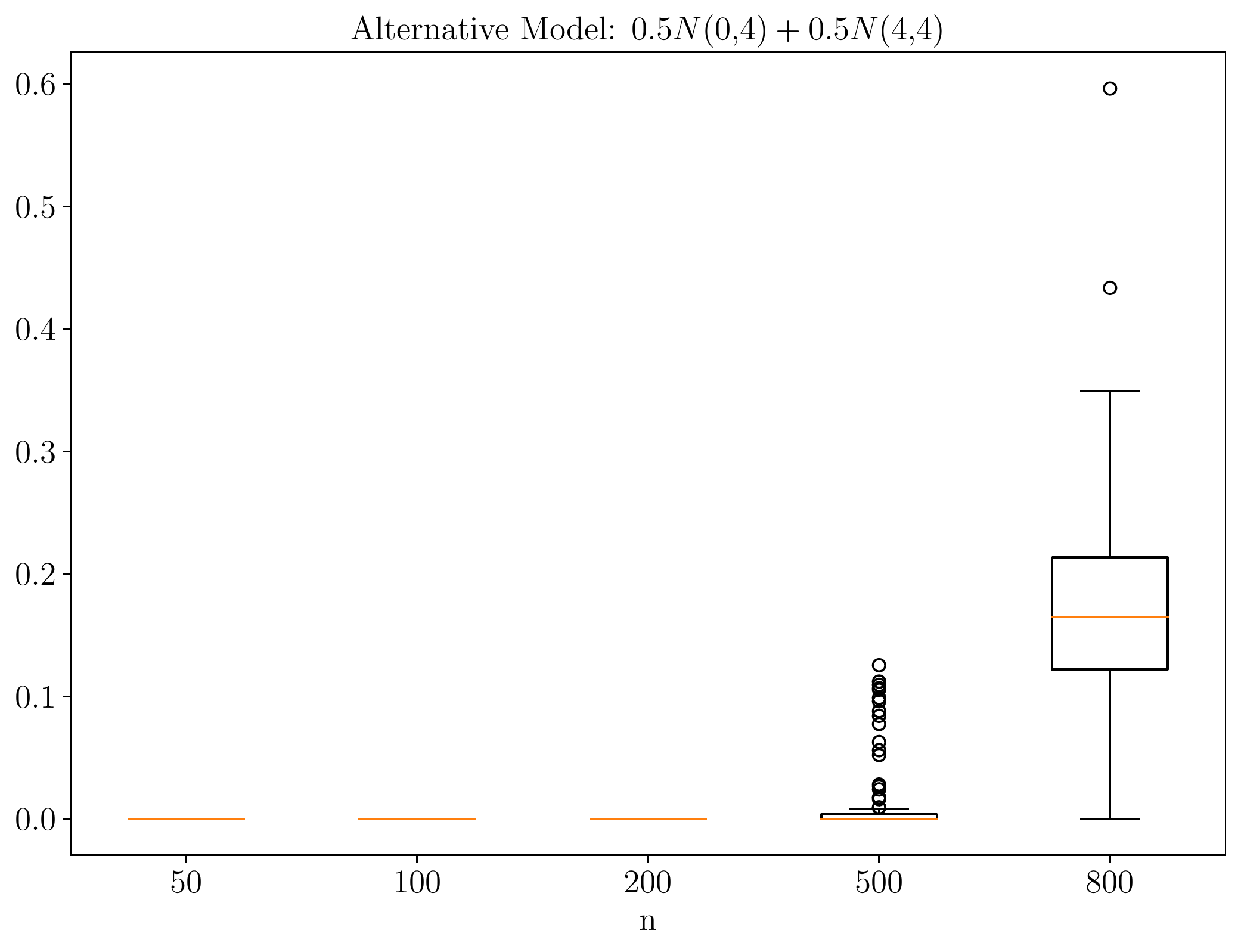}
    \end{subfigure}
    \begin{subfigure}{}
        \includegraphics[width=\textwidth]{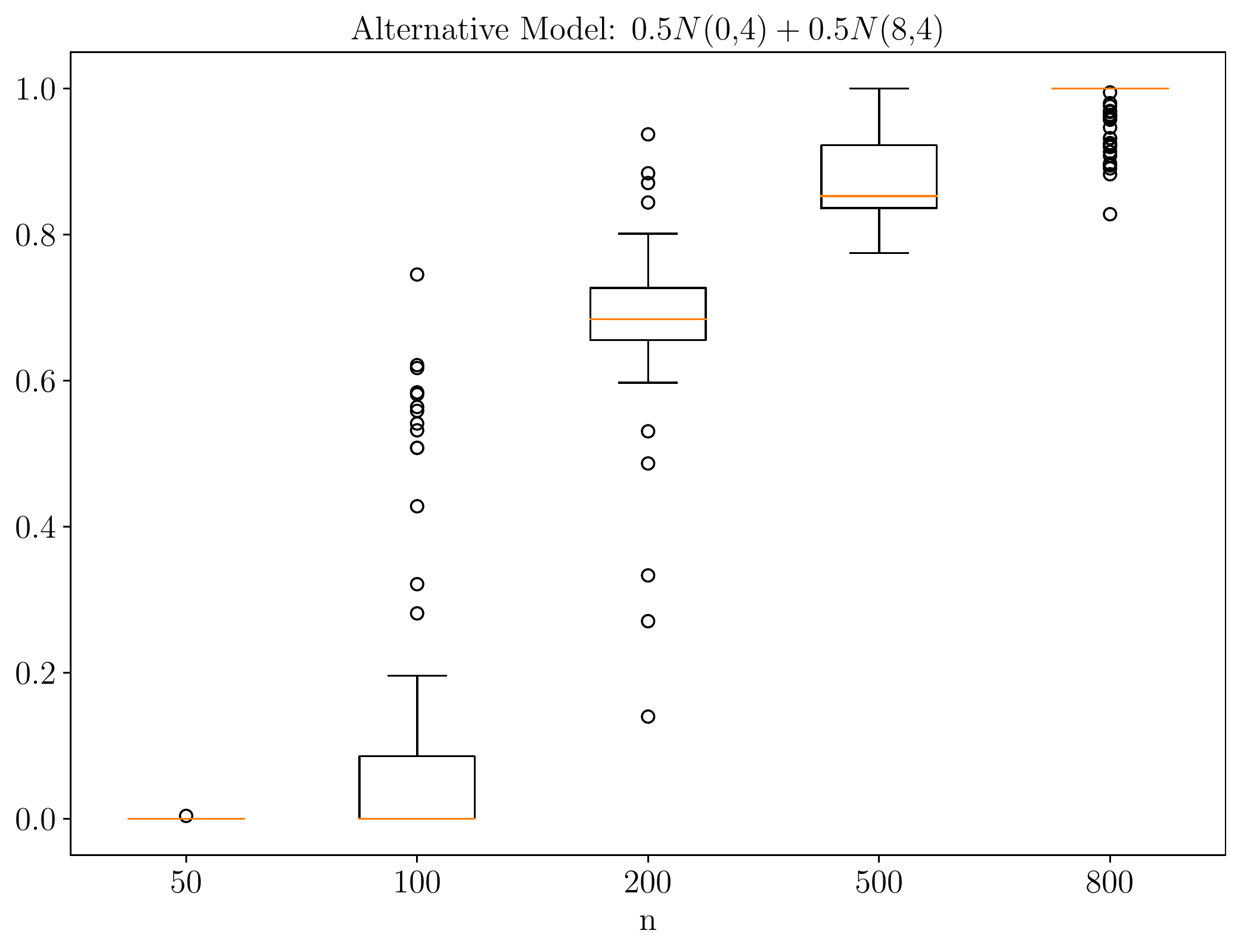}
    \end{subfigure}
\end{minipage}
\caption{1-dimensional mixture of Gaussian distributions: distribution (over 100 independent runs) of the probability of the alternative hypothesis $p(M=1|\mathcal{D})$ for a different number of observations $n$. The null model is $X\iid 0.5\mathcal{N}(0,1) + 0.5\mathcal{N}(4,1)$ and the alternative model is presented as the title of each plot. Continue onto the next page.} \label{fig6: 1Dmixture}
\end{figure}

\begin{figure}
\begin{minipage}{0.45\textwidth}
    \centering
    \begin{subfigure}{}
        \includegraphics[width=\textwidth]{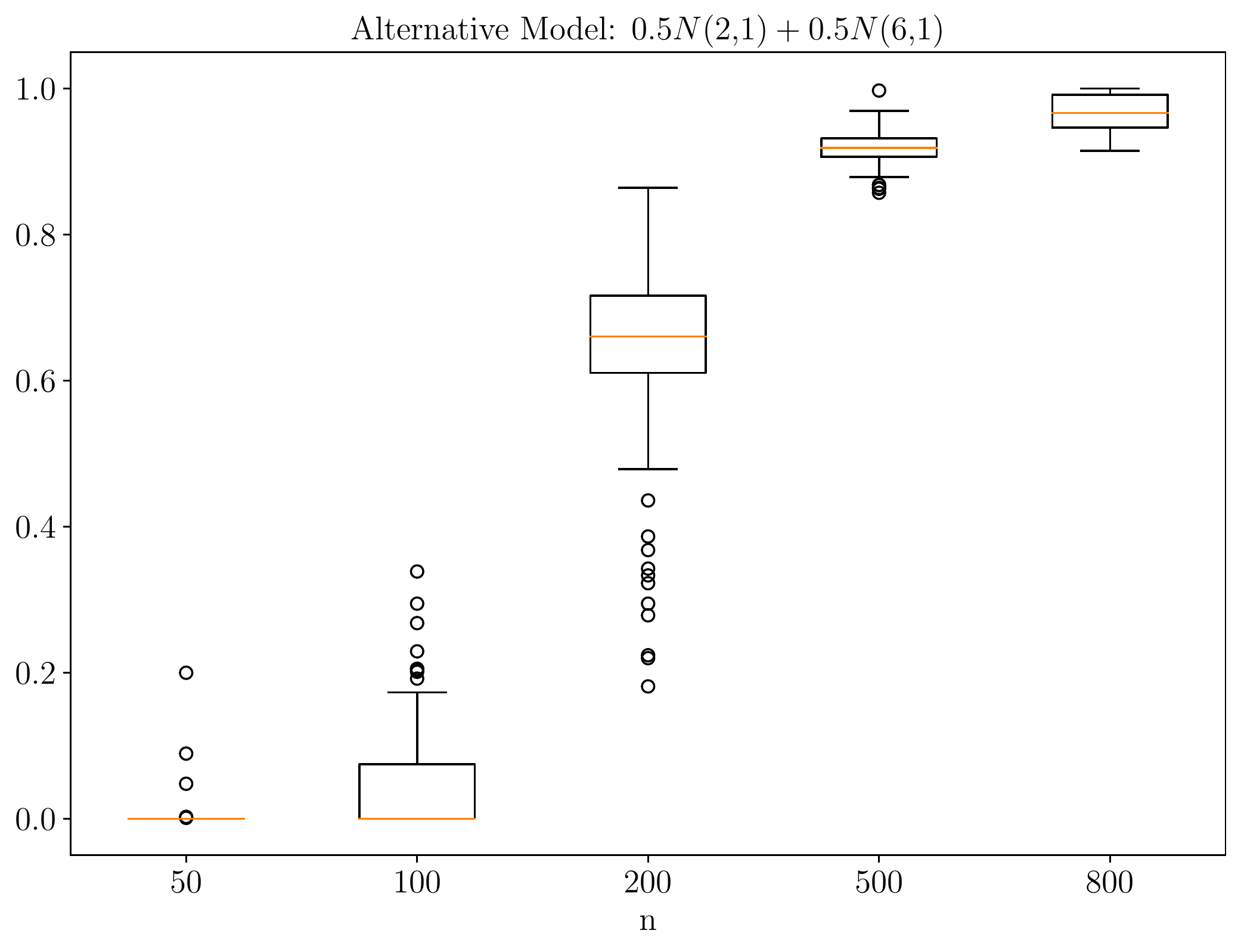}
    \end{subfigure}
    \begin{subfigure}{}
        \includegraphics[width=\textwidth]{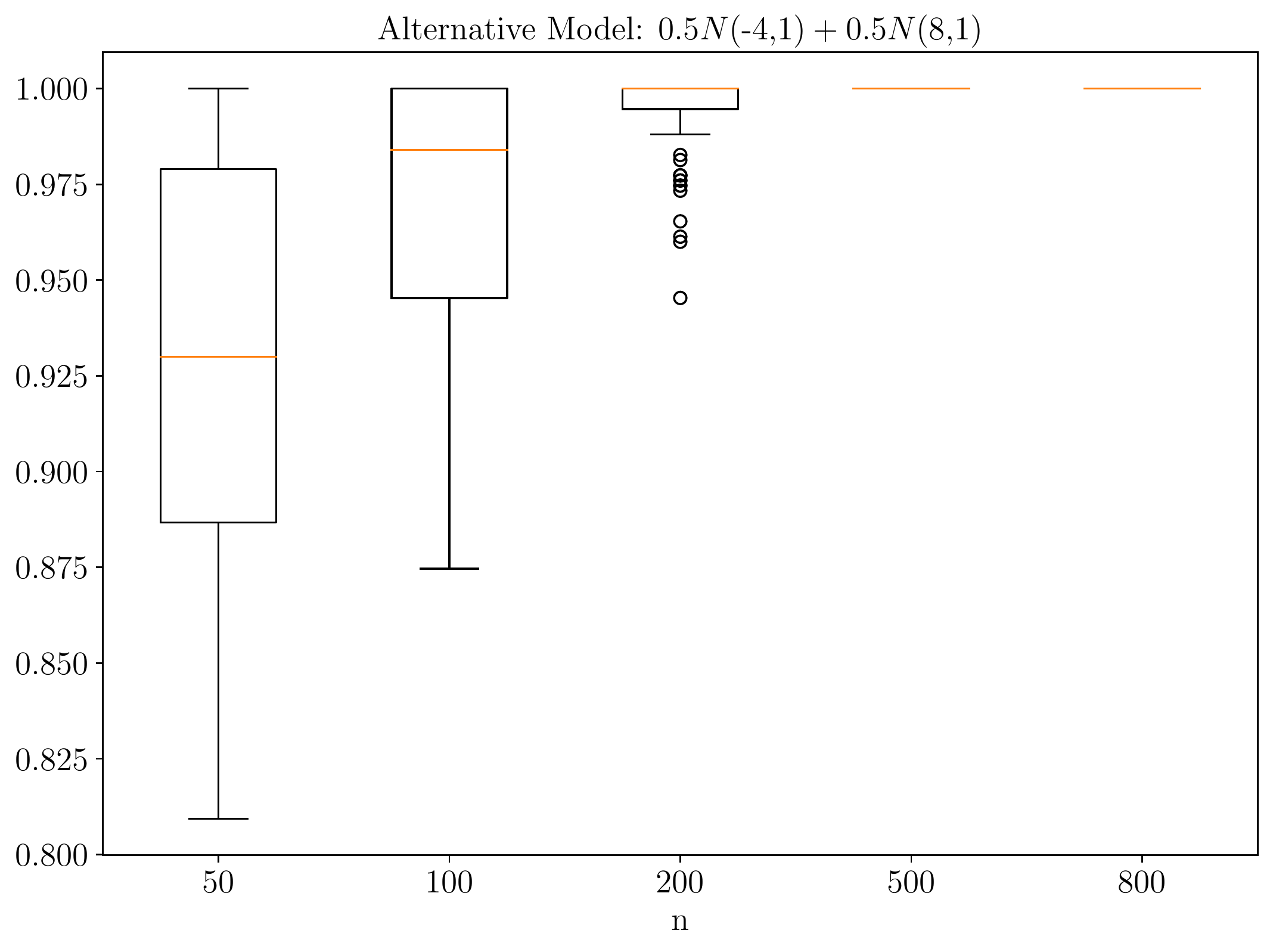}
    \end{subfigure}
\end{minipage}\hfill
\begin{minipage}{0.45\textwidth}
    \centering
    \begin{subfigure}{}
        \includegraphics[width=\textwidth]{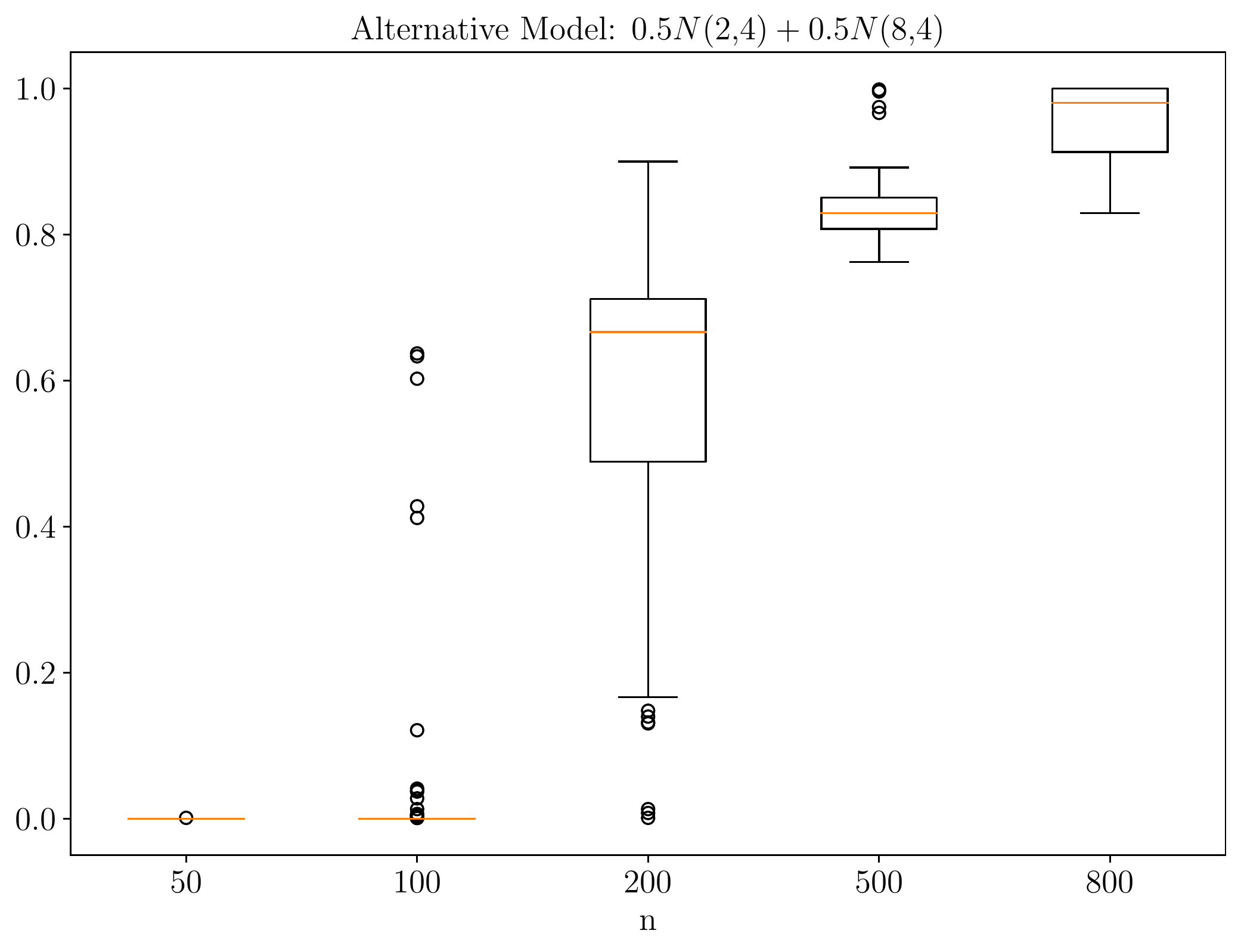}
    \end{subfigure}
    \begin{subfigure}{}
        \includegraphics[width=\textwidth]{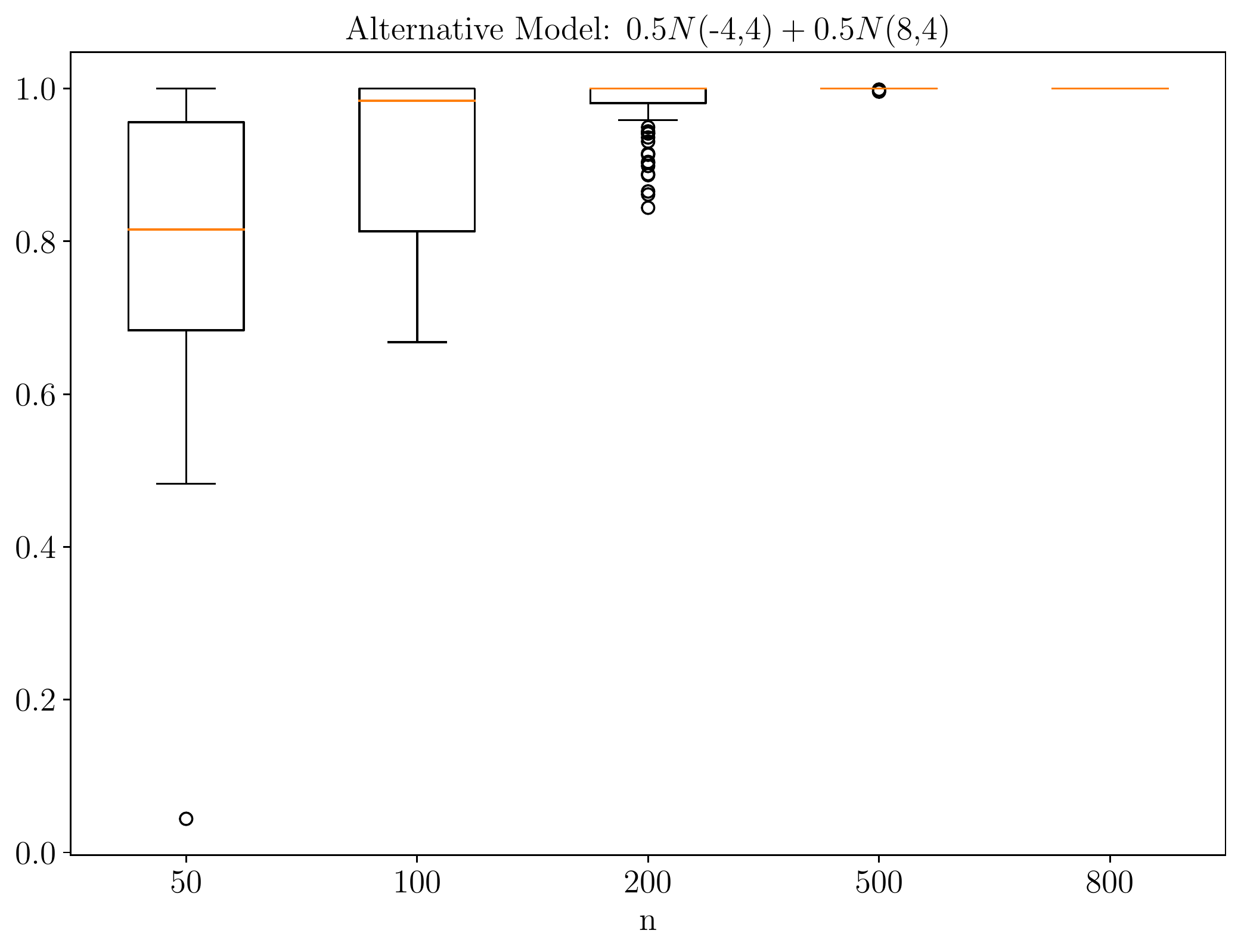}
    \end{subfigure}
\end{minipage}
\caption{1-dimensional mixture of Gaussian distributions: distribution (over 100 independent runs) of the probability of the alternative hypothesis $p(M=1|\mathcal{D})$ for a different number of observations $n$. The null model is $X\iid 0.5\mathcal{N}(0,1) + 0.5\mathcal{N}(4,1)$ and the alternative model is presented as the title of each plot.} \label{fig6: 1Dmixturecont}
\end{figure}

For each of the models, Figure~\ref{fig6: 1Dmixture} and \ref{fig6: 1Dmixturecont} show the distribution of $M=1| \mathcal{D}$ as a function of the number of samples $n$. For the first plot of Figure~\ref{fig6: 1Dmixture}, we observe that the probability consistently equals to zero for all sample sizes and for all simulated datasets. This is reassuring as the data $X$ and $Y$ are indeed simulated from the same distribution, i.e. we are under the null hypothesis. When the alternative model is $Y \sim 0.5\mathcal{N}(0,4) + 0.5\mathcal{N}(4,4)$, the problem is more challenging. At 800 samples, the test is uncertain, but preferring the null hypothesis. This comes back to the aforementioned idea of Bayesian modelling naturally prefers simple hypothesis. In the other plots of Figure \ref{fig6: 1Dmixture} and \ref{fig6: 1Dmixturecont}, we observed that the distribution of $M=1| \mathcal{D}$ gradually concentrates around 1 as the number of samples increases. This aligns with our expectations.

\section{2 Dimensional Gaussian Distributions} \label{sec6: 2DGaussian}
\begin{figure}[h]
\begin{minipage}{0.45\textwidth}
    \centering
    \begin{subfigure}{}
        \includegraphics[width=\textwidth]{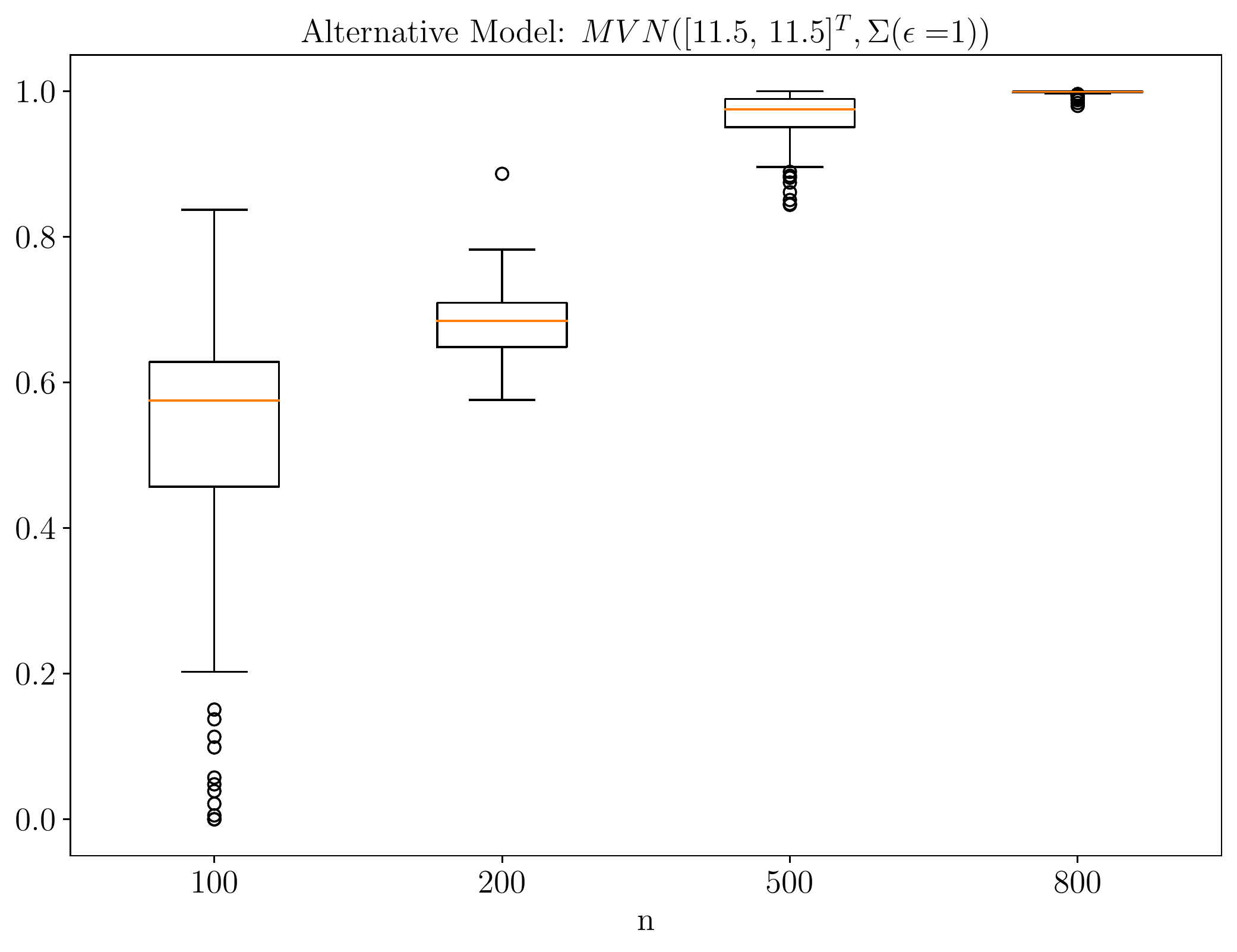}
    \end{subfigure}
\end{minipage}\hfill
\begin{minipage}{0.45\textwidth}
    \centering
    \begin{subfigure}{}
        \includegraphics[width=\textwidth]{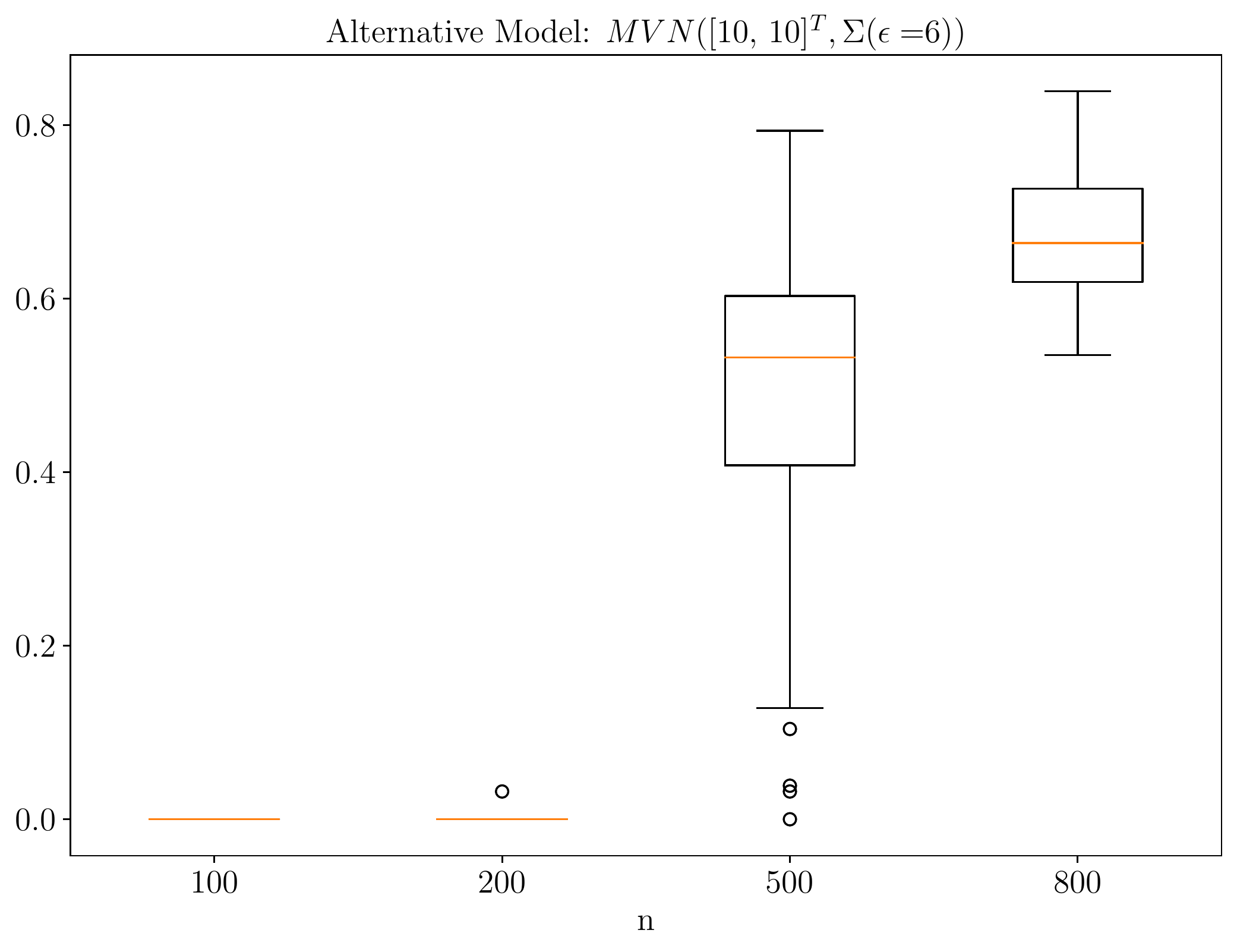}
    \end{subfigure}
\end{minipage}
\caption{2-dimensional Gaussian experiment: distribution (over 100 independent runs) of the probability of the alternative hypothesis $p(M=1|\mathcal{D})$ for a different number of observations $n$. The null and alternative hypotheses are as shown in Table \ref{tab6: 2DGaussian}. We illustrate the results from the first (Left) and forth (Right) alternative model in the Table while the other ones are presented in the Supplementary Material.} \label{fig6: 2DGaussian}
\end{figure}

In this section we consider several experiments to illustrate that the test is consistent in a two dimensional Gaussian setting. The distribution we compare are given as
\begin{align}
    X &\sim \mathcal{N}\left ( \begin{pmatrix} 10  \\ 10  \end{pmatrix}, \begin{pmatrix} 1 &0 \\ 0&1 \end{pmatrix}\right)   \\
    Y &\sim \mathcal{N} \left ( M_y, QS_\epsilon Q^\top \right )
\end{align}
where $M_y \in \mathbb{R}^2$ is the mean vector, $Q$ is the rotation matrix
\begin{equation}
    Q = \begin{pmatrix} \cos(\frac{\pi}{2}) & \sin (\frac{\pi}{2}) \\
    - \sin (\frac{\pi}{2}) & \cos(\frac{\pi}{2}) \end{pmatrix}
\end{equation}
and $S_\epsilon$ is the diagonal matrix of the form 
\begin{equation}
    S_\epsilon = \begin{pmatrix} \epsilon & 0 \\ 0& 1 \end{pmatrix}. 
\end{equation}
For the ease of notation, we denote $\Sigma(\epsilon) := Q S_\epsilon Q^\top$ highlighting the dependency on the parameter $\epsilon$. More specifically, we consider the following parameter settings for the simulated data as shown in Table \ref{tab6: 2DGaussian}. Only the results from the first and the fourth cases of the $Y$ distributions are presented in the main text while the others are in the Supplementary Materials. 
\begin{table}[ht]
\begin{tabular}{l|l|llllll}
 & Null           & Alternative         &                &                &                &                &                \\ \hline
$M_y^\top$            & $(10,10)$ & $(11.5, 11.5)$ & $(12,10)$ & $(10,10)$ & $(10,10)$ & $(10,10)$ & $(10,10)$ \\ \hline
$\epsilon$       & 1              & 1                   & 1              & 2              & 6              & 10             & 20   
\end{tabular}
\caption{The different parameters for the 2-dimensional Gaussian distribution experiment.} \label{tab6: 2DGaussian}
\end{table}
Under the null hypothesis, we simulated both $X$ and $Y$ independently from a 2 dimensional Gaussian distribution with $M_y = (10, 10)^\top$ and $\epsilon=1$. As expected, the distribution of the probability of $M=1$ is consistently zero for all sample sizes. Under the alternative hypothesis, as expected, we observe in Figure \ref{fig6: 2DGaussian} that the probability distribution of the alternative hypothesis becomes concentrated around 1 as the number of samples increases. In other words, as more samples are seen, the Bayes factor is able to favor the alternative hypothesis with more certainty. 

Note, for the case when $\epsilon=2$, the difference between the two distributions is not detected even at the maximum sample size $n=800$ considered and the Bayes factor consistently favours the null hypothesis. Since more samples are needed for the model to be able to detect the difference between the two distributions, we argue that large scale approximations such as random Fourier feature extension of the proposed method is needed. We will provide a brief discussion of this in Section \ref{sec6: futurework}.

\begin{figure}
\begin{minipage}{0.45\textwidth}
    \centering
    \begin{subfigure}{}
        \includegraphics[width=\textwidth]{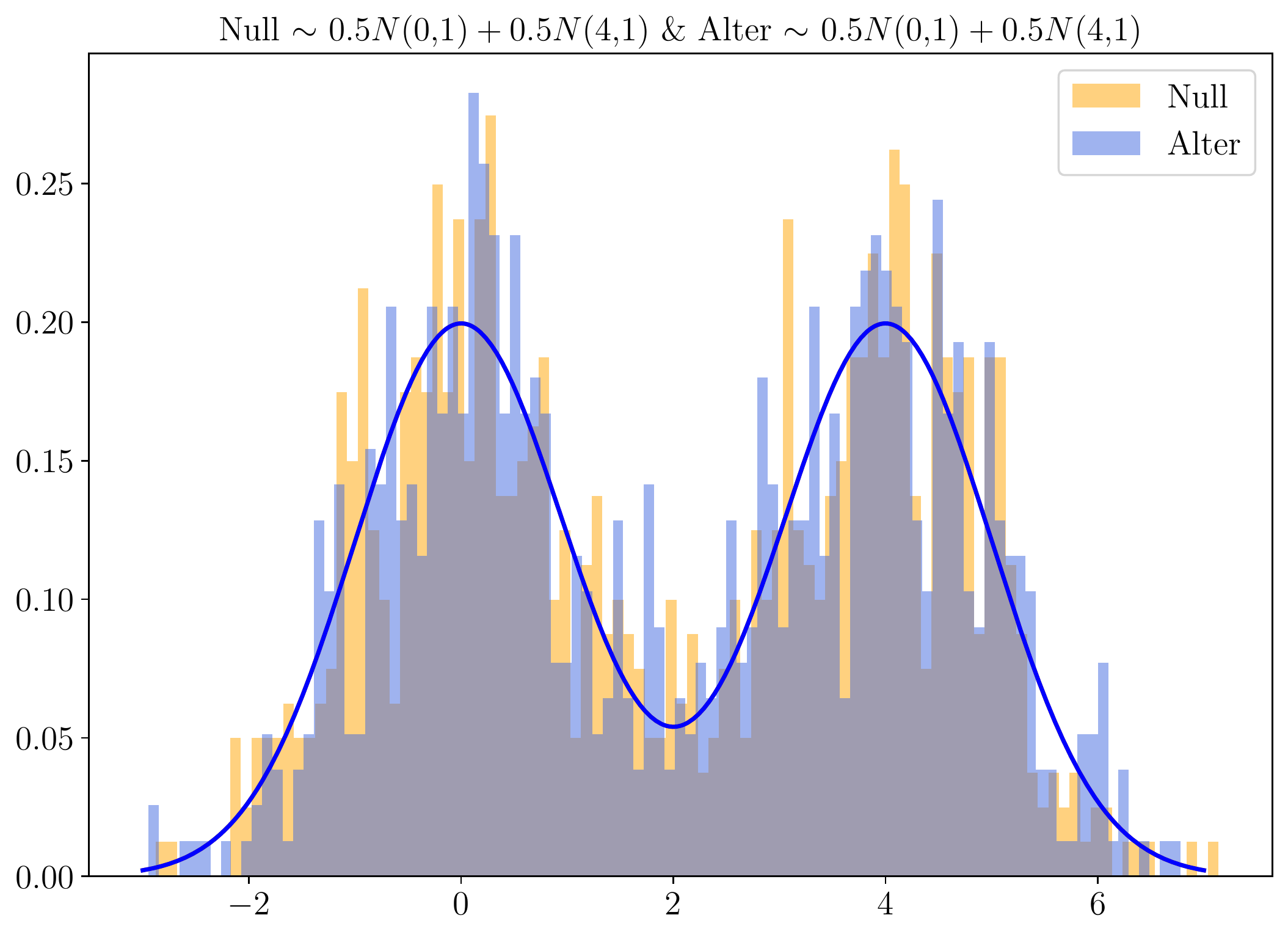}
    \end{subfigure}
    \begin{subfigure}{}
        \includegraphics[width=\textwidth]{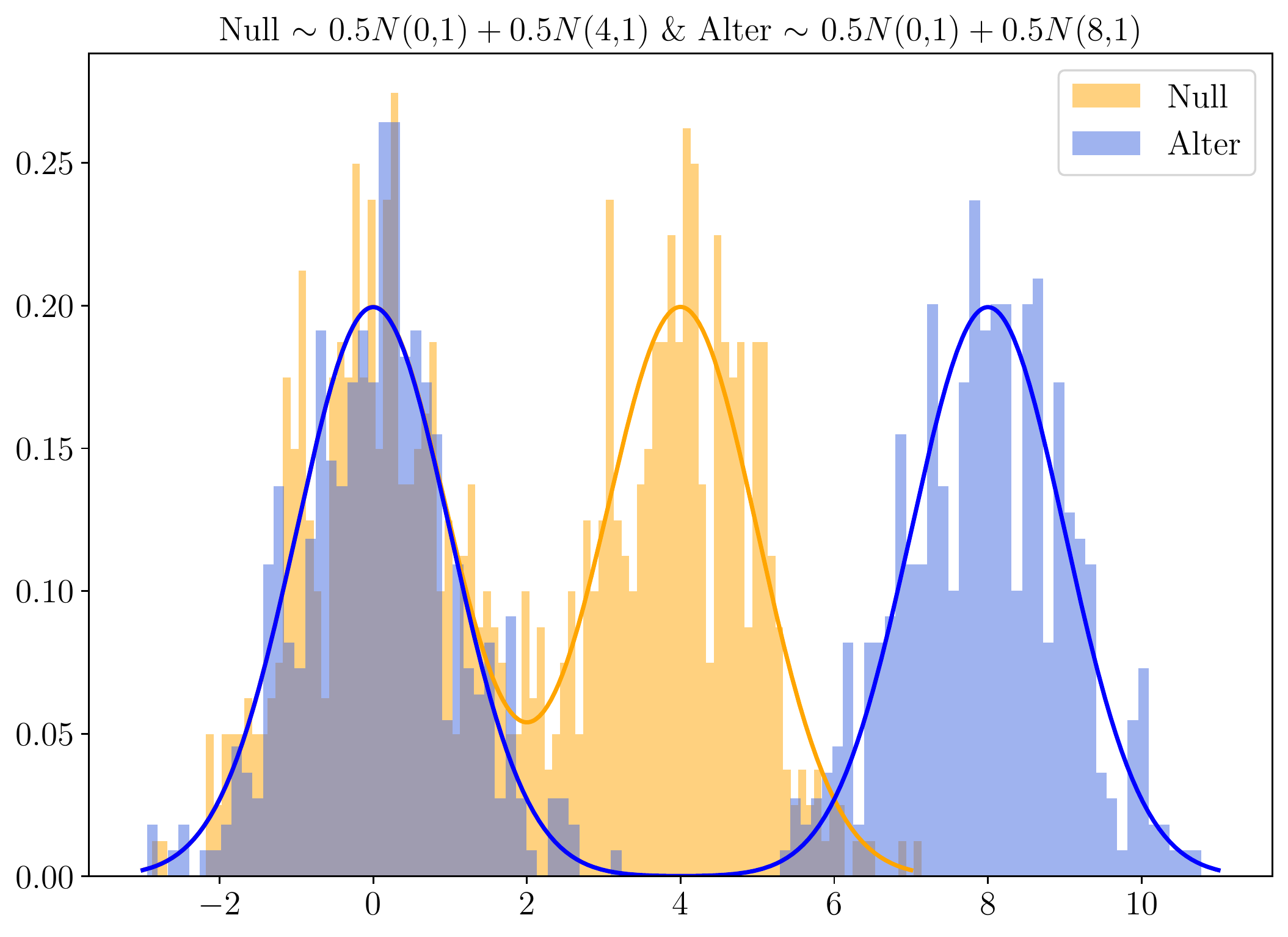}
    \end{subfigure}
\end{minipage}\hfill
\begin{minipage}{0.45\textwidth}
    \centering
    \begin{subfigure}{}
        \includegraphics[width=\textwidth]{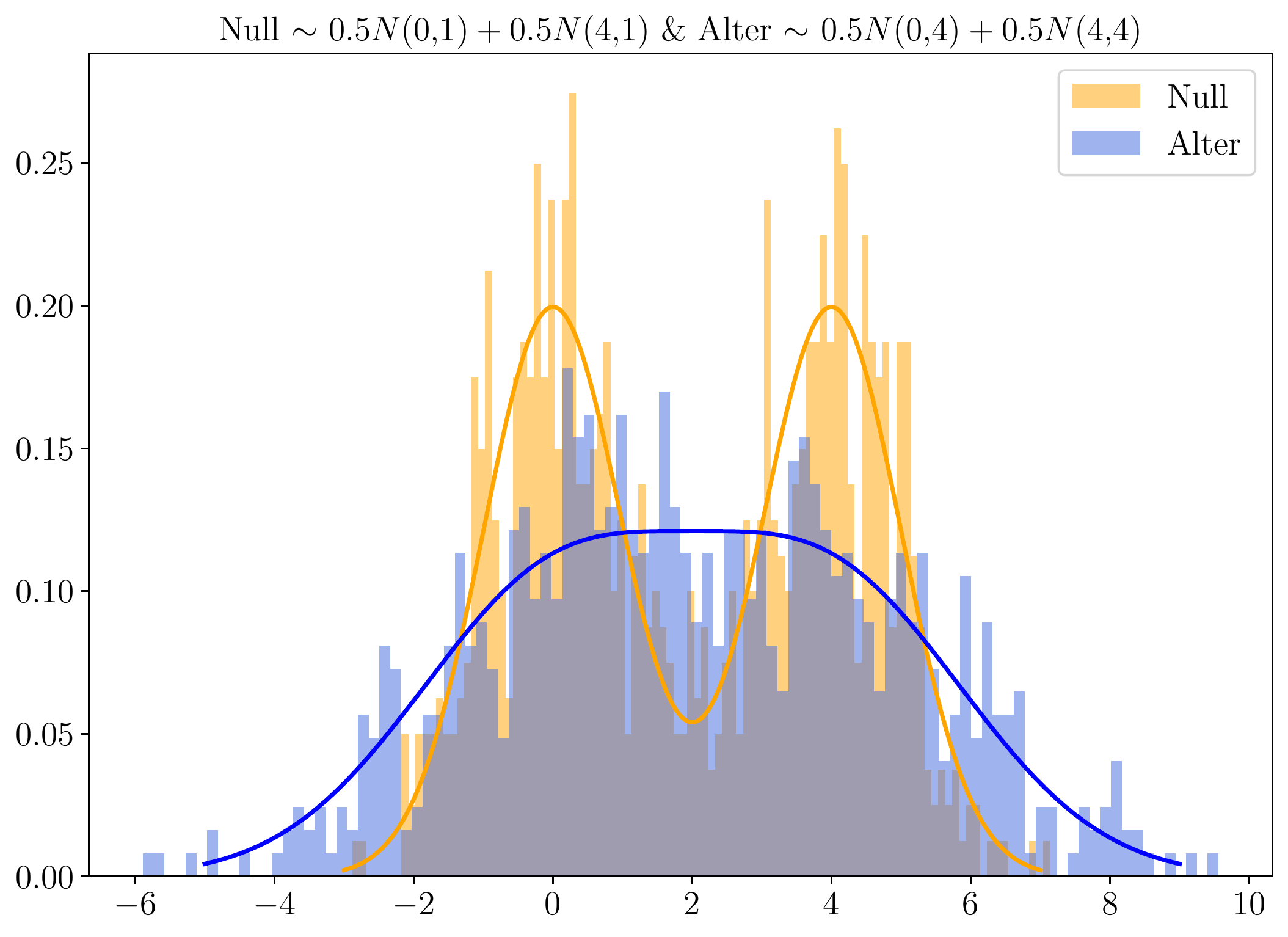}
    \end{subfigure}
    \begin{subfigure}{}
        \includegraphics[width=\textwidth]{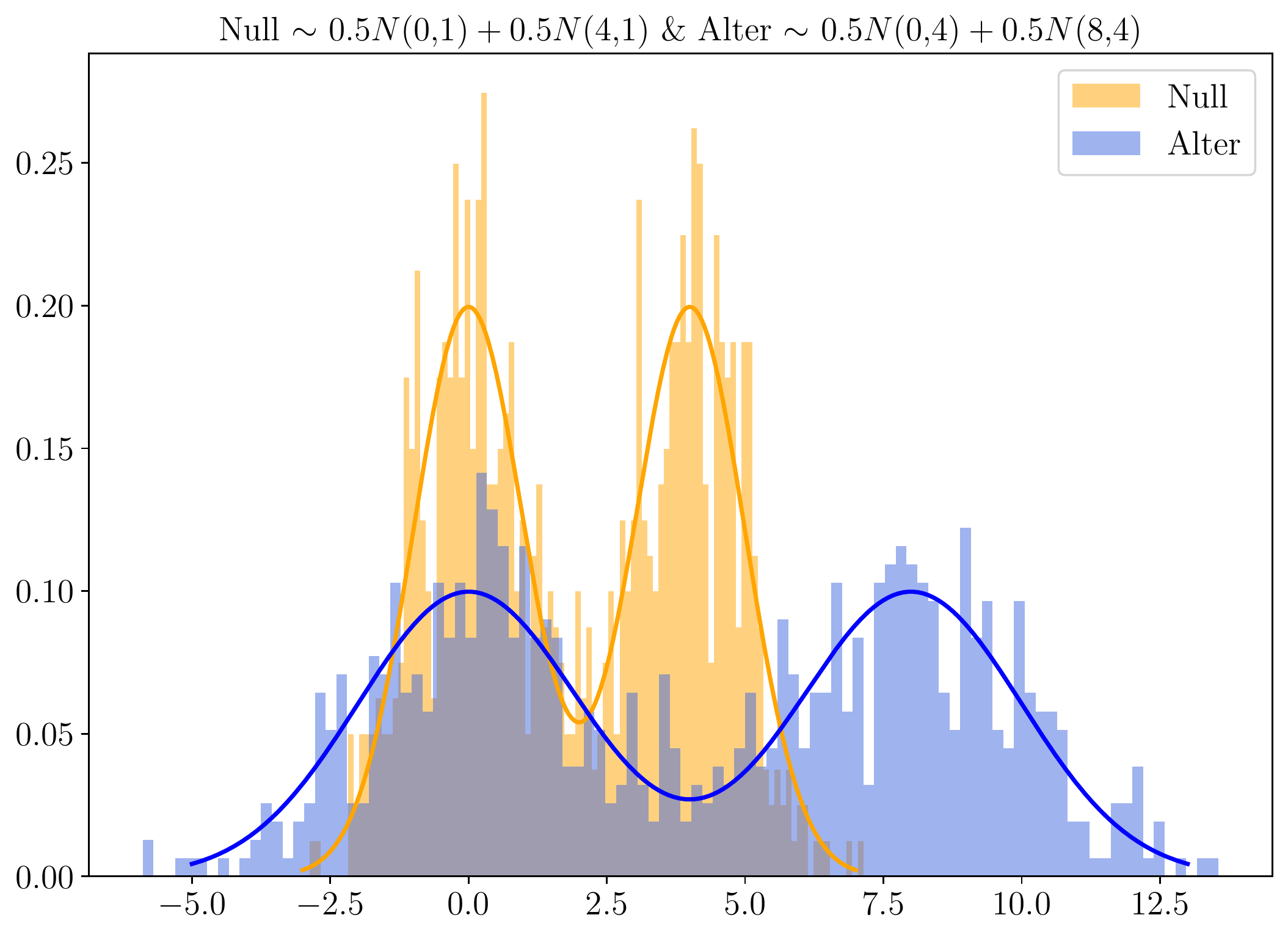}
    \end{subfigure}
\end{minipage}
\caption{Visualisation of 1-dimensional mixture of Gaussian distributions from Section \ref{sec6: 1DMixture} with 800 samples each. Continue on next page.} \label{fig6: visual_1Dmixture}
\end{figure}

\begin{figure}
\begin{minipage}{0.45\textwidth}
    \centering
    \begin{subfigure}{}
        \includegraphics[width=\textwidth]{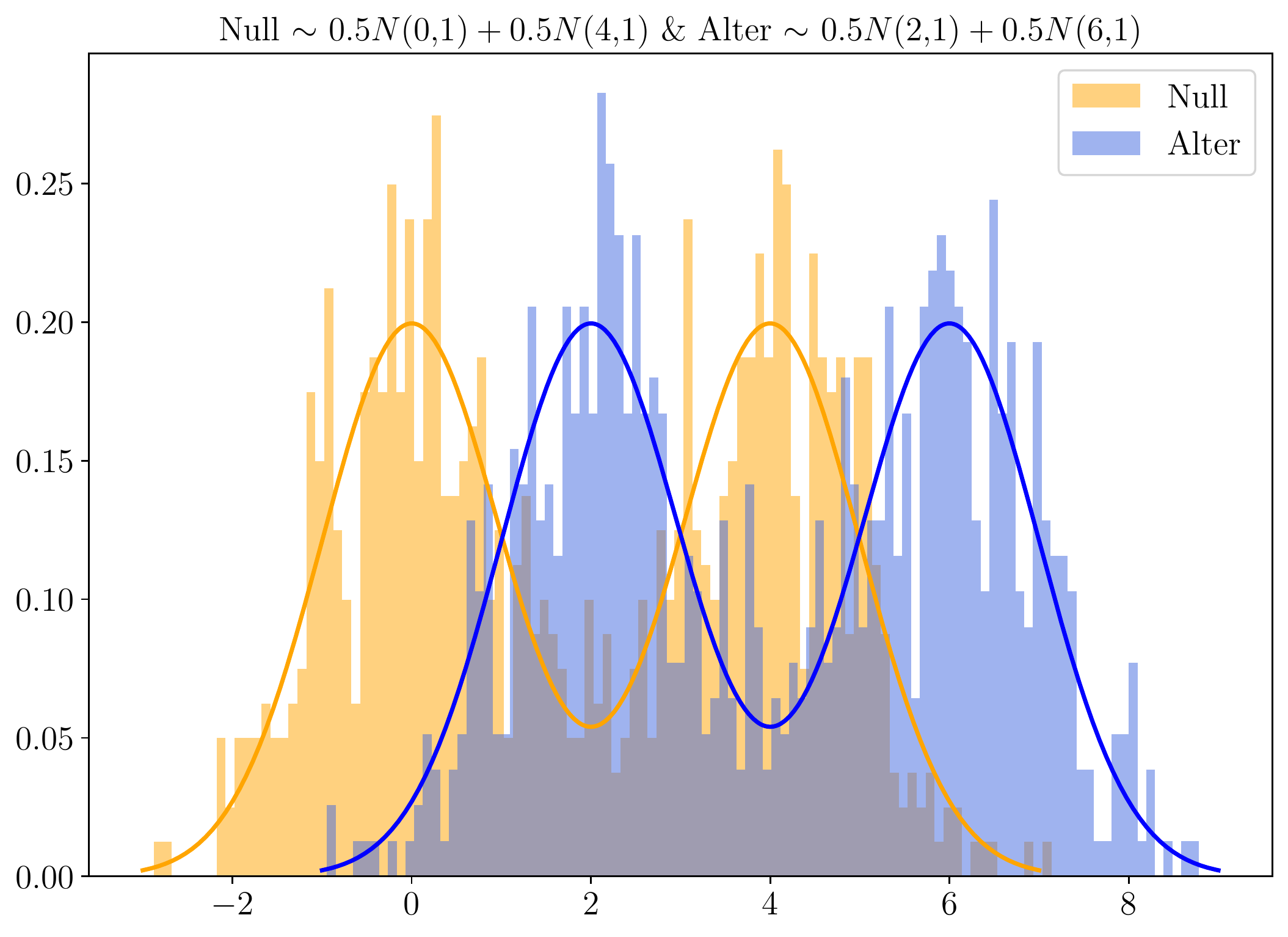}
    \end{subfigure}
    \begin{subfigure}{}
        \includegraphics[width=\textwidth]{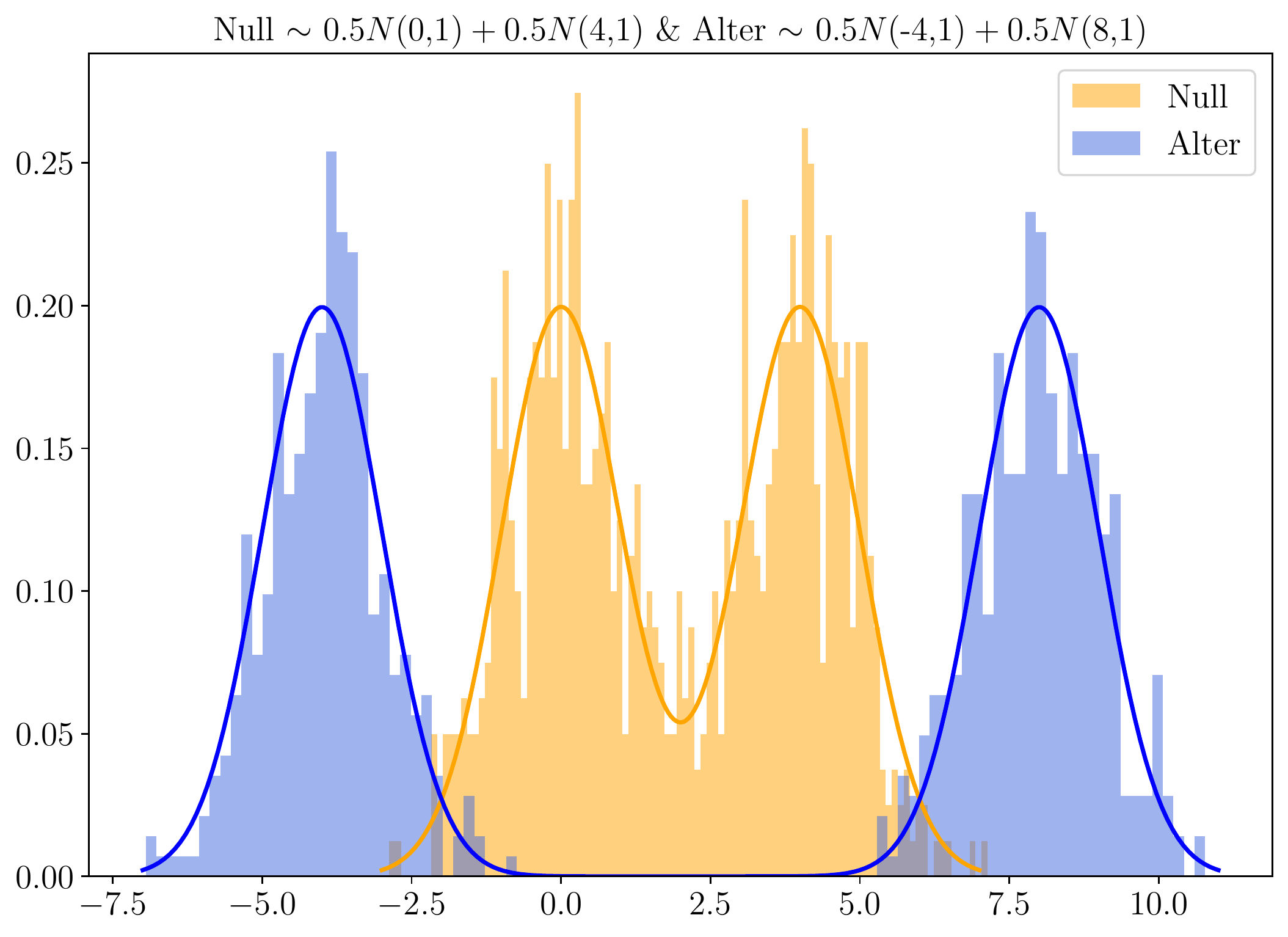}
    \end{subfigure}
\end{minipage}\hfill
\begin{minipage}{0.45\textwidth}
    \centering
    \begin{subfigure}{}
        \includegraphics[width=\textwidth]{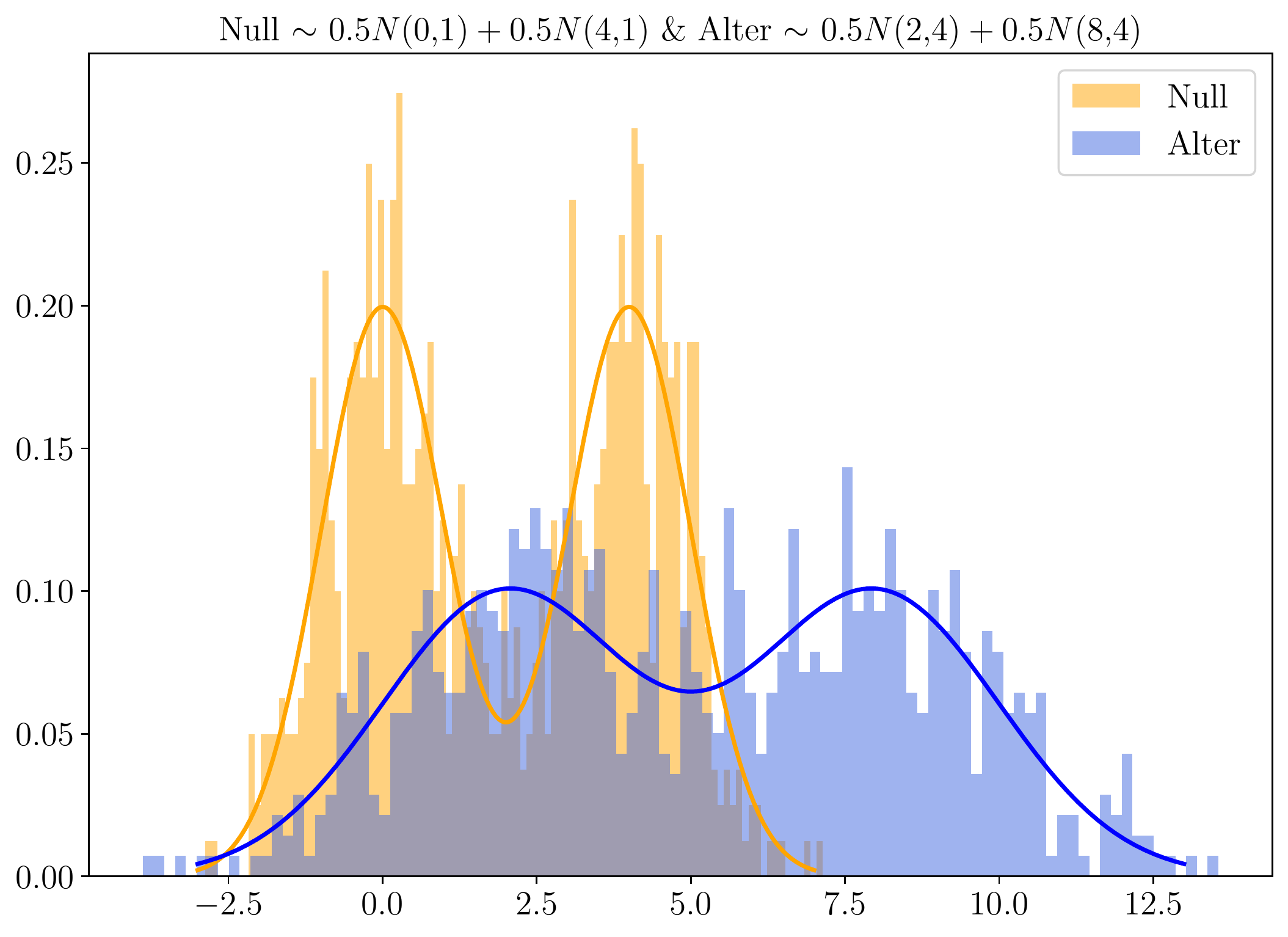}
    \end{subfigure}
    \begin{subfigure}{}
        \includegraphics[width=\textwidth]{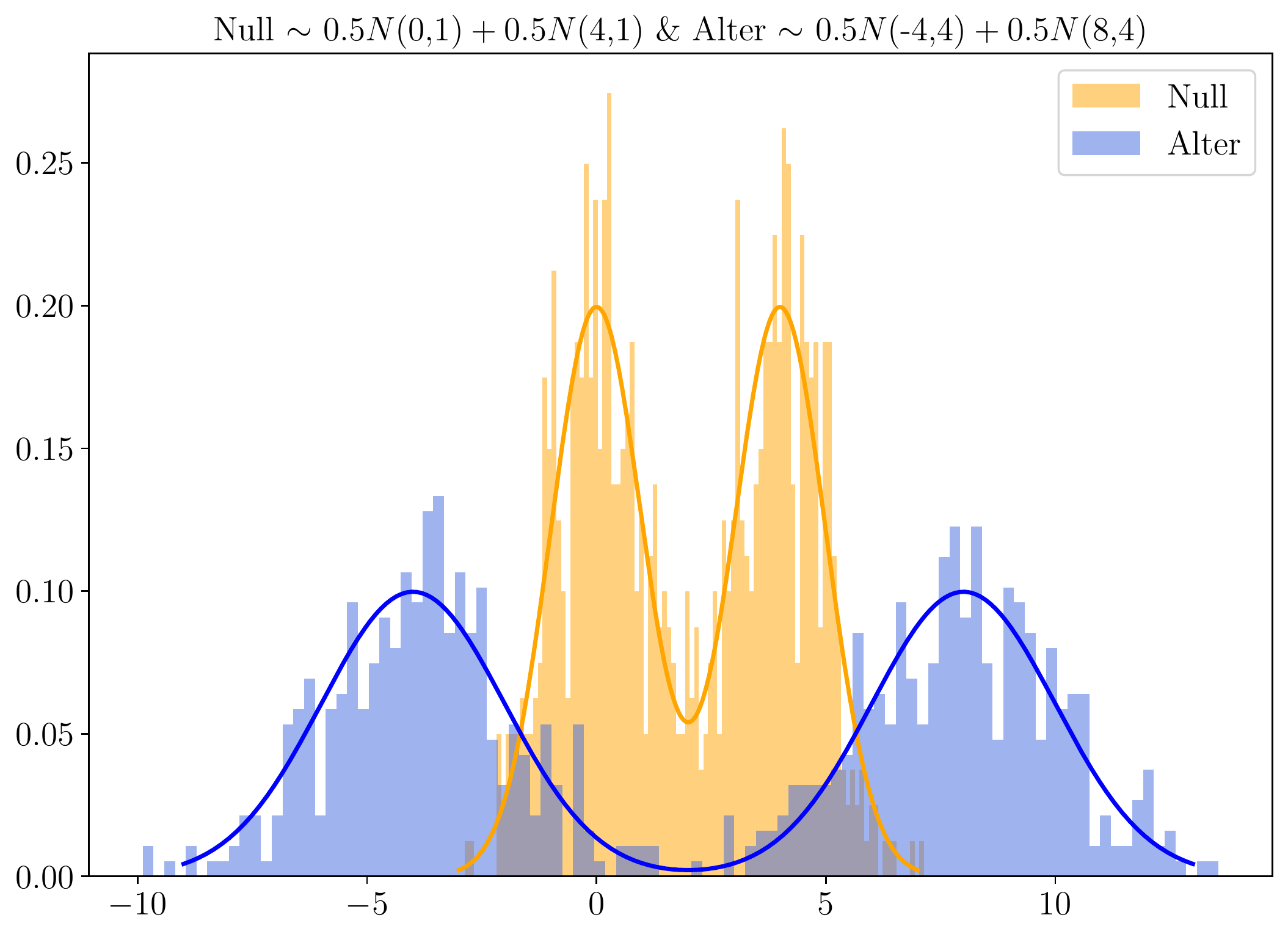}
    \end{subfigure}
\end{minipage}
\caption{Visualisation of 1-dimensional mixture of Gaussian distributions from Section \ref{sec6: 1DMixture} with 800 samples each.} \label{fig6: visual_1Dmixture}
\end{figure}

\begin{figure}
\begin{minipage}{0.45\textwidth}
    \centering
    \begin{subfigure}{}
        \includegraphics[width=\textwidth]{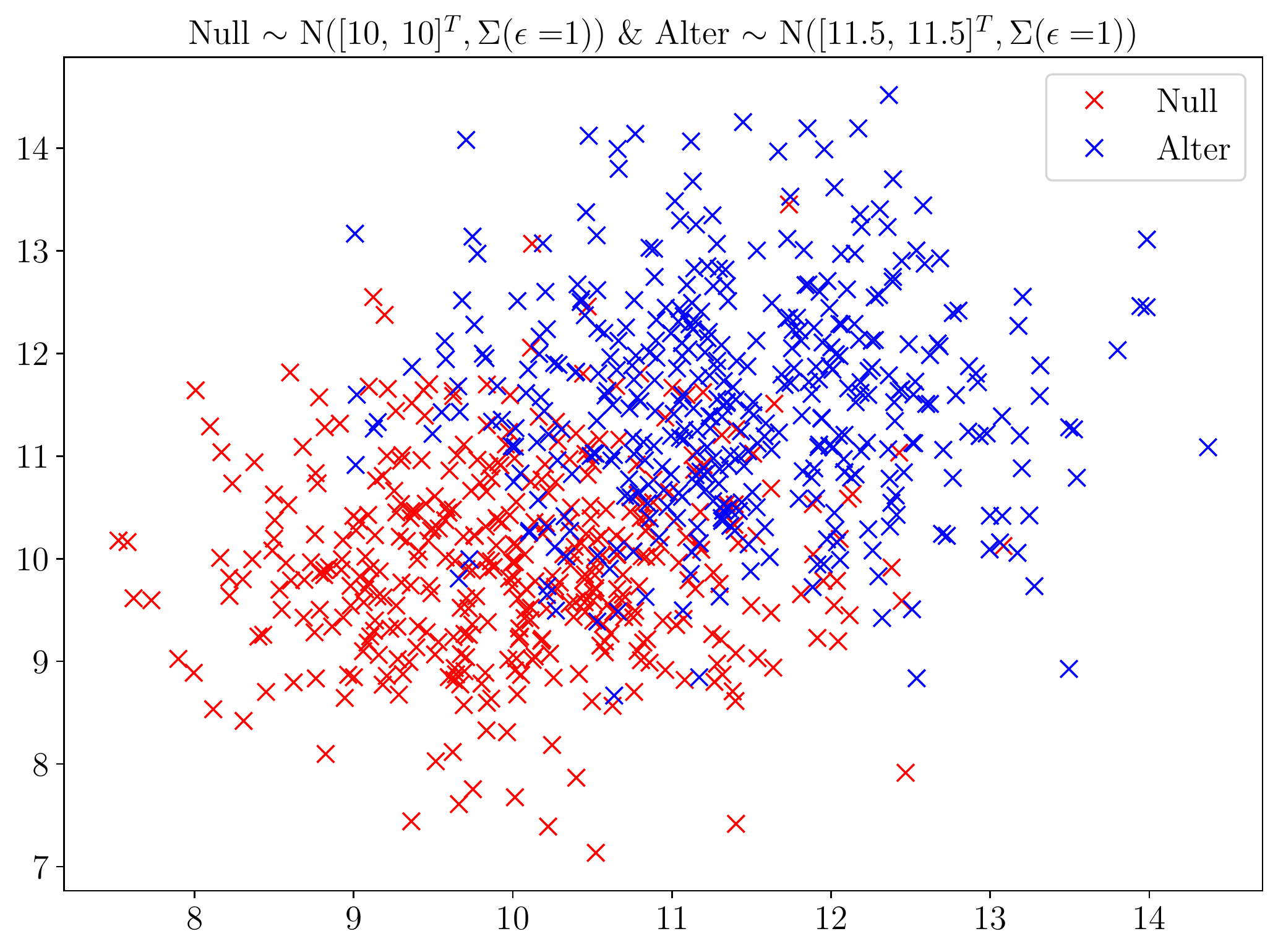}
    \end{subfigure}
    \begin{subfigure}{}
        \includegraphics[width=\textwidth]{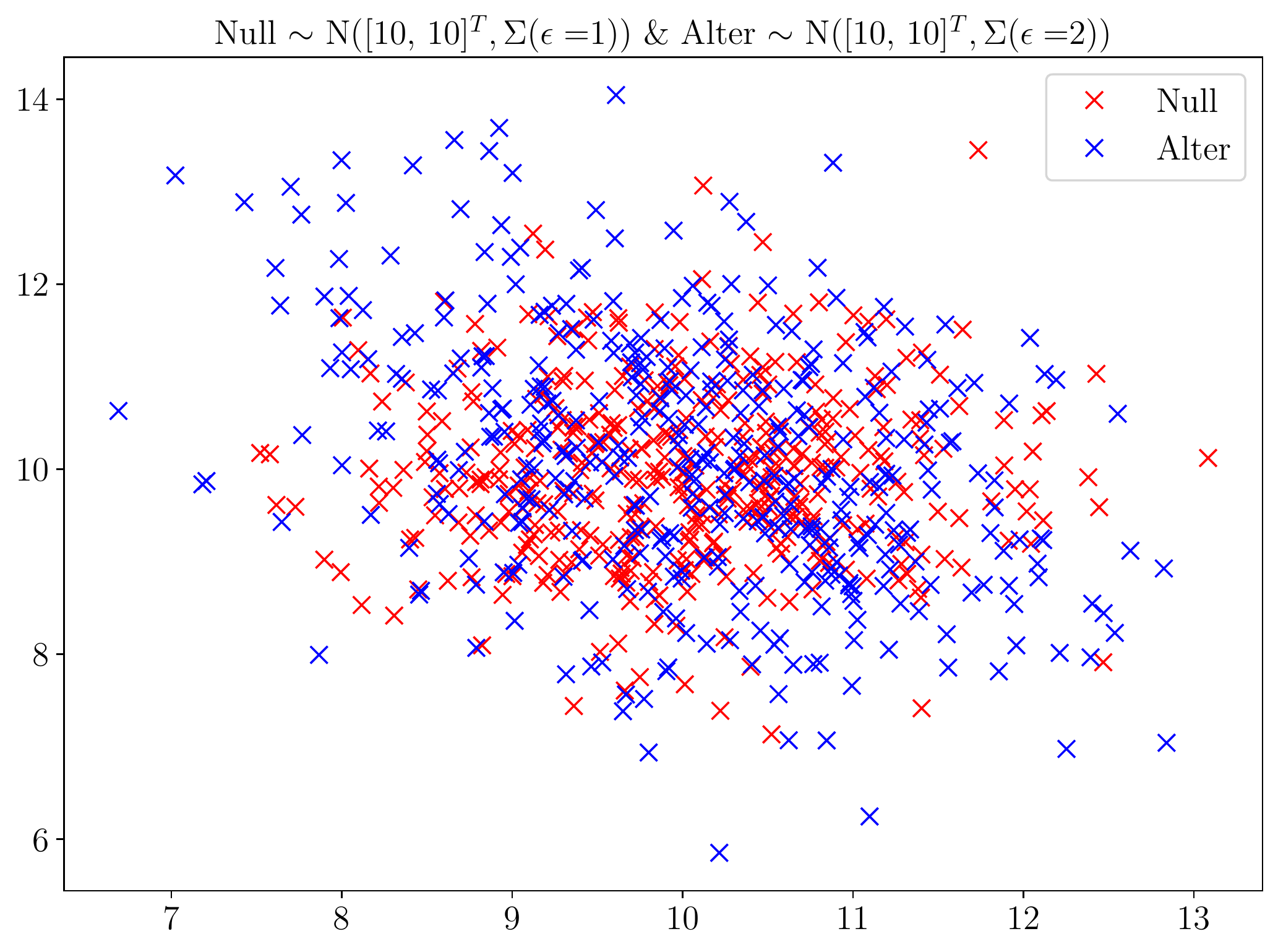}
    \end{subfigure}
    \begin{subfigure}{}
        \includegraphics[width=\textwidth]{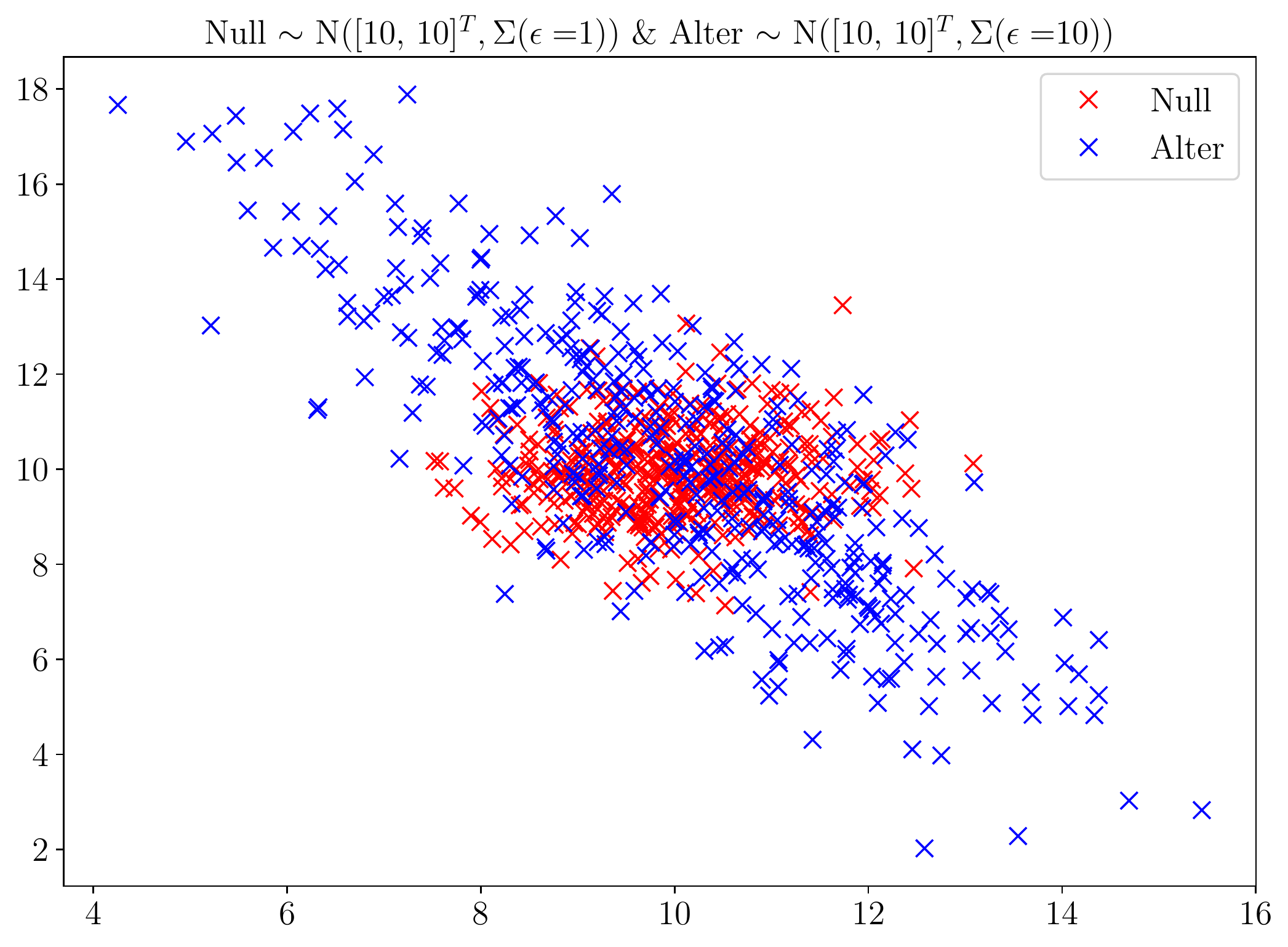}
    \end{subfigure}
\end{minipage}\hfill
\begin{minipage}{0.45\textwidth}
    \centering
    \begin{subfigure}{}
        \includegraphics[width=\textwidth]{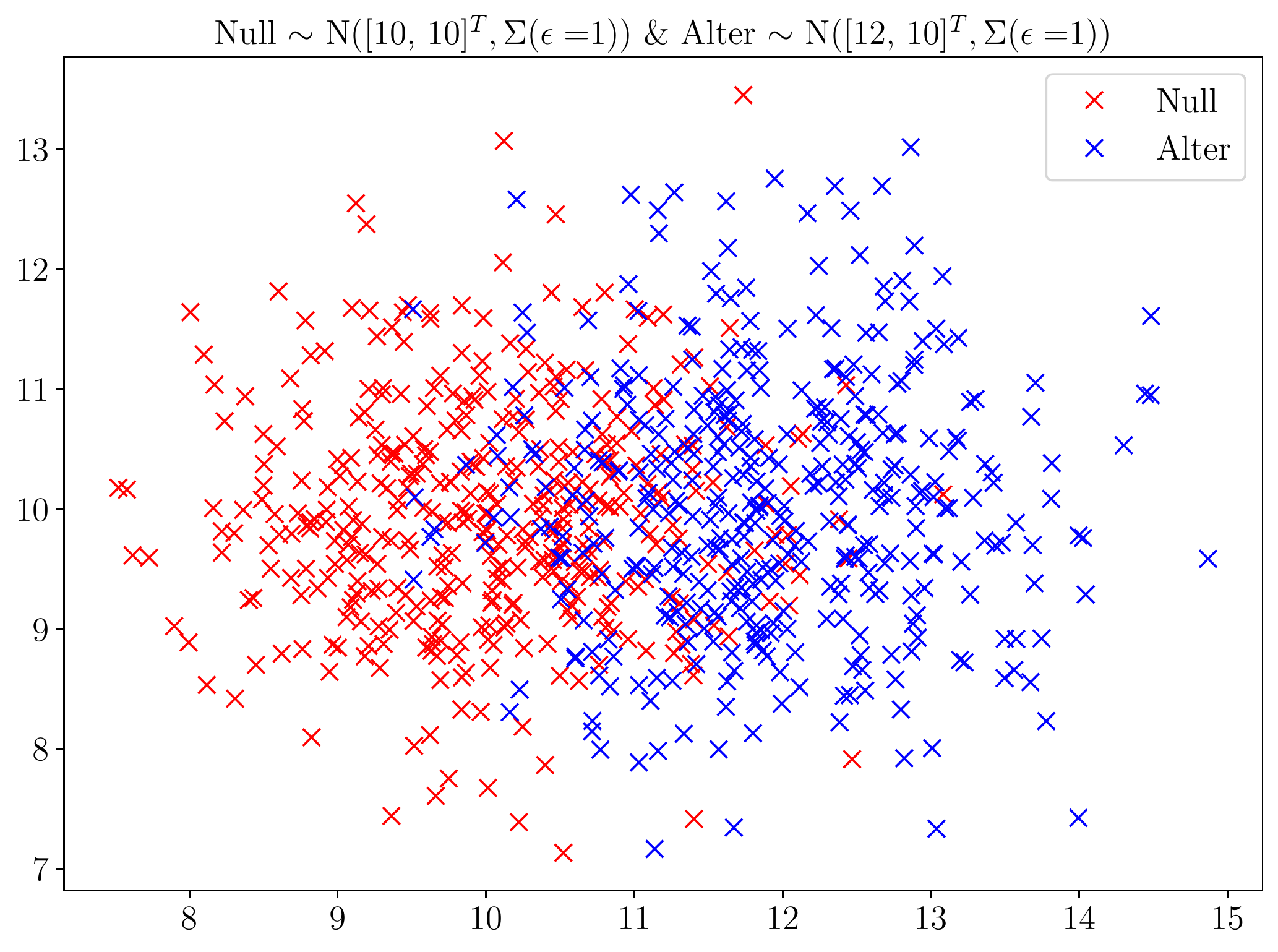}
    \end{subfigure}
    \begin{subfigure}{}
        \includegraphics[width=\textwidth]{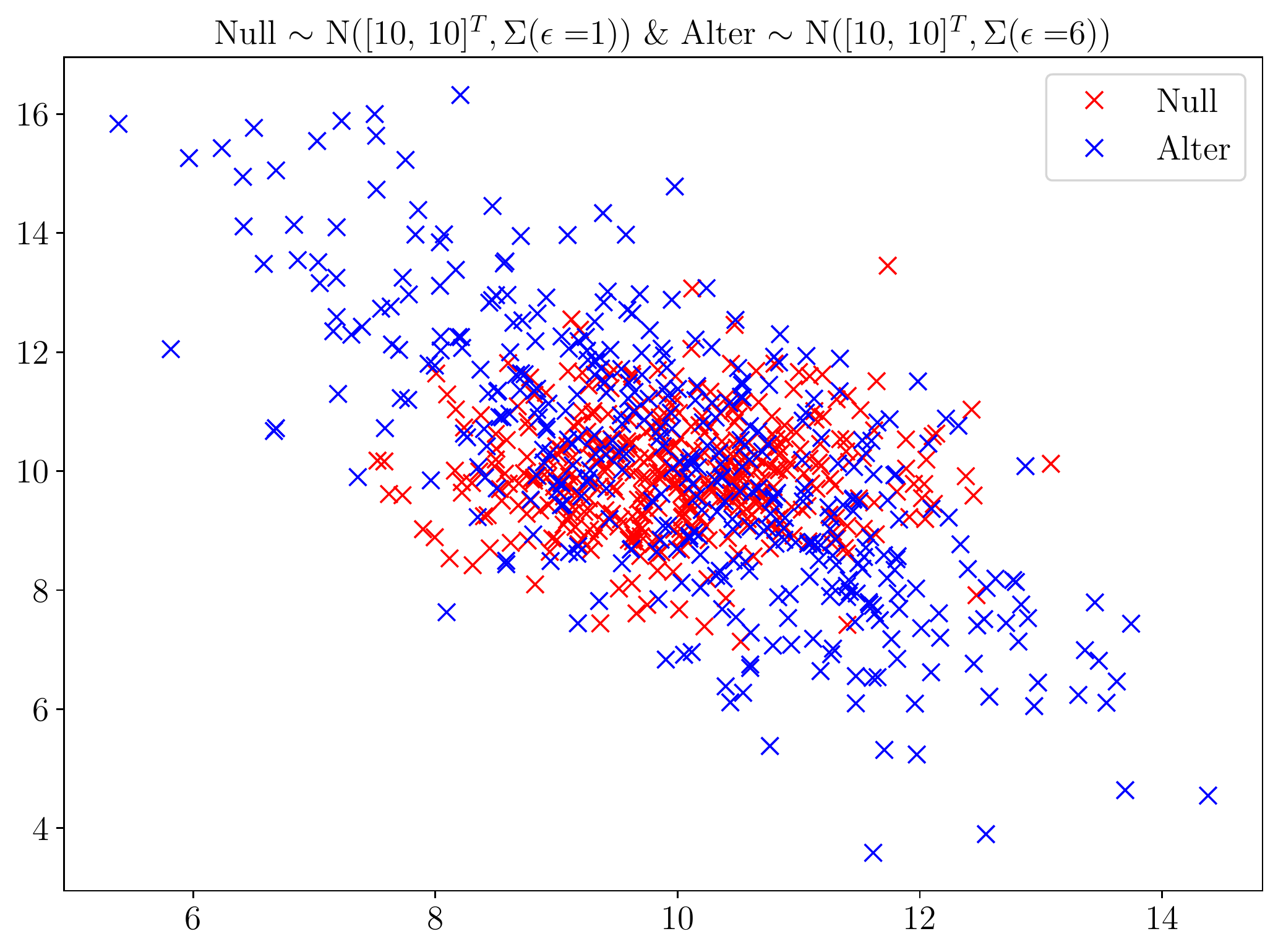}
    \end{subfigure}
    \begin{subfigure}{}
        \includegraphics[width=\textwidth]{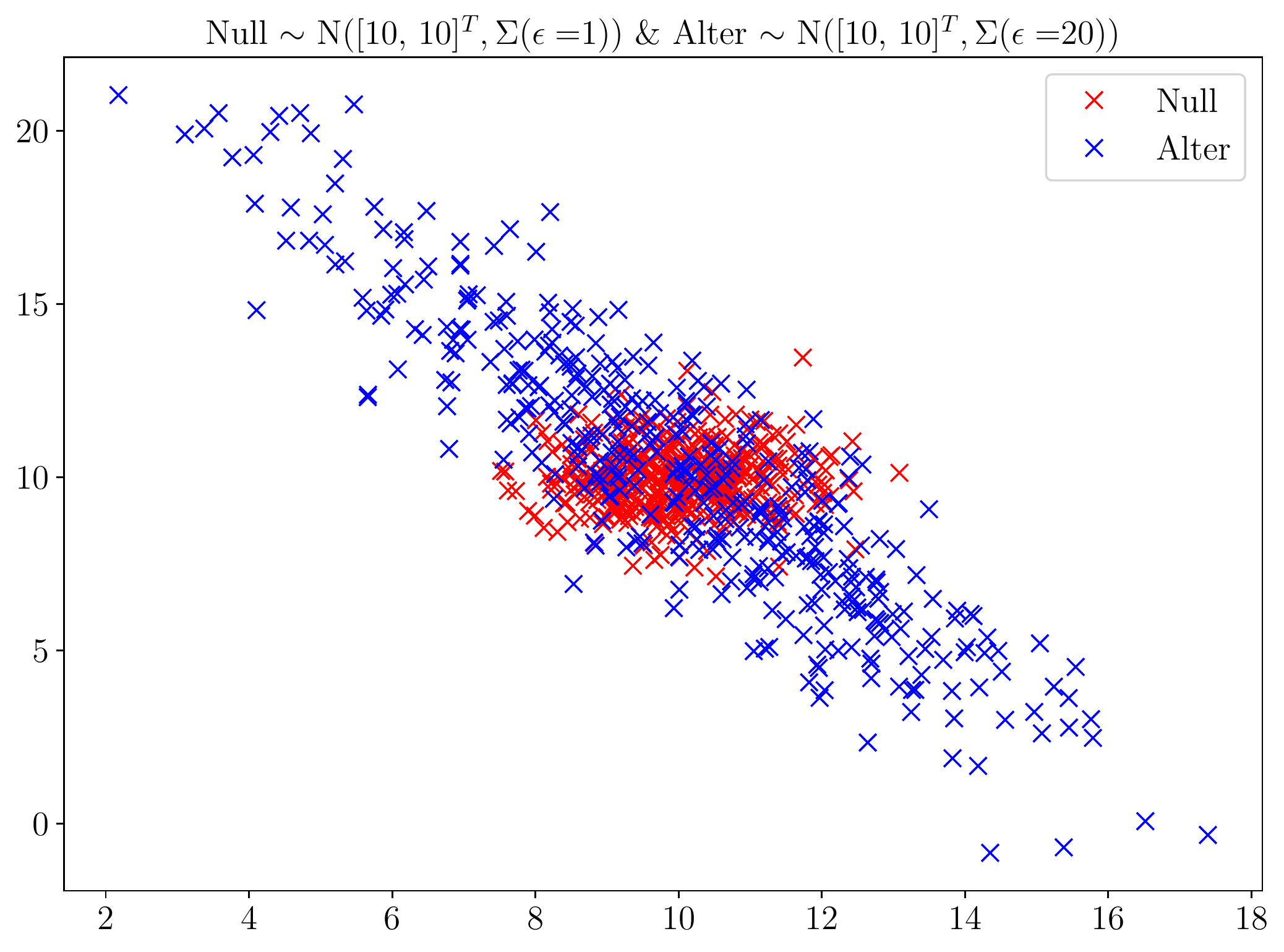}
    \end{subfigure}
\end{minipage}
\caption{Visualisation of 2-dimensional Gaussian distributions discussed in the main text with 400 samples each.} \label{fig6: visual_2DGaussian}
\end{figure}

\begin{figure}
\begin{minipage}{0.45\textwidth}
    \centering
    \begin{subfigure}{}
        \includegraphics[width=\textwidth]{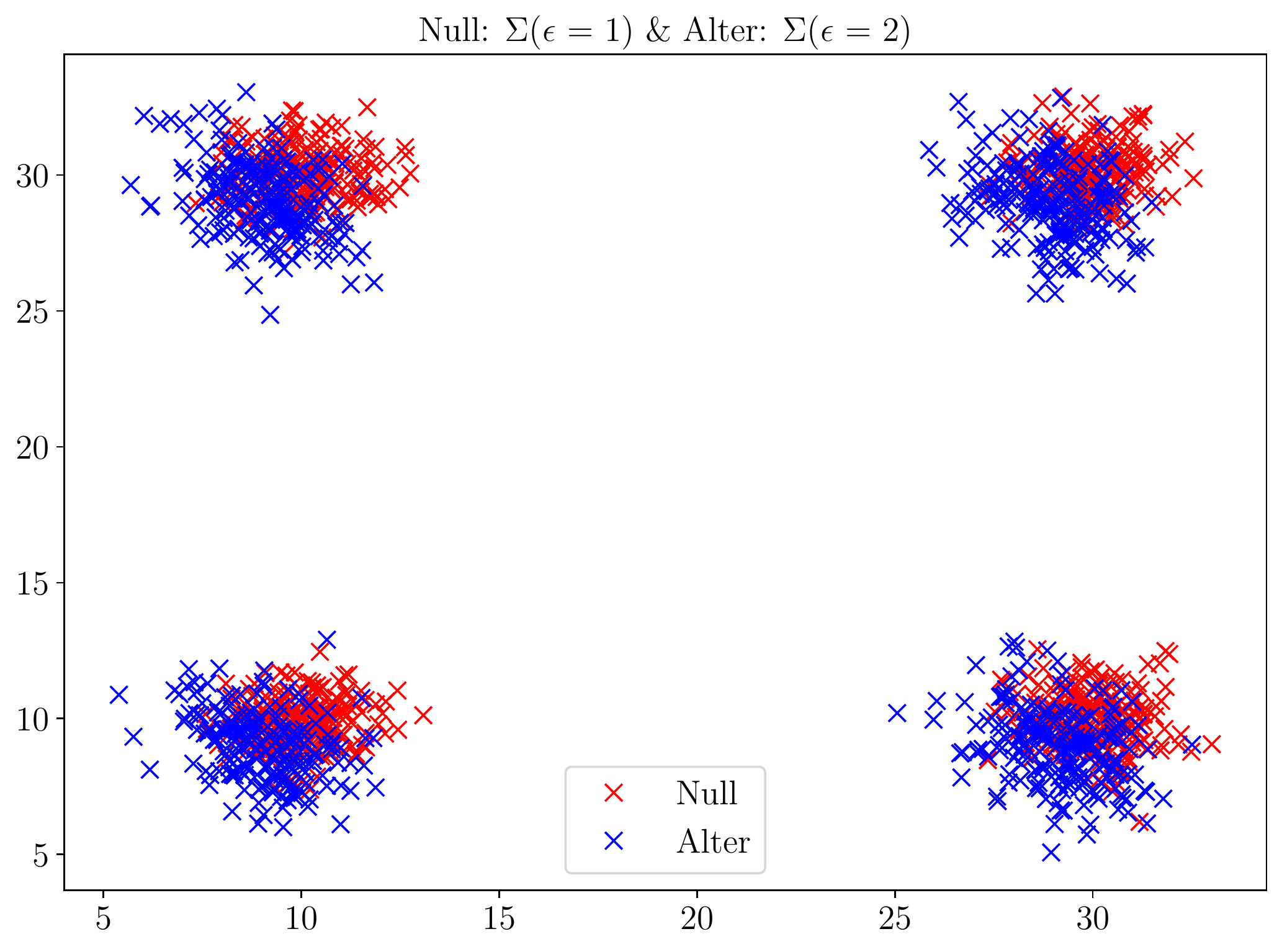}
    \end{subfigure}
    \begin{subfigure}{}
        \includegraphics[width=\textwidth]{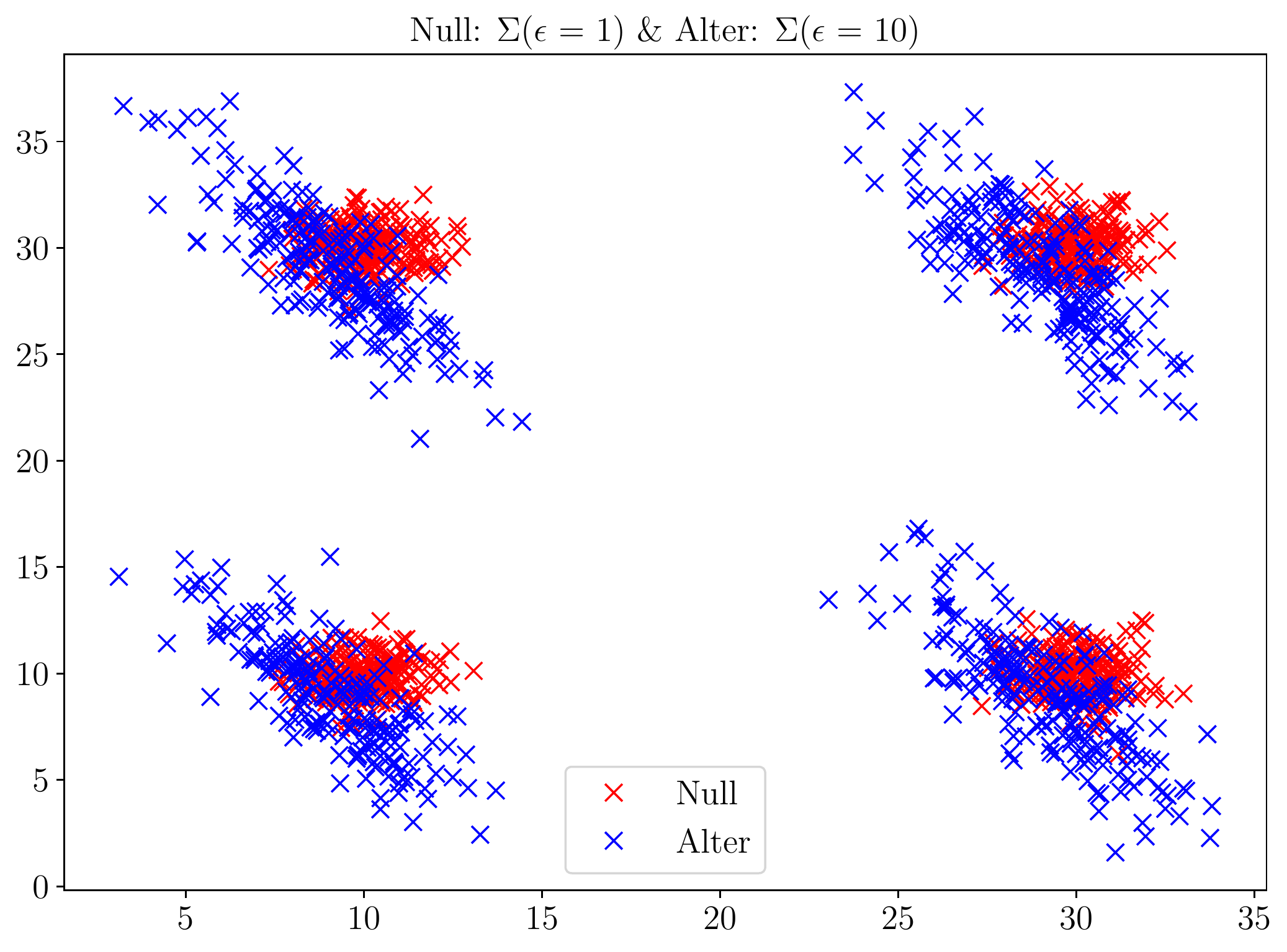}
    \end{subfigure}
\end{minipage}\hfill
\begin{minipage}{0.45\textwidth}
    \centering
    \begin{subfigure}{}
        \includegraphics[width=\textwidth]{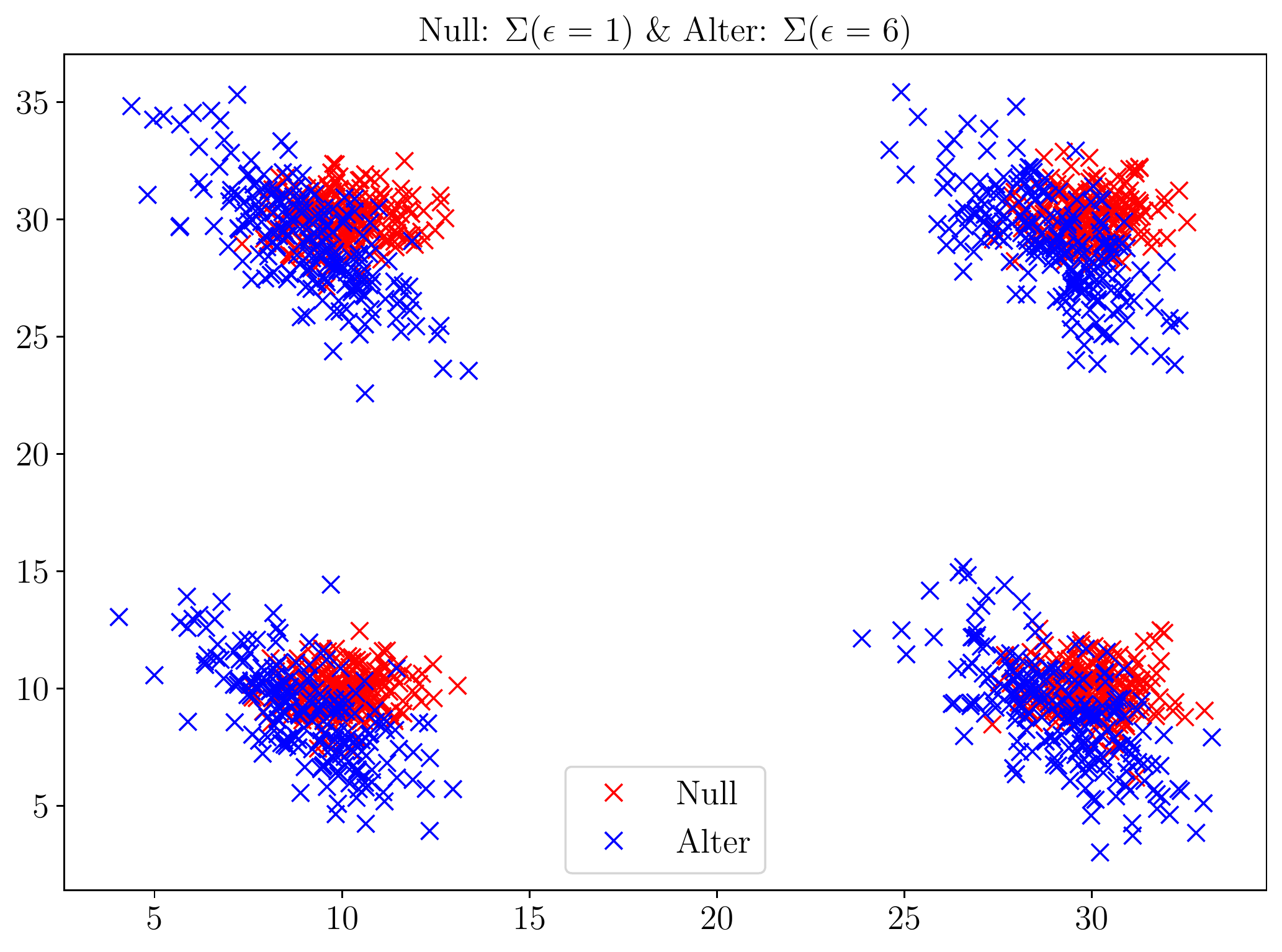}
    \end{subfigure}
    \begin{subfigure}{}
        \includegraphics[width=\textwidth]{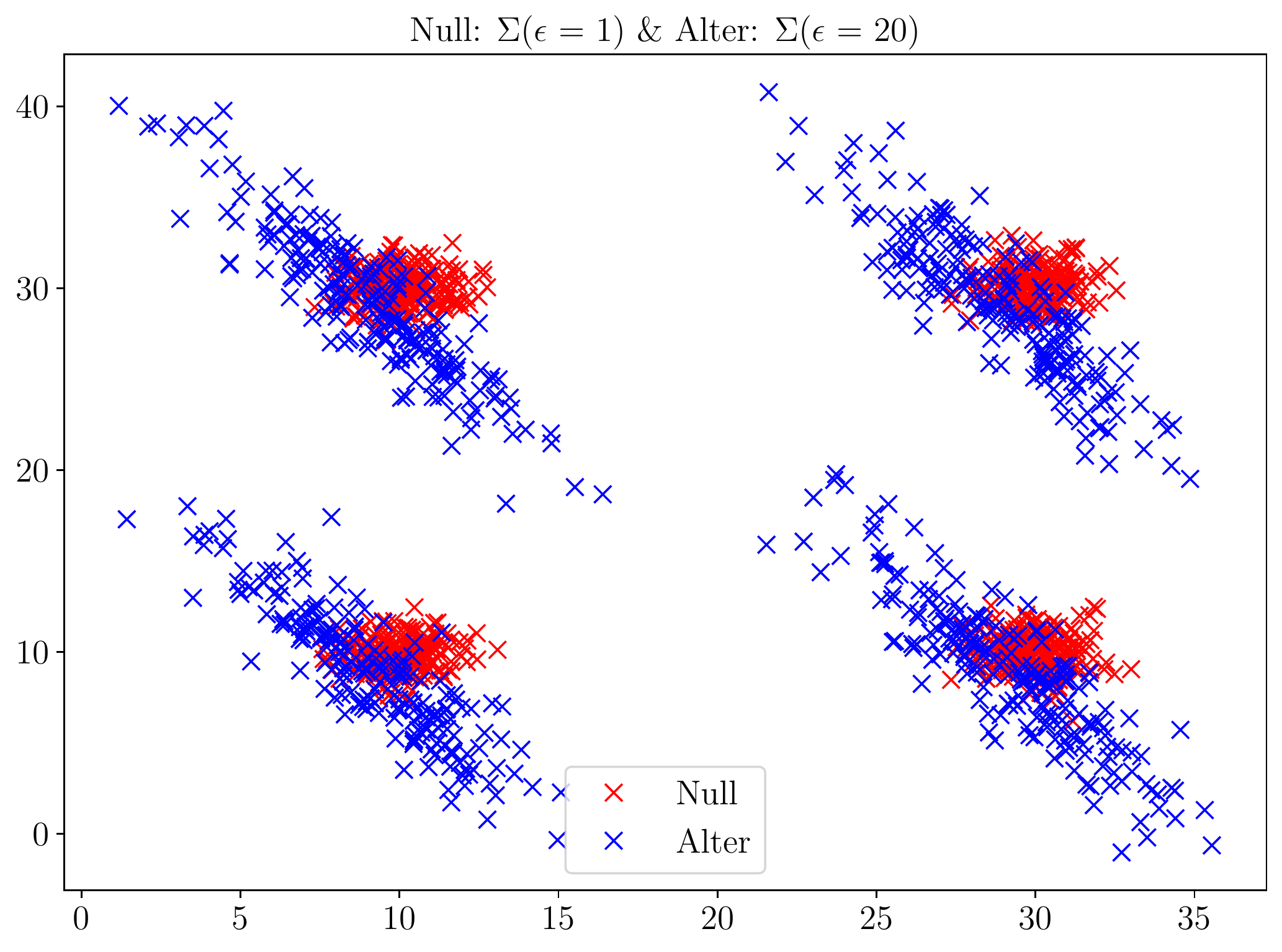}
    \end{subfigure}
\end{minipage}
\caption{Visualisation of 2 by 2 Blobs of 2-dimensional Gaussian distributions discussed in the main text with 200 samples in each blob.} \label{fig6: visual_2by2Blobs}
\end{figure}


\section{Additional Figures}
\begin{figure}[h]
\begin{minipage}{0.45\textwidth}
    \centering
    \begin{subfigure}{}
        \includegraphics[width=\textwidth]{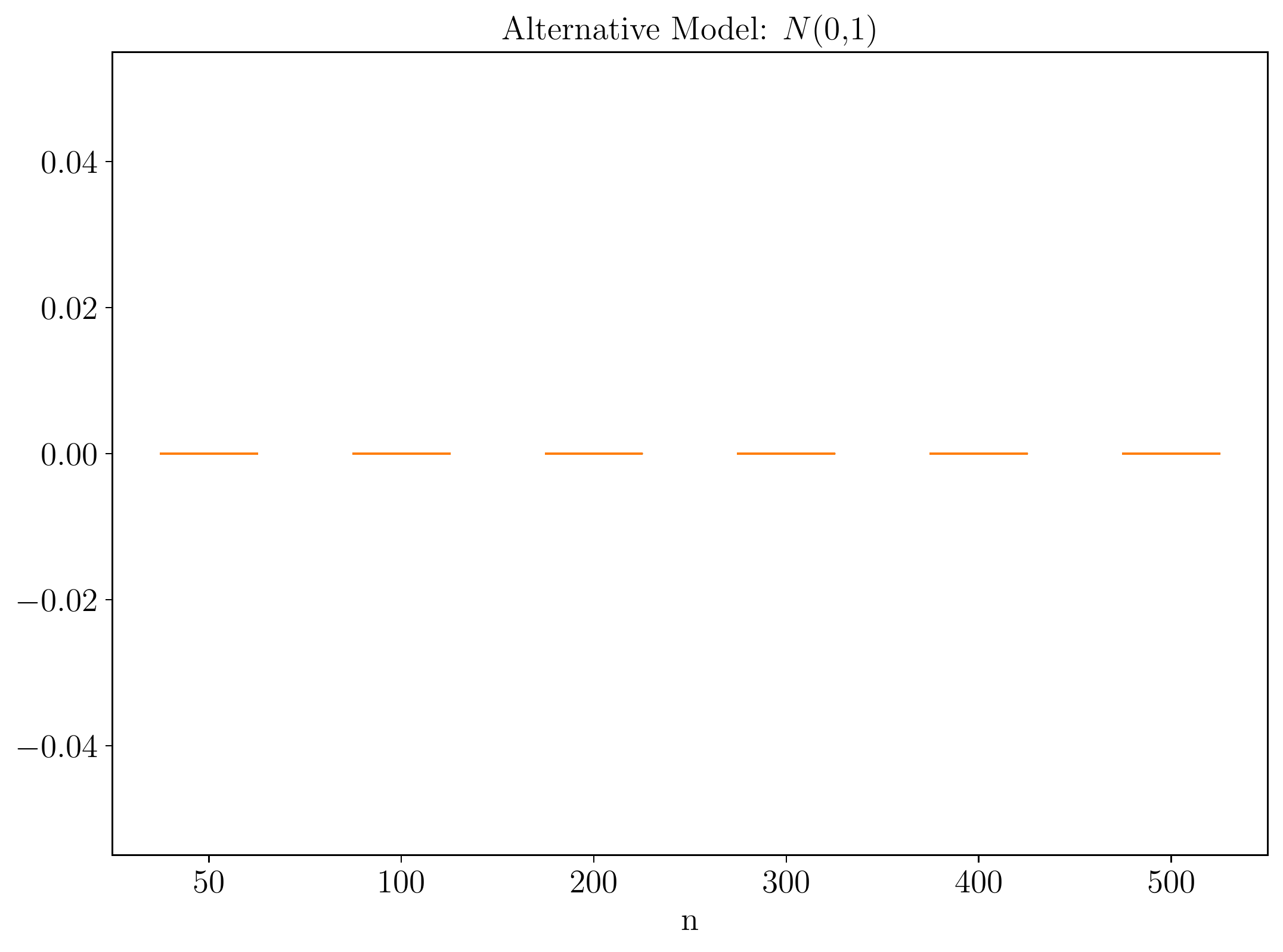}
    \end{subfigure}
    \begin{subfigure}{}
        \includegraphics[width=\textwidth]{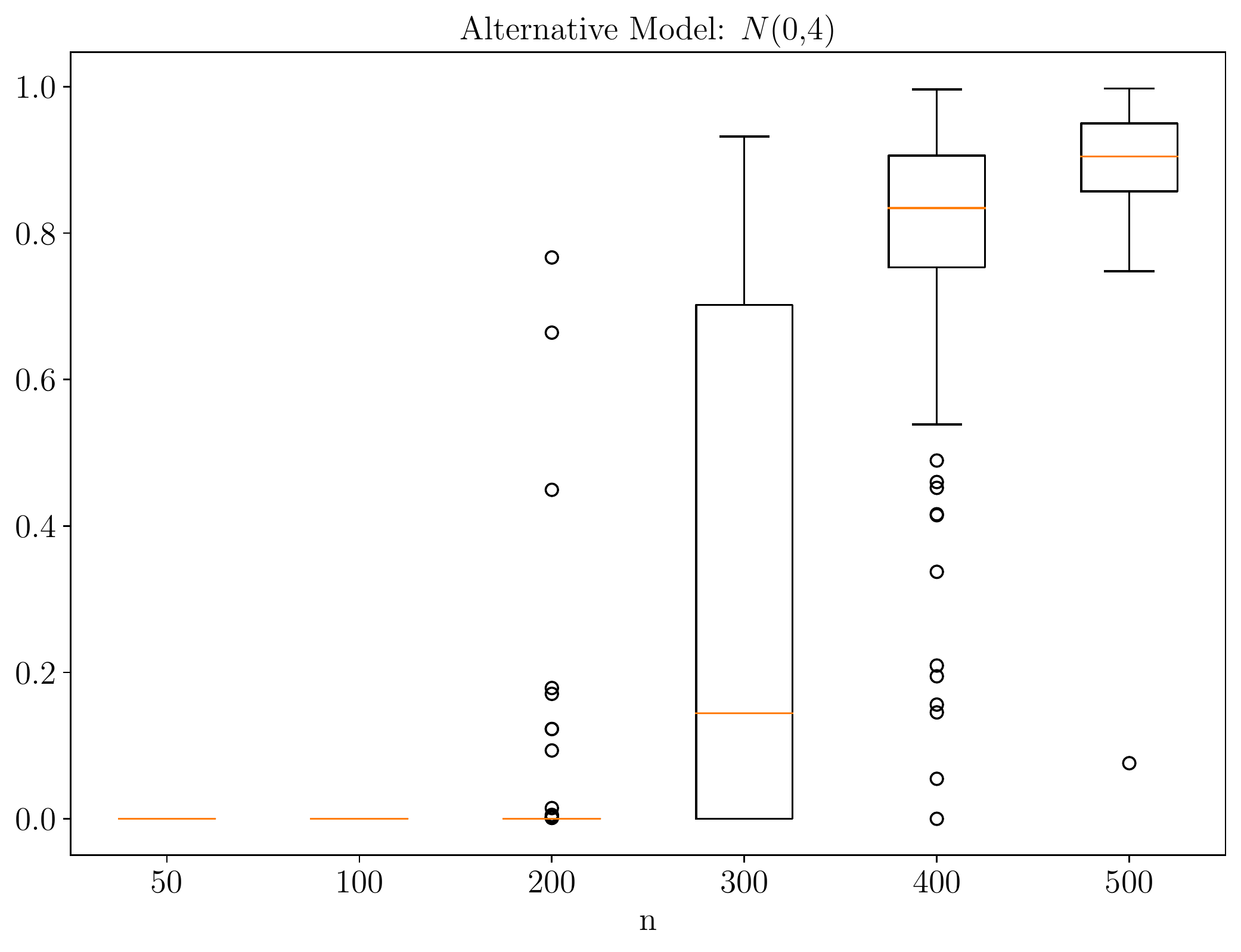}
    \end{subfigure}
    
\end{minipage}\hfill
\begin{minipage}{0.45\textwidth}
    \centering
    \begin{subfigure}{}
        \includegraphics[width=\textwidth]{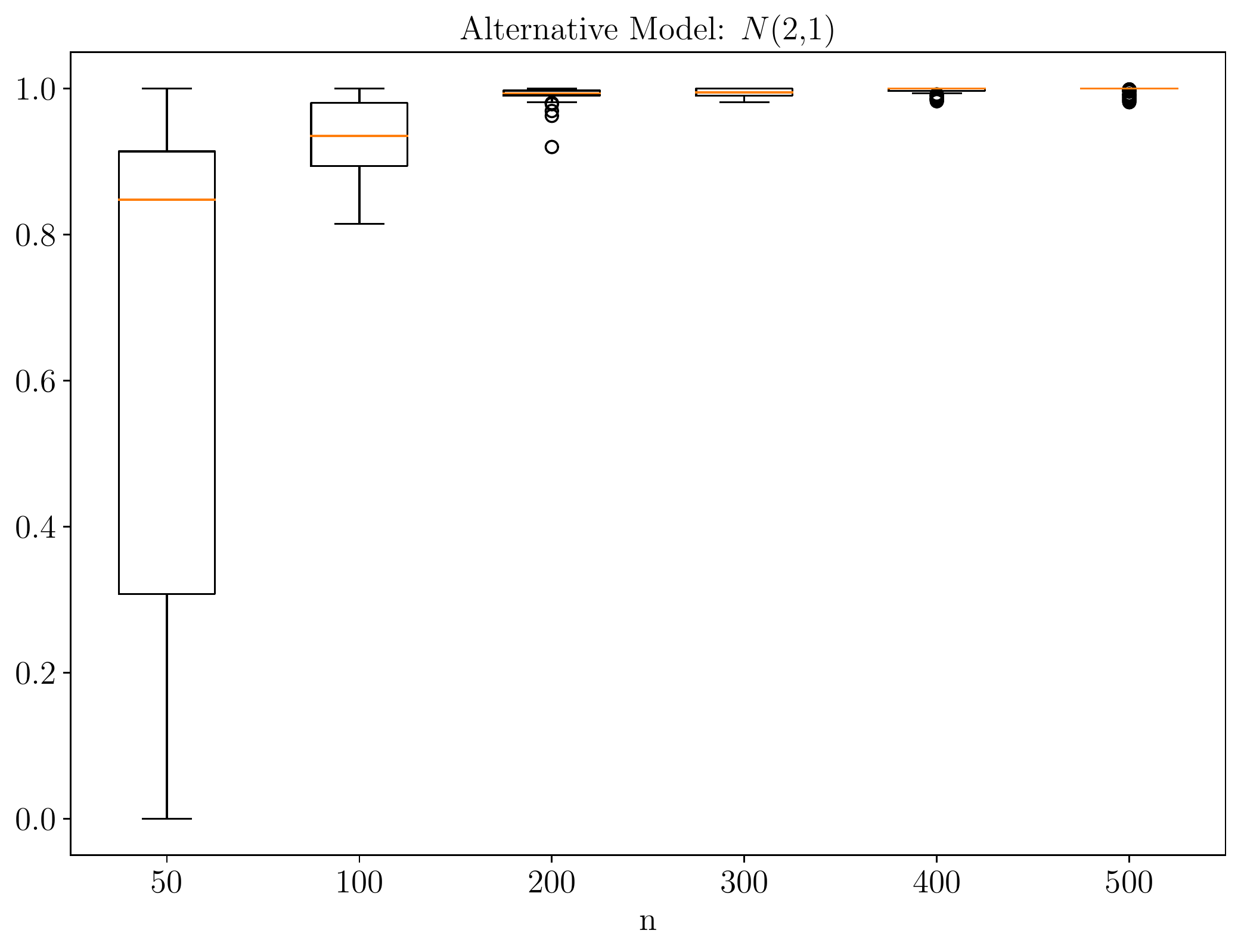}
    \end{subfigure}
    \begin{subfigure}{}
        \includegraphics[width=\textwidth]{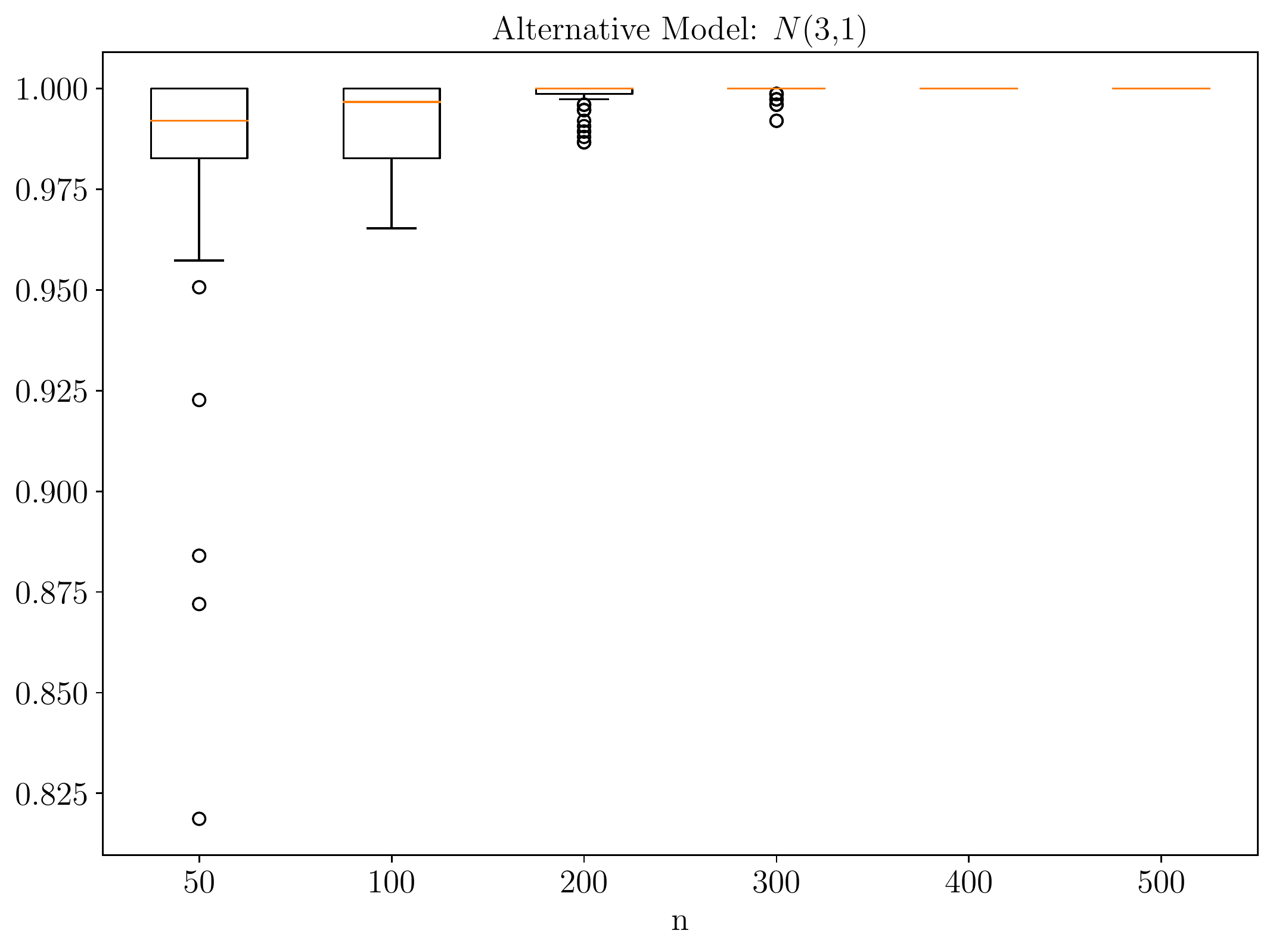}
    \end{subfigure}
\end{minipage}
\caption{1-dimensional Gaussian experiment: distribution (over 100 independent runs) of the probability of the alternative hypothesis $p(M=1|\mathcal{D})$ for a different number of observations $n$. The null hypothesis is both samples $X$ and $Y$ are i.i.d. $\mathcal{N}(0,1)$ (Top Left) and the alternative hypothesis is $Y\iid \mathcal{N}(2, 1)$ (Top Right), $Y \iid \mathcal{N}(0,4)$ (Bottom Left) and $Y \iid \mathcal{N}(3,1)$ (Bottom Right). As expected, the proposed method is able to detect the difference between the given samples. }
\end{figure}

\begin{figure}[h]
\begin{minipage}{0.45\textwidth}
    \centering
    \begin{subfigure}{}
        \includegraphics[width=\textwidth]{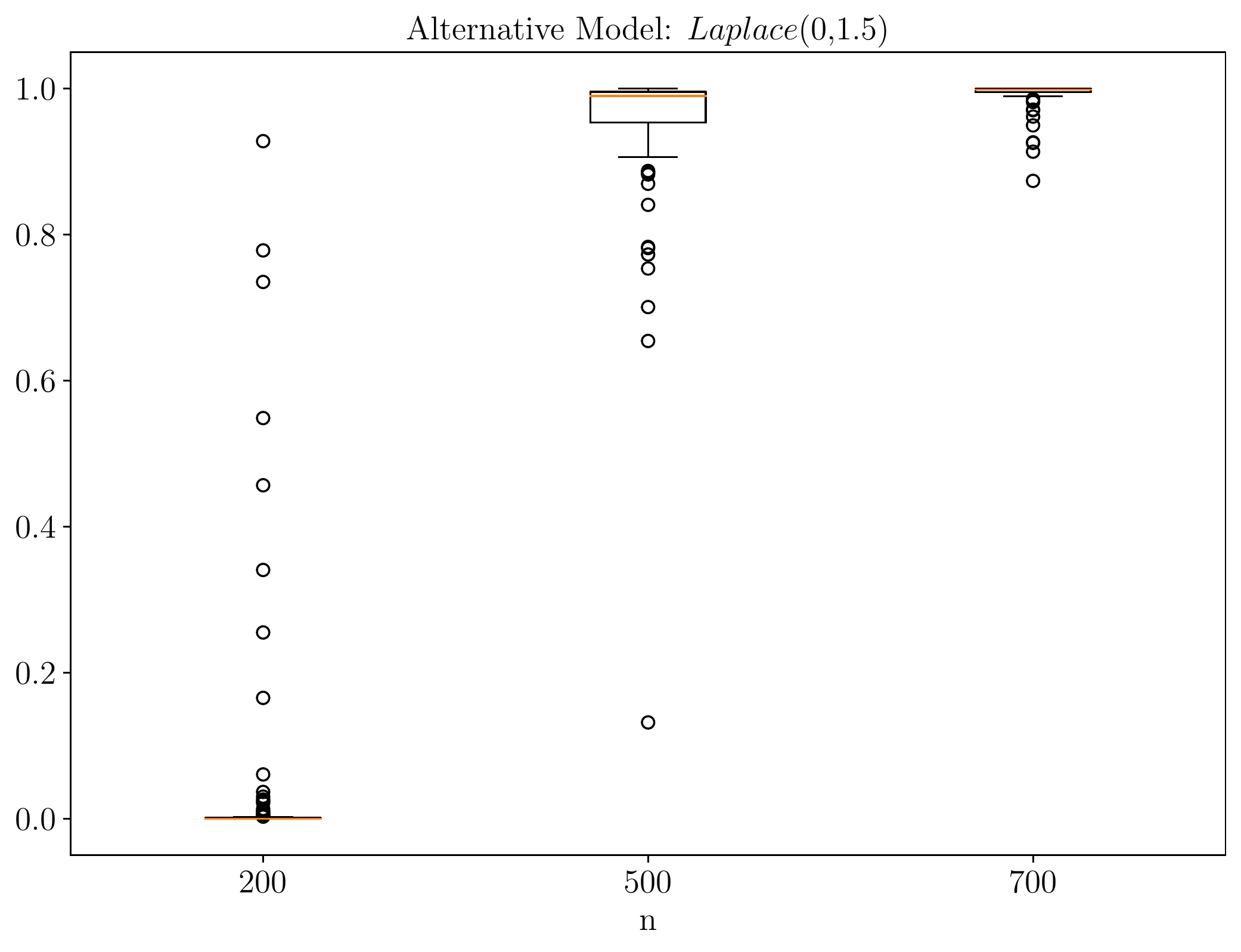}
    \end{subfigure}
\end{minipage}\hfill
\begin{minipage}{0.43\textwidth}
    \centering
    \begin{subfigure}{}
        \includegraphics[width=\textwidth]{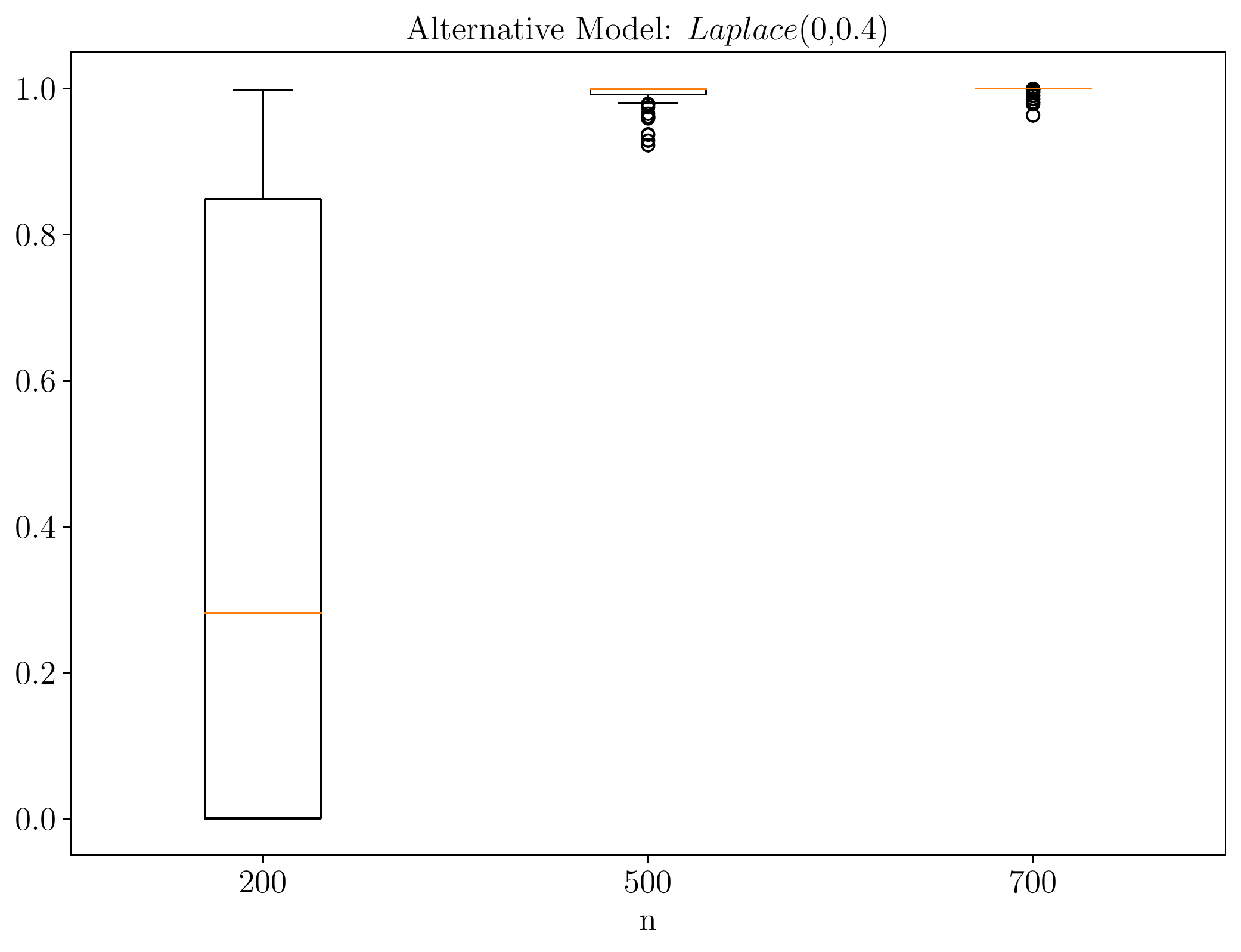}
    \end{subfigure}
\end{minipage}
\caption{1-dimensional experiment comparing correlated Gaussian and Laplace distributions: distribution (over 100 independent runs) of the probability of the alternative hypothesis $M=1|\mathcal{D}$ for a different number of samples $n$. (Left) $X\iid \mathcal{N}(0,1)$ and $Y\iid Laplace(0, 1.5)$. (Right) $X\iid \mathcal{N}(0,1)$ and $Y\iid Laplace(0, 0.4).$ The random variables $X$ and $Y$ are correlated with correlation 0.5. }
\label{fig6: Laplace_corr}
\end{figure}


\begin{figure}
\begin{minipage}{0.45\textwidth}
    \centering
    \begin{subfigure}{}
        \includegraphics[width=\textwidth]{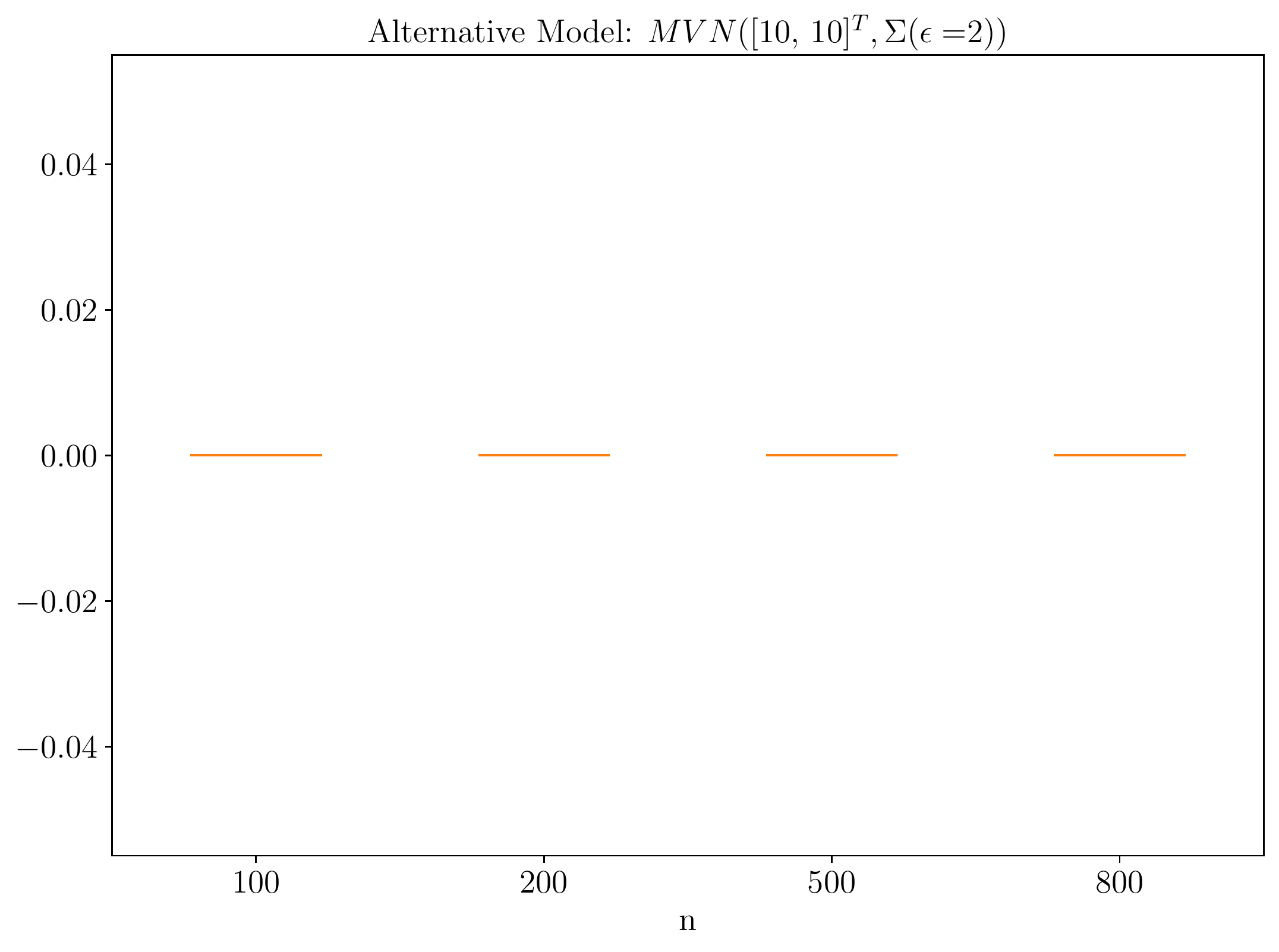}
    \end{subfigure}
    \begin{subfigure}{}
        \includegraphics[width=\textwidth]{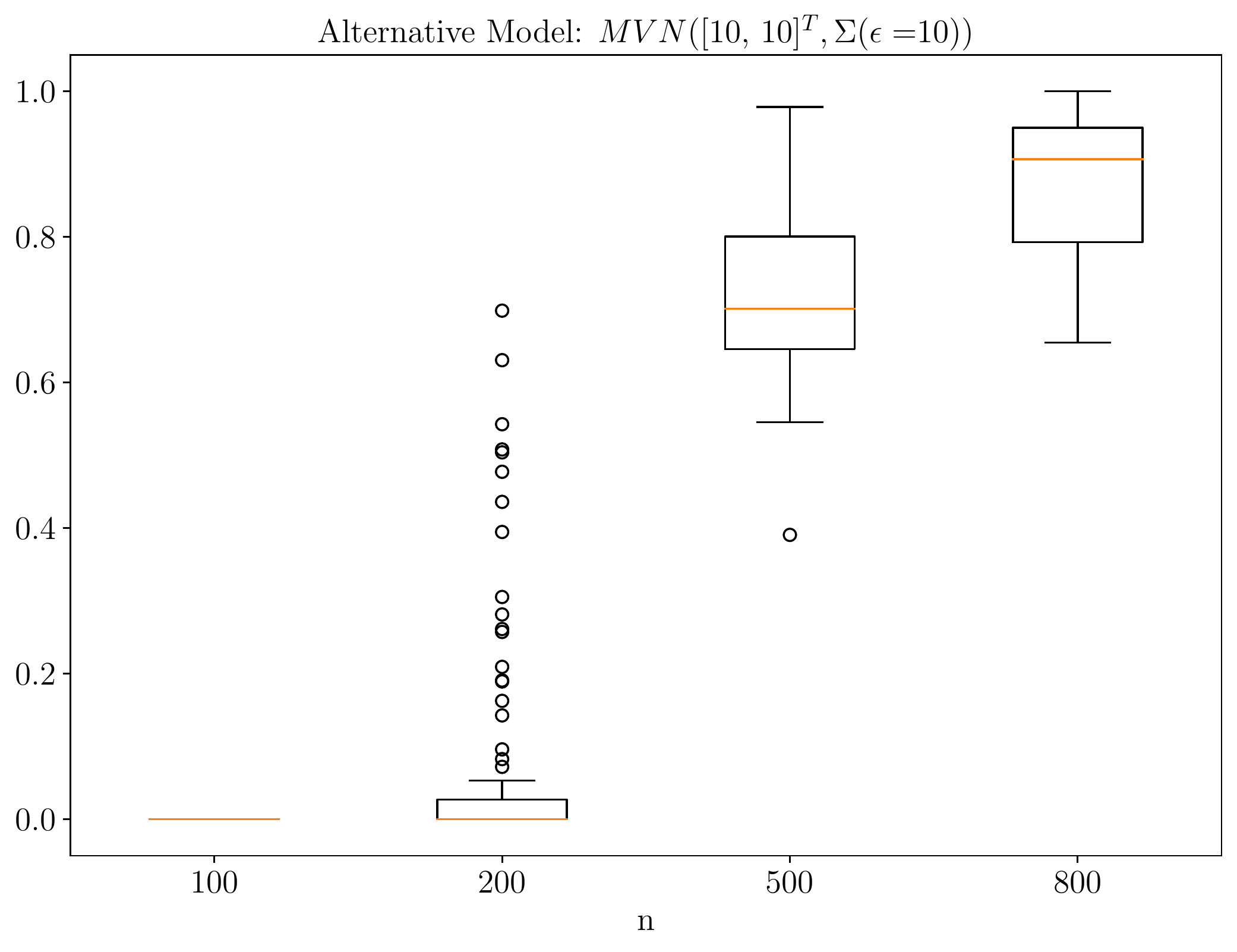}
    \end{subfigure}
\end{minipage}\hfill
\begin{minipage}{0.45\textwidth}
    \centering
    \begin{subfigure}{}
        \includegraphics[width=\textwidth]{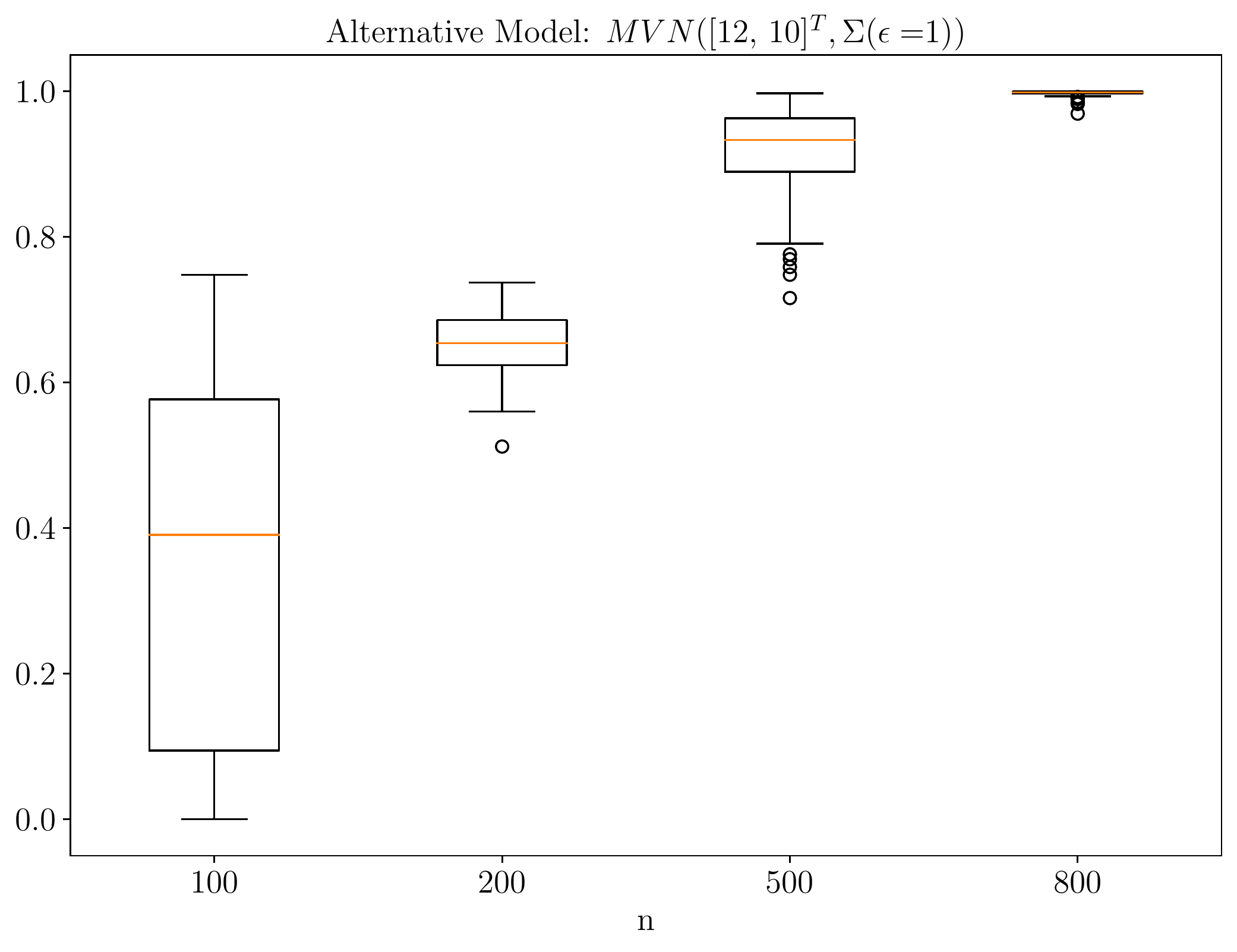}
    \end{subfigure}
    \begin{subfigure}{}
        \includegraphics[width=\textwidth]{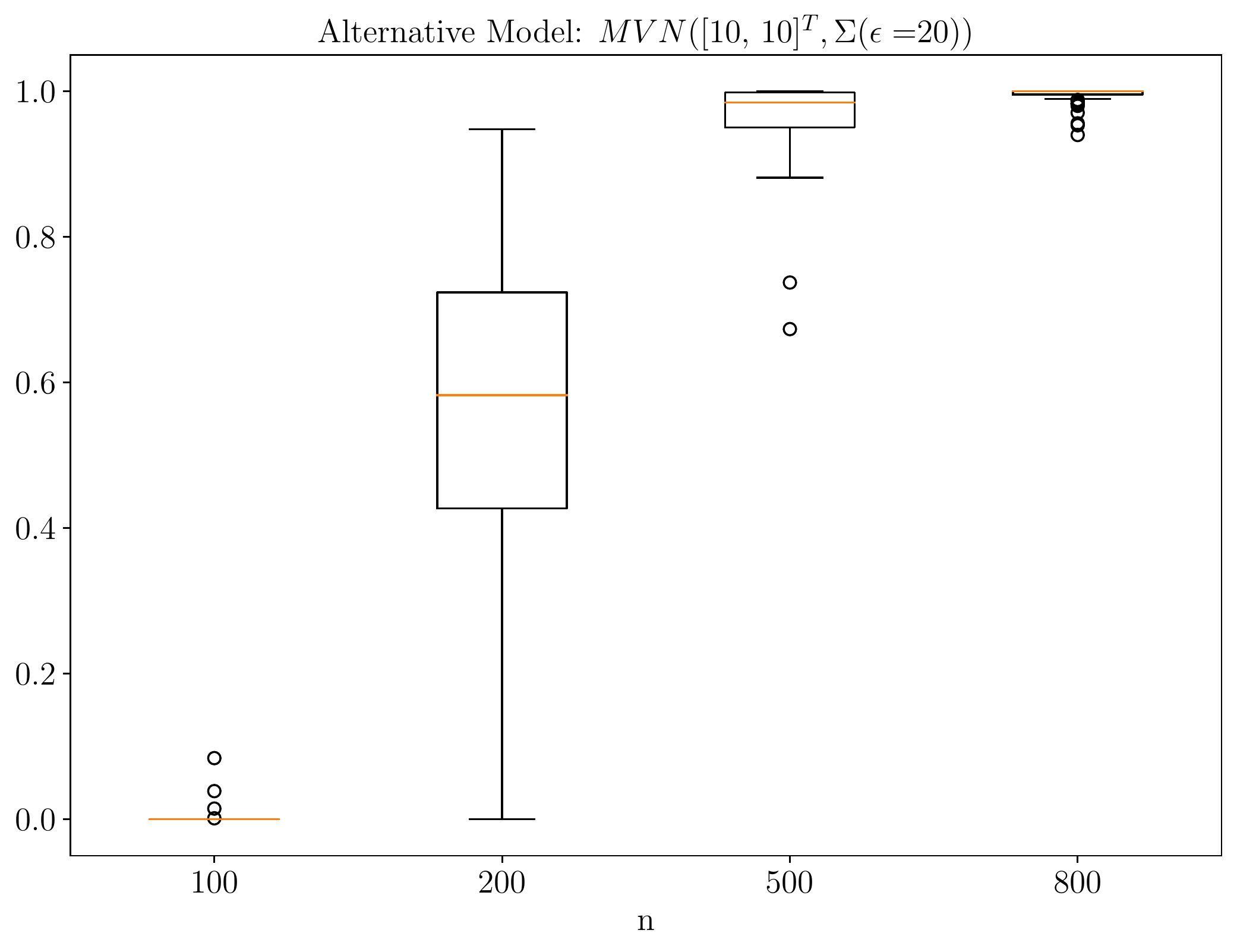}
    \end{subfigure}
\end{minipage}
\caption{2-dimensional Gaussian experiment: distribution (over 100 independent runs) of the probability of the alternative hypothesis $p(M=1|\mathcal{D})$ for a different number of observations $n$. The null hypothesis is that both random variables follows a bivariate Gaussian distribution centered at $[10,10]^\top$ with identity covariance matrix. Under the alternative hypothesis, the distribution of $Y$ stated on top of each plot.} 
\end{figure}

\end{appendix}


\end{document}